\documentclass[11pt]{article}

\usepackage[left=25mm,right=25mm,top=25mm,bottom=25mm]{geometry}
\usepackage[utf8]{inputenc}

\usepackage{multirow}
\usepackage{booktabs}
\usepackage{colortbl}
\usepackage{tabularx}
\usepackage{array}

\usepackage{amssymb}
\usepackage{amsmath}
\usepackage{amsthm}
\usepackage{nicefrac}
\usepackage{bbm}
\theoremstyle{definition}   
\usepackage{chngcntr}
\usepackage{apptools}
\usepackage{amsfonts}
\usepackage{amsbsy}
\usepackage{mathrsfs}
\usepackage{stmaryrd}

\usepackage[usenames,dvipsnames,svgnames]{xcolor}
\usepackage{tikz}
\usepackage{pgfplots}
\usepgfplotslibrary{groupplots}
\usepackage{pgfplotstable}
\usepackage{subcaption}
\usepackage[font=small]{caption}
\usepackage{wrapfig}
\usepackage{tikzscale}
\pgfplotsset{compat=1.13}
\usepgfplotslibrary{fillbetween}
\usetikzlibrary{patterns}
\usepgfplotslibrary{colormaps}

\graphicspath{{fig/}{fig/Numerical_Examples/}}

\usetikzlibrary{shapes,arrows,matrix,positioning,arrows.meta}

\tikzset{>=latex} 

\usepackage{tcolorbox}
\tcbuselibrary{skins}
\usepackage{color}

\usepackage[maxbibnames=20, maxcitenames=2, 
    natbib=true, style=numeric,
    isbn=false, url=false,giveninits=true,sortcites,date=year]{biblatex}   
\bibliography{bibliography_dk}
\bibliography{biblio}

\usepackage[german, main=english]{babel}
\usepackage[autostyle]{csquotes}
\usepackage[T1]{fontenc}
\usepackage{hyperref}
\usepackage[affil-it]{authblk}
\usepackage[capitalise]{cleveref}

\crefname{section}{Sect.}{Sect.}

\newcommand{\Cross}{\mathbin{\tikz [x=1.4ex,y=1.4ex,line width=.25ex] \draw (0.1,0.1) -- (0.9,0.9) (0.1,0.9) -- (0.9,0.1);}}

\newcommand{\vect}[1]{\boldsymbol{#1}}

\newcommand{\norm}[1]{\left\lVert#1\right\rVert}

\newcommand{\RL}{\mathcal{RL}}
\newcommand{\SP}{\mathcal{SP}}

\newcommand{\GL}{\text{GL}}
\newcommand{\SO}{\text{SO}}

\newcommand{\bnabla}{\vect{\nabla}}

\newcommand{\tr}{\ensuremath{\operatorname{tr}}}
\newcommand{\cof}{\ensuremath{\operatorname{cof}}}

\newcommand{\cI}{{\mathcal{I}}}
\newcommand{\bcI}{{\vect{\mathcal{I}}}}

\newcommand{\bbR}{{\mathbb{R}}}

\newcommand{\ba}{\boldsymbol{a}}
\newcommand{\bb}{\boldsymbol{b}}
\newcommand{\bd}{\boldsymbol{d}}

\newcommand{\bA}{\boldsymbol{A}}
\newcommand{\bB}{\boldsymbol{B}}

\newcommand{\bF}{\boldsymbol{F}}

\newcommand{\bH}{\boldsymbol{H}}
\newcommand{\bI}{\boldsymbol{I}}

\newcommand{\bX}{\boldsymbol{X}}

\newcommand{\IoneTI}{I_1^{\text{ti}}}
\newcommand{\ItwoTI}{I_2^{\text{ti}}}
\newcommand{\IthreeTI}{I_3^{\text{ti}}}
\newcommand{\IfourTI}{I_4^{\text{ti}}}
\newcommand{\IfiveTI}{I_5^{\text{ti}}}
\newcommand{\IsixTI}{I_6^{\text{ti}}}

\newcommand{\structTI}{\vect{G}^{\text{ti}}}

\newcommand{\ennISO}{e^{\text{NN}}}

\definecolor{CPSgreen}{RGB}{22,164,138}
\definecolor{CPSlightblue}{RGB}{104,143,198}
\definecolor{CPSdarkblue}{RGB}{67,83,132}
\definecolor{CPSgrey}{RGB}{204, 204, 204}
\definecolor{CPSorange}{RGB}{246,163,21}
\definecolor{CPSred}{RGB}{194,76,76}

\definecolor{blueR}{RGB}{0,114,189}
\definecolor{redR}{RGB}{206,59,20}

\definecolor{1_blue}{RGB}{0,119,187}
\definecolor{1_cyan}{RGB}{51,187,238}
\definecolor{1_teal}{RGB}{0,153,136}
\definecolor{1_orange}{RGB}{238,119,51}
\definecolor{1_red}{RGB}{204,51,17}
\definecolor{1_magenta}{RGB}{238,51,119}
\definecolor{1_grey}{RGB}{187,187,187}

\definecolor{color11a}{RGB}{236, 188, 0}
\definecolor{color12a}{RGB}{50, 67, 121}
\definecolor{color13}{RGB}{222, 222, 222}
\definecolor{color21a}{RGB}{92, 134, 196}
\definecolor{color22a}{RGB}{249, 156, 0}
\definecolor{color33a}{RGB}{191, 60, 60}
\definecolor{colorCPSa}{RGB}{0, 157, 129}

\colorlet{2yellow}{color11a!90!color13}
\colorlet{2darkblue}{color12a!90!color13}
\colorlet{2lightblue}{color21a!90!color13}
\colorlet{2orange}{color22a!90!color13}
\colorlet{2red}{color33a!90!color13}
\definecolor{2grey}{RGB}{222, 222, 222}
\colorlet{2green}{colorCPSa!90!color13}

\tikzstyle{node}=[thick,circle,draw=black,minimum size=22,inner sep=0.5,outer sep=0.6]
\tikzstyle{node icnn}=[color=2orange!10!black,node,draw=black,fill=2orange!25, text = black]
\tikzstyle{node in_out}=[node,2lightblue!10!black,draw=black,fill=2lightblue!25, text=black]
\tikzstyle{node inv_pot}=[rectangle, node,2red!10!black,draw=black,fill=2red!25, text=black]
\tikzstyle{connect}=[thick,black] 
\tikzstyle{connect arrow}=[-{Latex[length=4,width=3.5]},thick,black,shorten <=0.5,shorten >=1]

\pgfplotscreateplotcyclelist{colorlist_ROL_PE}{%
only marks,solid, mark options={solid,fill=1_blue,fill opacity=1,mark size=2.5pt},mark=halfsquare*,1_blue,mark repeat = 2, mark phase = 2\\
1_blue, very thick \\
only marks,solid, mark options={solid,fill=1_cyan,fill opacity=1,mark size=2.5pt},mark=halfsquare right*,1_cyan,mark repeat = 2, mark phase = 2\\
1_cyan, very thick\\
only marks,solid, mark options={solid,fill=1_teal,fill opacity=1,mark size=2.5pt},mark=halfsquare left*,1_teal,mark repeat = 2, mark phase = 2\\
1_teal, very thick \\
only marks,solid, mark options={solid,fill=1_red,fill opacity=1,mark size=2.5pt,rotate=-90},mark=halfcircle*,1_red,mark repeat = 2, mark phase = 2\\
1_red, very thick \\
only marks,solid, mark options={solid,fill=1_orange,fill opacity=1,mark size=2.5pt,rotate=90},mark=halfcircle*,1_orange,mark repeat = 2, mark phase = 2\\
1_orange, very thick \\
}

\pgfplotscreateplotcyclelist{colorlist_ROL_FE}{%
 thick, dashed,  mark options={solid,fill=1_red,fill opacity=1,mark size=2.5pt,rotate=-90},mark=halfcircle*,1_red,mark repeat = 2, mark phase = 2\\
 thick, dashed,  mark options={solid,fill=1_orange,fill opacity=1,mark size=2.5pt,rotate=90},mark=halfcircle*,1_orange,mark repeat = 2, mark phase = 2\\
 thick, dashed,  mark options={solid,fill=1_magenta,fill opacity=1,mark size=2.5pt,rotate=-180 },mark=halfcircle*,1_magenta,mark repeat = 2, mark phase = 2\\
 thick, dashed, mark options={solid,fill=1_blue,fill opacity=1,mark size=2.5pt},mark=halfsquare*,1_blue,mark repeat = 2, mark phase = 2\\
 thick, dashed,  mark options={solid,fill=1_cyan,fill opacity=1,mark size=2.5pt},mark=halfsquare right*,1_cyan,mark repeat = 2, mark phase = 2\\
 thick, dashed,  mark options={solid,fill=1_teal,fill opacity=1,mark size=2.5pt},mark=halfsquare left*,1_teal,mark repeat = 2, mark phase = 2\\
}

\usetikzlibrary{external}
\newtheorem{definition}{Definition}[section]
\newtheorem{theorem}[definition]{Theorem}
\newtheorem{corollary}[definition]{Corollary}

\title{Nonlinear electro-elastic finite element analysis\\with neural network constitutive models
\vspace{1ex}}

\author[1,*]{Dominik~K.~Klein}
\author[2]{Rogelio~Ortigosa}
\author[2]{\\Jes\'us~Mart\'inez-Frutos}
\author[1]{Oliver~Weeger}

\affil[1]{\footnotesize Cyber-Physical Simulation, 
Department of Mechanical Engineering,\protect\\Technical University of Darmstadt, 64293 Darmstadt, Germany}
\affil[2]{\footnotesize Technical University of Cartagena\\
Multiphysics Simulation and Optimization,\protect\\
Campus~Muralla~del~Mar, 30202, Cartagena (Murcia), Spain}
\affil[*]{\footnotesize Corresponding author, email: klein@cps.tu-darmstadt.de}

\date{January 3, 2024}

\begin{document}
\maketitle
\par\noindent\rule{\textwidth}{0.4pt}
\begin{abstract}

\noindent
In the present work, the applicability of physics-augmented neural network (PANN) constitutive models for complex electro-elastic finite element analysis is demonstrated.
For the investigations, PANN models for electro-elastic material behavior at finite deformations are calibrated to different synthetically generated datasets, including an analytical isotropic potential, a homogenised rank-one laminate, and a homogenised metamaterial with a spherical inclusion. %
Subsequently, boundary value problems inspired by engineering applications of composite electro-elastic materials are considered. Scenarios with large electrically induced deformations and instabilities are particularly challenging and thus necessitate extensive investigations of the PANN constitutive models in the context of finite element analyses. 
First of all, an excellent prediction quality of the model is required for very general load cases occurring in the simulation. Furthermore, simulation of large deformations and instabilities poses challenges on the stability of the numerical solver, which is closely related to the constitutive model.
In all cases studied, the PANN models yield excellent prediction qualities and a stable numerical behavior even in highly nonlinear scenarios. This can be traced back to the PANN models excellent performance in learning both the first and second derivatives of the ground truth electro-elastic potentials, even though it is only calibrated on the first derivatives.
Overall, this work demonstrates the applicability of PANN constitutive models for the efficient and robust simulation of engineering applications of composite electro-elastic materials.

\end{abstract}
\vspace*{2ex}
\par\noindent\rule{\textwidth}{0.4pt}\vspace*{2pt} 
{\small
Pre-print under review. \\
Date submitted: January 3, 2024
}
\vspace*{-1.6mm}
\par\noindent\rule{\textwidth}{0.4pt}

\section{Introduction}

Electro-active polymers (EAPs) are a class of multiphysical materials which experience mechanical deformation when subjected to an electric field \cite{RPelrine_98_01,Pelrine_2003}, with one of the most popular materials being dielectric elastomers (DEs) \cite{Pelrine_2000,Skov_review_2016,Feifei_2020}. 
DEs usually comprise of a microstructure where an ultra-soft and low-permittivity material is combined with a stiffer and high-permittivity material, e.g., in the form of random fibres or particles \cite{siboni2014,siboni2015} or laminated structures \cite{gei2018}. These composite DEs not only drastically reduce the operational voltage required for actuation compared to single-phase DEs \cite{huang2004a,huang2004b}, but also enable the design of DEs with microstructures tailored to different applications \cite{clarke_stiffeners,clarkenature,ortigosa2023,goshkoderia2020}.

Outstanding properties such as light weight and fast response time make composite DEs promising for applications such as microfluidic pumps \cite{tavakol2014}, soft robots \cite{Pelrine_2002,Chen_Review_SoftRobots,DE_review_2008,gu2017}, or electrically tunable lenses \cite{Park_17,Shian_13_lenses}. 
In their intended fields of application, DEs can experience very large deformations, including wrinkling \cite{Zhou_Zhao_Suo_2008} and instabilities \cite{keplinger2012,li2013_instab}. While in the past, instabilities such as buckling were mostly considered as failure modes, in the recent years, a paradigm shift has happened towards the design of microstructured materials which exploit such phenomena to achieve outstanding (electro-)mechanical properties \cite{bertoldi2017,kochmann2017,reis2015,li2021,tarantino2019}. For this, a lot of effort was put into understanding the relationship between microstructure and the macroscopic instability behavior of a material \cite{bahreman2022,oneill2022,chen2022_buckling,arora2022}. With tailored instabilities of microstructures being a promising field of research, it is very likely that further applications of DEs in this direction will emerge.

\medskip

In order to fully exploit the capabilities of these materials, efficient and stable simulation methods are required. However, instability phenomena pose a particular challenge since they easily lead to numerical convergence issues in the simulation.
Furthermore, the simulation methods should be able to take the influence of the microstructure on the macrostructure into account. 
In order to avoid computationally expensive methods which rely on the simultaneous treatment of boundary value problems across scales such as the FE$^2$ method \cite{keip2014}, sequential multiscale simulations can be applied. For this, a constitutive model is formulated which represents the effective (or homogenised) behavior of the microstructure, and is then used for simulations on the macroscale.
For some microstructures such as rank-one laminates or porous materials, it is possible to analytically homogenise the microstructure \cite{KBertoldi_11_01, shrimali2019}. However, for the majority of microstructures, analytical expressions are not available. Then, the homogenisation is done numerically, and constitutive models are calibrated to these homogenisation data.
With the effective behavior of microstructures easily being very complex, analytical constitutive models can often not be calibrated to them, since they are not flexible enough for this application \cite{klein2022a}.
 
\medskip

To address this lack of flexibility associated with analytical constitutive models, the extraordinary flexibility of neural networks (NNs) \cite{Hornik1991} can be exploited. In particular, by formulating physics-augmented neural network (PANN) constitutive models, the flexibility of NNs can be combined with a sound mechanical basis, e.g., by including constitutive conditions such as thermodynamic consistency in the model formulation. Furthermore, including mechanical conditions in NN models serves as an \emph{inductive bias} \cite{haussler1988} which enables the calibration of models with relatively small datasets usually available from experimental investigations \cite{linka2023a,fuhg2023extreme}. Combining scientific knowledge with machine learning methods is not restricted to constitutive modeling, but a concept with many promising applications \cite{rueden2021,peng2021,kumar2021,karniadakis2021}.

In multiphysical elasticity theory, basically, NNs are used as very flexible ansatz functions for the corresponding energy potentials.
Inspired by analytical constitutive modeling \cite{Ebbing2010,Schroeder2003}, by using invariants as inputs for the NN, both objectivity and material symmetry conditions can be fulfilled by construction \cite{kalina2023,linka2023a,tac2023a,linka2023b,fuhg2023stress}.
Even more, it is also possible to incorporate the polyconvexity condition \cite{Ball_1976} into NNs. Polyconvexity is a sufficient condition complying with the rank-one convexity condition, intimately related to the concept of material stability \cite{neff2015}. Specifically for NNs, different approaches exist \cite{klein2022a,tac2022,Chen2022}, e.g., based on input convex neural network architectures \cite{Amos2016}. Further conditions can be included by adding growth and normalisation terms to the \textit{physics-augmented} NN potential \cite{klein2022a,linden2023}. 
In the literature, various multiphysical PANN constitutive models were already proposed, including electro-elasticity \cite{klein2022b}, magneto-elasticity \cite{kalina2023b}, and thermo-elasticity \cite{zlatic2023}.
Furthermore, NN-based constitutive models have been successfully applied in finite element analysis (FEA) in purely mechanical 2D \cite{weber2021,weber2023,linka2020} and 3D simulations including complex deformation modes \cite{kalina2023,linden2023,asad2022}.
Moreover, advanced discretization techniques such as mixed finite element methods for PANNs have been developed \cite{franke2023} and thermo-elastic simulations with PANNs were demonstrated \cite{zlatic2023}.

The literature on NN-based constitutive models including their application in FEA is extensive. However, to the best of the authors knowledge, no investigation has been done on the behavior of electro-elastic PANN models in complex FEA including large deformations and instabilities. This is of particular interest, since from an engineering point of view, these are the load scenarios relevant in real-world applications of composite DEs, while from a mathematical point of view, they are very demanding to simulate and pose challenges to the numerical methods and PANN constitutive models. 
First of all, the very general load cases occuring in the simulations require an excellent prediction quality of the PANN model. Furthermore, simulation of large deformation and instabilities poses challenges on the stability of the numerical solver, which is closely related to the robustness and stability of the constitutive model.

\medskip

In the present work, extensive investigations are carried out on the performance of electro-elastic PANN constitutive models for highly challenging FEA scenarios. For this, the electro-elastic PANN constitutive model proposed by the authors in a recent contribution \cite{klein2022b} is extended by polyconvex invariant-based stress normalisation terms as proposed by \cite{linden2023}. 
The material models are calibrated to different datasets, including an analytical isotropic potential, an analytically homogenised transversely isotropic rank-one laminate, and a numerically homogenised metamaterial with a spherical inclusion.
The load scenarios considered in this work are inspired by real-world engineering applications of composite DEs, and include very large electrically induced deformations and instability phenomena.
In further investigations, it is demonstrated that the PANN model shows excellent performance in learning both the first and second derivatives of the ground truth electro-elastic potentials, even though being calibrated only on the first derivatives.

\medskip

The outline of the manuscript is as follows. In \cref{sec:basics}, the fundamentals of finite strain electro-elasticity are introduced, followed by the PANN constitutive model in \cref{sec:PANN}. In \cref{sec:num_1}, the PANN model is calibrated to different datasets, and analysis on its predictions of the first and second derivatives of the ground truth electro-elastic potentials are conducted. Then, the applicability of the PANN models in challenging FEA is investigated in \cref{sec:FEA}. After the conclusion in \cref{sec:conc}, some additional information on the constitutive models are presented in the appendices.

\paragraph{Notation}
Throughout this work, tensor compositions and contractions are denoted by $\left(\bA\,\bB\right)_{ij}=A_{ik}B_{kj}$, $\ba\cdot\bb=a_ib_i$, $\bA:\bB=A_{ij}B_{ij}$ and $\left(\mathbb{A}:\bA\right)_{ij}=\mathbb{A}_{ijkl}A_{kl}$, respectively, with vectors $\ba$ and $\bb$, second order tensors $\bA$ and $\bB$, and fourth order tensor $\mathbb{A}$. The tensor product is denoted by~$\otimes$, the second order identity tensor by $\bI$.
The gradient operator is denoted by $\bnabla$ and the divergence operator by $\bnabla \cdot$ . A zero as the lower index next to the nabla symbol indicates that the operation is carried out in the referential configuration.
The cross-product operator $\Cross$ is defined as $(\vect{A}\Cross\vect{B})_{iI}=\mathcal{E}_{ijk}\mathcal{E}_{IJK}A_{jJ}B_{kK}$, where $\mathcal{E}_{ijk}$ symbolises the third-order permutation tensor. 
For tensors in index notation, lower case indices $\{i, j, k\}$ will be used to represent the spatial configuration, whereas capital case
	indices $\{I, J, K\}$ will be used to represent the material description. 
The first Fr\'echet derivative of a function $f$ w.r.t.\ $\bX$ is denoted by $\nicefrac{\partial f}{\partial \bX}$.
The set of invertible second order tensors with positive determinant is denoted by $\text{GL}^+(3):=\big\{\bX \in\allowbreak \;\mathbb{R}^{3\times 3}\,\rvert\,\allowbreak \det \bX > 0\big\}$ and the special orthogonal group in $\mathbb{R}^3$ by $\SO(3):=\big\{\bX \in\allowbreak \mathbb{R}^{3\times 3}\;\rvert\allowbreak \;\bX^T\bX=\bI,\;\det \bX =1\big\}$.
The Frobenius norm is denoted by $\|*\|$.

\section{Finite strain electro-elasticity}\label{sec:basics}

In this section, the fundamentals of finite strain electro-elasticity are introduced. In \cref{sec:field_eq}, the governing continuum field equations are introduced, followed by general information about the internal energy density and the free energy density in \cref{sec:energies}, and the constitutive conditions relevant for the internal energy density in \cref{sec:const_cond}. Finally, in \cref{sec:fe}, the finite element discretization of the corresponding boundary value problem is described.

\subsection{Continuum field equations}\label{sec:field_eq}

Let $\mathcal{B}_0\subset\bbR^3$ represent the material, reference, or undeformed configuration of an electro-elastic solid body, and $\mathcal{B}\subset\bbR^3$ its current or deformed configuration. The injective function $\vect{\phi}:\mathcal{B}_0\to\mathcal{B}$ maps material points $\vect{X}\in\mathcal{B}_0$ to their counterparts $\vect{x}\in\mathcal{B}$ of the deformed configuration, i.e., $\vect{x}=\vect{\phi}(\vect{X})$. Associated with the mapping $\vect{\phi}$, the deformation gradient is defined as $\vect{F} = \boldsymbol{\nabla}_0\vect{\phi}\in\text{GL}^+(3)$.
The behaviour of an electro-elastic solid is represented by the boundary value problem
\begin{equation}\label{eqn:BVP}
\left.
\begin{aligned}
\vect{F} &= \boldsymbol{\nabla}_0\vect{\phi},&\quad &\text{in }\mathcal{B}_0\\
\vect{\nabla}_0\vect{P}  &= -\vect{f}_0,&\quad &\text{in }\mathcal{B}_0\\
\vect{\phi}&=\vect{\phi}^{\star},&\quad &\text{on }\partial_{\vect{\phi}}\mathcal{B}_0\\
\vect{P}\vect{N}&=\vect{t}_0,&\quad &\text{on }\partial_{\vect{t}}\mathcal{B}_0
\end{aligned}\right\}\,,	\qquad  \qquad \qquad
\left.
\begin{aligned}
\vect{e}_0 &= -\vect{\nabla}_0{\varphi},&\quad &\text{in }\mathcal{B}_0\\\vect{\nabla}_0\cdot\vect{d}_0  &= \rho_0,&\quad &\text{in }\mathcal{B}_0\\
{\varphi}&={\varphi}^{\star},&\quad &\text{on }\partial_{{\varphi}}\mathcal{B}_0\\
\vect{d}_0\cdot\vect{N}&=-{\omega}_0,&\quad &\text{on }\partial_{\omega}\mathcal{B}_0
\end{aligned}\right\}\,.
\end{equation}
Within \cref{eqn:BVP}, the left-hand side terms pertain to the domain of elastostatics, while those on the right-hand side correspond to the realm of electrostatics.

On the left-hand side of \cref{eqn:BVP}, $\vect{P}$ and $\vect{f}_0$ denote the first Piola-Kirchhoff stress and the force exerted per unit volume within $\mathcal{B}_0$, respectively. The imposition of Dirichlet boundary conditions upon the field $\vect{\phi}$ takes place along $\partial_{\vect{\phi}}\mathcal{B}_0$, with $\vect{t}_0$ representing a force applied per unit undeformed area. In this context, $\vect{N}$ signifies the outward normal vector at $\vect{X}\in\partial_{\vect{t}}\mathcal{B}_0$. The boundaries satisfy $\partial\mathcal{B}_0=\partial_{\vect{\phi}}\mathcal{B}_0\cup \partial_{\vect{t}}\mathcal{B}_0$, while also adhering to $\emptyset=\partial_{\vect{\phi}}\mathcal{B}_0\cap \partial_{\vect{t}}\mathcal{B}_0$.

Furthermore, on the right-hand side of \cref{eqn:BVP}, $\varphi$, $\vect{e}_0$, and $\vect{d}_0$ denote the electric potential, the material electric field, and the material electric displacement field, respectively. 
The electric charge per unit undeformed volume within $\mathcal{B}_0$ is denoted by $\rho_0$. Dirichlet boundary conditions are prescribed upon $\partial_{{\varphi}}\mathcal{B}_0$ for the field $\varphi$, with $\omega_0$ representing an electric charge per unit undeformed area located at $\partial{{\omega}}\mathcal{B}_0$. Similar to the mechanical context, $\vect{N}$ signifies the outward normal vector at $\vect{X}\in\partial_{{\omega}}\mathcal{B}_0$. The boundaries satisfy $\partial\mathcal{B}_0=\partial{{\varphi}}\mathcal{B}_0\cup \partial_{\omega}\mathcal{B}_0$, along with the constraint $\emptyset=\partial_{\varphi}\mathcal{B}_0\cap \partial_{\omega}\mathcal{B}_0$.%

\subsection{Internal energy density and free energy density}\label{sec:energies}

For the closure of the BVP introduced in \cref{eqn:BVP}, and to establish a connection between the purely mechanical and the purely electrostatic physics embodied in the field equations, a constitutive model is required to connect the deformation gradient $\vect{F}$, the first Piola-Kirchhoff stress $\vect{P}$, the electric field $\vect{e}_0$, and the electric displacement field $\vect{d}_0$ with each other. 
In finite electro-elasticity, this can be achieved in different ways. First of all, the constitutive model can be defined in terms of the internal energy density 
\begin{equation}
e:\text{GL}^+(3)\times \mathbb{R}^3\rightarrow \mathbb{R},\qquad (\vect{F},\,\vect{d}_0)\mapsto e(\vect{F},\,\vect{d}_0)\,,
\end{equation}
where $\vect{P}$ and $\vect{e}_0$ are given as the gradient fields
\begin{equation}\label{eqn:Piola}
\vect{P}=\partial_{\vect{F}}e(\vect{F},\vect{d}_0)\,, \qquad \vect{e}_0=\partial_{\vect{d}_0}e(\vect{F},\vect{d}_0)\,.
\end{equation}
Alternatively, the free energy density 
\begin{equation}
\Psi:\text{GL}^+(3)\times \mathbb{R}^3\rightarrow \mathbb{R},\qquad (\vect{F},\vect{e}_0)\mapsto \Psi(\vect{F},\vect{e}_0) \,,
\end{equation}
can be considered, from which $\vect{P}$ and $\vect{d}_0$ follow as 
\begin{equation}\label{eqn:Piola free energy}
\vect{P}=\partial_{\vect{F}}\Psi(\vect{F},\vect{e}_0)\,,\qquad \vect{d}_0=-\partial_{\vect{e}_0}\Psi(\vect{F},\vect{e}_0)\,.
\end{equation}
The internal energy density $e(\vect{F},\vect{d}_0)$ is subject to essential convexity conditions, which will be elaborated on further in \cref{sec:const_cond}. The simplest among those conditions dictates that $e(\vect{F},\,\vect{d}_0)$ must exhibit a convex behaviour with respect to both $\vect{F}$ and $\vect{d}_0$ in the vicinity of the origin, namely when $\vect{F}\approx \vect{I}$ and $\vect{d}_0\approx \vect{0}$. In contrast, the free energy density $\Psi(\vect{F},\vect{e}_0)$ imposes distinct convexity constraints compared to its dual counterpart
$e(\vect{F},\,\vect{d}_0)$.  Consequently, in the origin, $\Psi(\vect{F},\,\vect{e}_0)$ assumes the character of a saddle point
function, exhibiting convexity with respect to $\vect{F}$ while displaying concavity concerning $\vect{e}_0$. 

This disparity in convexity/concavity attributes in the context of both mechanics and electro physics can introduce challenges in the
definition of a free energy density $\Psi(\vect{F},\,\vect{e}_0)$ that inherently complies \textit{ab initio} with these simultaneous convexity/concavity properties. Hence, following previous work by the authors in the field of nonlinear electro-mechanics \cite{Ortigosa_ElectroMechanics,Gil_electro_partI_2016,Gil_electro_partII_2016,Gil_electro_partIII_2016,Ortigosa_Gil_Lee_incompressibility}, we advocate for the definition of constitutive models grounded on the internal energy $e(\vect{F},\,\vect{d}_0)$.  However, the most widely used variational formulation of the BVP in \cref{eqn:BVP}, described in \cref{sec:fe}, is typically articulated in terms of the free energy density  $\Psi(\vect{F},\,\vect{e}_0)$, which can be derived from $e(\vect{F},\,\vect{d}_0)$ by making use of the following Legendre transformation \cite{Gil_electro_partI_2016} as
\begin{equation}\label{eqn:legendre transformation}
\Psi(\vect{F},\vect{e}_0):=\inf_{\vect{d}_0}\{e\left(\vect{F},\vect{d}_0\right) - \vect{d}_0\cdot\vect{e}_0\} \, .
\end{equation}
It should be noted that not only the free energy density can be derived through the Legendre transformation in \cref{eqn:legendre transformation}, but also its first and second derivatives, which are both required when applying the constitutive model in numerical applications. In this scenario, $\vect{P}$ and $\vect{d}_0$ can be sequentially obtained. Initially, for a given value of $\vect{F}$ and $\vect{e}_0$,\, $\vect{d}_0$ is obtained by utilizing \cref{eqn:Piola}$_b$, leading to the following potentially nonlinear equation
\begin{equation}
    \vect{e}_0  =  \partial_{\vect{d}_0}e(\vect{F},\,\vect{d}_0(\vect{F},\,\vect{e}_0)) \, .
\end{equation}
Introduction of $\vect{d}_0(\vect{F},\,\vect{e}_0)$ into  \eqref{eqn:Piola}$_a$ permits to finally obtain $\vect{P}$ as
\begin{equation}
   \vect{P}=\partial_{\vect{F}}e(\vect{F},\,\vect{d}_0(\vect{F},\,\vect{e}_0))\,.
\end{equation}
Finally, the second derivatives of $\psi(\vect{F},\,\vect{e}_0)$, from which the second order dielectric tensor, the third order piezoelectric tensor, and the fourth order elasticity tensor emerge, follow as
\begin{equation}\label{eqn:free energy dielectric}
\begin{aligned}	
\partial_{\vect{e}_0\vect{e}_0}^2\Psi&=\left(\partial^2_{\vect{d}_0\vect{d}_0}e\right)^{-1}\,,
\\
\partial^2_{\vect{F}\vect{e}_0}\Psi&=-\Big(\partial^2_{\vect{F}\vect{d}_0}e\Big)\Big(\partial_{\vect{e}_0\vect{e}_0}^2\Psi\Big)\,,
\\
\partial^2_{\vect{F}\vect{F}}\Psi&=\partial^2_{\vect{F}\vect{F}}e - \Big(\partial^2_{\vect{F}\vect{d}_0}e\Big)\Big(\partial_{\vect{e}_0\vect{F}}^2\Psi\Big)\,.
\end{aligned}
\end{equation}

\vspace{10mm}

\subsection{Constitutive conditions for the internal energy density}\label{sec:const_cond}

The thermodynamic consistency of the internal energy density $e(\vect{F},\,\vect{d}_0)$ is established through the definition of both the first Piola-Kirchhoff stress tensor $\vect{P}$ and the electric field $\vect{e}_0$ in \cref{eqn:Piola} \cite{Dormann_Odgen_2005,Bustamante_Merodio_2011}.
Furthermore, $e(\vect{F},\vect{d}_0)$ should comply with  two physically-motivated invariance conditions. The first condition dictates that a constitutive model must be independent on the choice of observer, encapsulated by the objectivity condition
\begin{equation}\label{eqn:material frame indifference}
e(\vect{Q}\vect{F},\vect{d}_0) = e(\vect{F},\vect{d}_0)\qquad \forall \,\vect{F}\in\text{GL}^+(3),\,\vect{d}_0\in\mathbb{R}^3,\,\vect{Q}\in\text{SO}(3)\,.
\end{equation}
The second invariance condition, known as the material symmetry condition, reads as follows
\begin{equation}\label{eqn:material symmetry}
e(\vect{F}\vect{Q},\vect{Q}\vect{d}_0) = e(\vect{F},\vect{d}_0)\qquad \forall \,\vect{F}\in\text{GL}^+(3),\,\vect{d}_0\in\mathbb{R}^3,\,\vect{Q}\in\mathcal{G}\subseteq \text{O}(3)\,,
\end{equation}
and takes into account the materials (an-)isotropy, where $\mathcal{G}$ denotes the symmetry group of the material under consideration. Moreover, both $\vect{P}$ and $\vect{e}_0$ must vanish in the undeformed configuration, namely
\begin{equation}\label{eqn:stress free}
\left.\vect{P}(\vect{F},\,\vect{d}_0)\right\vert_{\vect{F}=\vect{I},\,\vect{d}_0=\vect{0}}=\vect{0}\,,\qquad 
\left.\vect{e}_0(\vect{F},\,\vect{d}_0)\right\vert_{\vect{F}=\vect{I},\,\vect{d}_0=\vect{0}}=\vect{0}\,,
\end{equation}
These stress- and electric field-free conditions are referred to as normalisation conditions. Furthermore, the volumetric growth condition
\begin{equation}\label{eq:growth}
    e(\vect{F},\vect{d}_0)\rightarrow\infty \qquad \text{as} \qquad (\det\vect{F}\rightarrow 0^+ \quad\vee\quad \det\vect{F}\rightarrow\infty)
\end{equation}
reflects the observation that a material body cannnot be compressed to zero volume nor dilatated to an infinitely large volume.

\medskip

In addition, there exist further conditions that can be imposed on $e(\vect{F},\,\vect{d}_0)$, which are grounded in the concept of convexity. It is well-known that convexity of $e(\vect{F},\,\vect{d}_0)$ with respect to both $\vect{F},\,\vect{d}_0$ is extremely stringent, as it fails to encompass physically plausible material phenomena such as buckling, beside other limitations \cite{Ebbing2010}. On the other hand, quasiconvexity of $e(\vect{F},\,\vect{d}_0)$ is an integral condition and thus of very limited practical use. Instead, rank-one convexity (or ellipticity) of $e(\vect{F},\,\vect{d}_0)$ seems a more sensible condition, since it is related to the concept of material stability \cite{Zee1983,neff2015}, which ensures a stable numerical behavior when applying the constitutive model in numerical applications. In electro-elasticity, rank-one convexity is defined as
\begin{equation}\label{eqn:rank one conexity}
\begin{aligned}
&D^2e(\boldsymbol{F},\vect{d}_0)[\delta \vect{\mathcal{U}};\delta \vect{\mathcal{U}}]=\delta  \vect{\mathcal{U}} \bullet [\mathbb{H}_e] \bullet \delta \vect{\mathcal{U}}
\geq 0 &\quad &\forall \,\{\boldsymbol{F},\vect{d}_0\}\in\text{GL}^+(3)\times \mathbb{R}^3,\\
&&\quad&\forall\delta \boldsymbol{\mathcal{U}}=\{\vect{u}\otimes\vect{V},\vect{V}_{\perp}\},\,\,\vect{u},\vect{V}\in\mathbb{R}^3,\vect{V}_{\perp}\cdot \vect{V}=0
\end{aligned}
\end{equation}
This condition is intimately linked with the propagation of planar waves within the material domain. 
The existence of real wave speeds for the specific governing equations presented in \cref{eqn:BVP} is guaranteed when the rank-one convexity condition is fulfilled. Rank-one convexity is equivalent to the positive definiteness of the electromechanical acoustic tensor $\vect{Q}_{ac,e}\in\mathbb{R}^{3\times 3}$, which is defined as
\begin{equation}\label{eq:acoustic_tensor}
[\vect{Q}_{ac,e}]_{ij}=[\widetilde{\vect{\mathcal{C}}}_e]_{iIjJ}\,V_{I}V_{J}\,,
\quad \forall \mathbf{V}\in\mathbb{R}^3,
\end{equation}
where
\begin{equation}
\widetilde{\vect{\mathcal{C}}}_e=\partial^2_{\vect{FF}}e + \partial^2_{\vect{F}\vect{d}_0}e\left(\partial^2_{\vect{d}_0\vect{d}_0}e \right)^{-1}\Bigg(\frac{\vect{V}\otimes \left(\partial^2_{\vect{d}_0\vect{d}_0}e \right)^{-1}\vect{V}}{\vect{V}\cdot \left(\partial^2_{\vect{d}_0\vect{d}_0}e \right)^{-1}\vect{V}} - \vect{I}\Bigg)\partial^2_{\vect{d}_0\vect{F}}e.
\end{equation}

A sufficient condition that ensures the rank-one condition in \cref{eqn:rank one conexity} is polyconvexity of $e(\vect{F},\,\vect{d}_0)$. 
This concept was initially introduced by Ball \cite{Ball1976,Ball1977} for finite elasticity theory and later extended to finite electro-elasticity \cite{Gil_electro_partI_2016,Silhavy_electro}. Together with additional coercivity conditions \cite{kruzik2019}, polyconvexity ensures existence of solutions for the BVPs of (electro-)elasticity. In electro-elasticity, polyconvexity  implies the existence of an equivalent (and possibly non-unique) representation of $e(\vect{F},\vect{d}_0)$ as
\begin{align}\label{eq:pc}
e(\boldsymbol{F},\vect{d}_0) = \mathcal{P}(\boldsymbol{\mathcal{V}}), \qquad \boldsymbol{\mathcal{V}}=(\vect{F},\vect{H},J,\vect{d}_0,\vect{d}),
\end{align}
where $\mathcal{P}$ is a convex function in each of its arguments. Here, $\vect{H}$ and $J$ represent the cofactor and determinant of $\vect{F}$, while $\vect{d}$ denotes the electric displacement field in the spatial configuration, i.e.,
\begin{equation}
\vect{H}=\cof\vect{F}=\left(\text{det}\vect{F}\right)\vect{F}^{-T}=\frac{1}{2}\left(\vect{F}\Cross\vect{F}\right)\,,\qquad 
J=\det\vect{F}=\frac{1}{6}\left(\vect{F}\Cross\vect{F}\right):\vect{F}\,,
\qquad 
\vect{d}=\vect{Fd}_0.
\end{equation}

\subsection{Finite element discretization}\label{sec:fe}

The static behavior of electro-elastic solids can be described by the field equations introduced in \cref{eqn:BVP} in combination with a constitutive model, cf.~\cref{sec:energies}. Both can be encapsulated in the following variational principle \cite{Gil_electro_partII_2016} as
\begin{equation}\label{eqn:mixed variational principle II}
{\Pi}\left(\vect{\phi},\varphi\right) = 
\inf_{\vect{\phi}}\sup_{\varphi}\left\{
\int_{\mathcal{B}_0}{\Psi(\vect{F},\vect{e}_0)}\,dV - \Pi_{\text{ext}}^m\left(\vect{\phi}\right) - \Pi_{\text{ext}}^e\left(\varphi\right)
\right\} \,,
\end{equation}
where $\Pi_{\text{ext}}^m$ and $\Pi_{\text{ext}}^e$ represent the external energy contributions, defined as 
\begin{equation}\label{eqn:external terms in the potential}
\Pi_{\text{ext}}^m\left(\vect{\phi}\right) = \int_{\mathcal{B}_0}\vect{f}_0\cdot\vect{\phi}\,dV + 
\int_{\partial_{\boldsymbol{t}}\mathcal{B}_0}\vect{t}_0\cdot\vect{\phi}\,dA \,,\qquad  
\Pi_{\text{ext}}^e\left({\varphi}\right) = -\int_{\mathcal{B}_0}\rho^e_0{\varphi}\,dV -
\int_{\partial_\omega\mathcal{B}_0}\omega_0^e{\varphi}\,dA.
\end{equation}
We discretise the previous variational principle of \cref{eqn:mixed variational principle II} making use of the finite element method, where the domain $\mathcal{B}_0$ described in \cref{sec:field_eq} is (potentially) approximated by $\mathcal{B}_0^h$ as a collection of $N$ distinct and non-overlapping elements $\mathcal{B}^e_{0}$ as
\begin{equation}
\mathcal{B}_0\approx \mathcal{B}_0^h = \bigcup_{e=1}^N   \mathcal{B}^e_{0}.
\end{equation}
The unknown fields $\left\{\vect{\phi},\varphi\right\}$, along with their corresponding test functions $\left\{\delta\vect{\phi},\delta\varphi\right\}$, are discretized utilizing the functional spaces $\mathbb{V}^{\vect{\phi}^h}\times\mathbb{V}^{{\varphi}^h}$ and $\mathbb{V}_0^{\vect{\phi}^h}\times\mathbb{V}_0^{{\varphi}^h}$, respectively, defined as
\begin{equation}\label{eqn:functional spaces for discretisation}
\begin{aligned}
\mathbb{V}^{\vect{\phi}^h}& = \left\{\vect{\phi}\in \mathbb{V}^{\vect{\phi}}\,,\,\,\,\,\left.\vect{\phi}^h\right\vert_{\mathcal{B}_0^e} = \sum_{a=1}^{n_{\text{node}}^{\vect{\phi}}}N^{\vect{\phi}}_a\vect{\phi}_a\right\}\,,&\qquad
\mathbb{V}^{{\varphi}^h} &= \left\{{\varphi}\in \mathbb{V}^{{\varphi}};\,\,\,\,\left.{\varphi}^h\right\vert_{\mathcal{B}_0^e} = \sum_{a=1}^{n_{\text{node}}^{{\varphi}}}N^{{\varphi}}_a{\varphi}_a\right\}\,,\\
\mathbb{V}_0^{{\vect{\phi}}^h} & = \left\{\forall \vect{\phi}\in\mathbb{V}^{\vect{\phi}^h}\,,\,\,\,\,\, \vect{\phi} = \vect{0} \,\,\text{on}\,\,\partial_{\vect{\phi}}\mathcal{B}_0\right\}\,.&\qquad
\mathbb{V}_0^{{\varphi}^h}&  = \left\{\forall\varphi\in\mathbb{V}^{\varphi^h};\,\,\,\,\,\,\,{\varphi} = {0} \,\,\text{on}\,\,\partial_{{\varphi}}\mathcal{B}_0\right\},
\end{aligned}
\end{equation}
In the context of any given field $\vect{\mathcal{Y}}$ from the set $\left\{\vect{\phi},\varphi\right\}$, the quantity $n_{\text{node}}^{\vect{\mathcal{Y}}}$ signifies the count of nodes within each element of the discretization pertaining to the field $\vect{\mathcal{Y}}$. Furthermore, let $N^{\vect{\mathcal{Y}}}_{a}:\mathcal{B}_0^e\rightarrow \mathbb{R}$ be representative of the $a$-th shape function employed for the purpose of interpolating the field $\vect{\mathcal{Y}}$. Additionally, $\vect{\mathcal{Y}}_a$ denotes the value attributed to the field $\vect{\mathcal{Y}}$ at the $a$-th node of a specific finite element.
By incorporating the functional spaces as defined in \cref{eqn:functional spaces for discretisation}, it becomes feasible to express the stationary conditions of \cref{eqn:mixed variational principle II} in relation to their corresponding residual contributions arising at the elemental level, namely
\begin{equation}\label{eqn:stationary conditions}
D{\Pi}[\delta\vect{\phi}] = \sum_{e=1}^N\delta\vect{\phi}_a\cdot\vect{R}_{a,e}^{\vect{\phi}}=0\,,\qquad
D{\Pi}[\delta{\varphi}]= \sum_{e=1}^N\delta\varphi_a \, {R}_{a,e}^{\varphi}=0\,.
\end{equation}
Notice that both expressions in \cref{eqn:stationary conditions} represent the weak forms of both mechanical and electrical parts of the BVP in \cref{eqn:BVP}.
Additionally, each of the distinct residual contributions denoted as $\vect{R}^{\vect{\phi}}_{a,e}$ and ${R}^{{\varphi}}_{a,e}$ can be expressed as
\begin{equation}\label{eqn:the residuals}
\begin{aligned}
\vect{R}_{a,e}^{\vect{\phi}}  &=  \int_{\mathcal{B}_0^e}\left(\partial_{\vect{F}}\Psi\right)\vect{\nabla}_0N^{\vect{\phi}}_a\,dV + \int_{\mathcal{B}_0^e}N^{\vect{\phi}}_{a}\vect{f}_{0}\,dV\,,\\
R_{a,e}^{\varphi}  &=-\int_{\mathcal{B}^e_0}\left(\partial_{\vect{e}_0}\Psi\right)\cdot\vect{\nabla}_0N^{\varphi}_a\,dV + \int_{\mathcal{B}^e_0}N^{\varphi}_a\rho^e\,dV\,,
\end{aligned}
\end{equation}
where, for the sake of simplicity, the external contributions pertaining to $\vect{t}_0$ and $\omega_0^e$ at the boundary have been omitted. 

A Newton-Raphson scheme can be used for the solution of the nonlinear system of equations arising from the weak forms in \cref{eqn:stationary conditions}, which implies the linearisation with respect to incremental fields $\Delta\vect{\phi}\in\mathbb{V}_0^{\vect{\phi}}$ and $\Delta{\varphi}\in\mathbb{V}_0^{{\varphi}}$ as
\begin{equation}\label{eqn:NR}
\begin{aligned}
0=D{\Pi}[\delta\vect{\phi}] + D{\Pi}[\delta{\varphi}] + 
D^2{\Pi}[\delta\vect{\phi};\Delta\vect{\phi}] + D^2{\Pi}[\delta\vect{\phi};\Delta{\varphi}] + 
D^2{\Pi}[\delta{\varphi};\Delta\vect{\phi}] + 
D^2{\Pi}[\delta{\varphi};\Delta{\varphi}]\,,
\end{aligned}
\end{equation}
with
\begin{equation}
\begin{aligned}
D^2{\Pi}[\delta\vect{\phi},\Delta\vect{\phi}] &= \sum_{e=1}^N\delta\vect{\phi}_a\cdot\vect{K}_{ab,e}^{\vect{\phi}\vect{\phi}} \, \Delta\vect{\phi}_b\,,\qquad
&&D^2{\Pi}[\delta\vect{\phi},\Delta{\varphi}] = \sum_{e=1}^N\delta\vect{\phi}_a\cdot\vect{K}_{ab,e}^{\vect{\phi}{\varphi}} \, \Delta{\varphi}_b\,,\\
D^2{\Pi}[\delta{\varphi},\Delta{\vect{\phi}}]& = \sum_{e=1}^N\delta{\varphi}_a \, \vect{K}_{ab,e}^{{\varphi}\vect{\phi}} \, \Delta{\vect{\phi}}_b\,,
\qquad
&&D^2{\Pi}[\delta{\varphi},\Delta{{\varphi}}] = \sum_{e=1}^N\delta{\varphi}_a  \,  K_{ab,e}^{{\varphi}{\varphi}} \, \Delta{{\varphi}}_b\,,\\
\end{aligned}
\end{equation}
where each of the stiffness contributions is expressed as
\begin{equation}\label{eqn:tangent matrices}
\begin{aligned}
\left(\vect{K}_{ab,e}^{\vect{\phi}\vect{\phi}}\right)_{ij}&=\int_{\mathcal{B}^e_0}\left(\vect{\nabla}_0N_a^{\vect{\phi}}\right)_I\left(\vect{\nabla}_0N_b^{\vect{\phi}}\right)_J\left(\partial^2_{\vect{F}\vect{F}}\Psi\right)_{iIjJ}\,dV\,,\\
\left(\vect{K}_{ab,e}^{\vect{\phi}{\varphi}}\right)_{i}&=-\int_{\mathcal{B}^e_0}\left(\vect{\nabla}_0N_a^{\vect{\phi}}\right)_I\left(\vect{\nabla}_0N_b^{{\varphi}}\right)_J\left(\partial^2_{\vect{Fe}_0}\Psi\right)_{iIJ}\,dV\,,\\
K_{ab,e}^{{\varphi}{\varphi}}&=\int_{\mathcal{B}^e_0}\Big(\vect{\nabla}_0N_a^{{\varphi}}\Big)_I\Big(\partial^2_{\vect{e}_0\vect{e}_0}\Psi\Big)_{IJ}\Big(\vect{\nabla}_0N_b^{{\varphi}}\Big)_{J}\,dV\,, \\
\vect{K}_{ab,e}^{\varphi\vect{\phi}}&=\left(\vect{K}_{ab,e}^{\varphi\vect{\phi}}\right)^T\,.
\end{aligned}
\end{equation}
Finally, standard global assembly of the residual and stiffness contributions in \cref{eqn:the residuals} and \cref{eqn:tangent matrices}, respectively, permits to obtain the discrete form of  \cref{eqn:NR} in terms of the nodal incremental vector fields $\{\widetilde{\vect{\Delta\phi}},\widetilde{\vect{\Delta\varphi}}\}$ as
\begin{equation}
\begin{bmatrix}
\vect{K}^{\vect{\phi}\vect{\phi}}  &  \vect{K}^{\vect{\phi}{\varphi}}\\
\vect{K}^{{\varphi}\vect{\phi}}  &  \vect{K}^{{\varphi}{\varphi}}
\end{bmatrix}
\begin{bmatrix}
\widetilde{\vect{\Delta\phi}}\\
\widetilde{\vect{\Delta\varphi}}
\end{bmatrix}=-\begin{bmatrix}
\vect{R}^{\vect{\phi}}\\
\vect{R}^{{\varphi}}
\end{bmatrix}.
\end{equation}
Solution of this linear system of equations for $\{\widetilde{\vect{\Delta\phi}},\widetilde{\vect{\Delta\varphi}}\}$
permits the update of the nodal solution fields at every node of the underlying finite element discretisation at a given Newton-Raphson iteration $k+1$ as
\begin{equation}
\begin{aligned}
\widetilde{\vect{\phi}}^{k+1}=\widetilde{\vect{\phi}}^{k} + \widetilde{\vect{\Delta\phi}}\,,\qquad
\widetilde{\vect{\varphi}}^{k+1}=\widetilde{\vect{\varphi}}^{k} + \widetilde{\vect{\Delta\varphi}}.
\end{aligned}
\end{equation}

\section{Physics-augmented neural network constitutive model}\label{sec:PANN}

In this section, the physics-augmented neural network (PANN) constitutive model applied in this work is introduced. After a general introduction to invariant-based modeling in \cref{sec:invs}, the constitutive equations of the invariant-based PANN model are provided in \cref{sec:PANN_ces}.

\subsection{Invariant-based representation of the internal energy density}\label{sec:invs}

By formulating the internal energy density $e(\vect{F},\,\vect{d}_0)$ in terms of electro-mechanical invariants, both the objectivity and the material symmetry condition can be fulfilled, cf.~\cref{eqn:material frame indifference,eqn:material symmetry}.
With $n$ group-specific invariants of the symmetry group under consideration contained in the vector $\bcI\in\bbR^n$, this allows for a representation of the internal energy as
\begin{equation}\label{eq:pot_inv}
    \widetilde{\mathcal{P}}:\bbR^n\rightarrow\bbR\,,\qquad \bcI\mapsto \widetilde{\mathcal{P}}(\bcI)=e(\vect{F},\,\vect{d}_0)\,.
\end{equation} 
When the invariants in \cref{eq:pot_inv} are polyconvex and the internal energy density $\widetilde{\mathcal{P}}(\bcI)$ is a convex and non-decreasing function, the overall potential is polyconvex, cf.~\cite{klein2022a} for an explicit proof and \cite{klein2023a} for an introductory 1D example on convexity of function compositions.  
In the present work, both polyconvex and non-polyconvex potentials are considered.

In the following, polyconvex invariants for isotropy and transverse isotropy are introduced. Note that further non-polyconvex invariants  can be constructed for these symmetry groups, which could generally be applied in the non-polyconvex models, see e.g.~\cite[Remark~2.4]{klein2022b}.
However, since polyconvex potentials require polyconvex invariants, for the sake of simplicity, only polyconvex invariants are considered throughout this work, which are applicable to all models.

\subsubsection{Isotropic material behavior}
Starting with isotropy, i.e., $\mathcal{G}=\text{O}(3)$, polyconvex invariants are given by
\begin{equation}\label{pg_iso}
\begin{aligned}
   {I_1}&= \|\vect{F}\|^2\,,\qquad 
   {I_2}= \|\vect{H}\|^{2}\,,\qquad
    I_4 = \|\bd_0\|^2\,,\qquad 
    I_5 = \|\bd\|^2\,,\qquad J=\det\vect{F}\,.
    \end{aligned}
\end{equation}
Thus, in \cref{eq:pot_inv} we employ the sets of invariants
\begin{equation}\label{eq:invs_iso}
\vect{\mathcal{I}}^{\text{iso}} = 
\left\{\begin{aligned}
&\big\{I_1,\,I_2,\,J,\,I_4,\,I_5,\,I_6\big\}&&\in\bbR^6\,, &&& (\text{Non-polyconvex models})\\
&\big\{I_1,\,I_2,\,J,\,I_4,\,I_5,\,I_6,\,I_7\big\}&&\in\bbR^7\,, &&& (\text{Polyconvex models})\\
\end{aligned}\right. \,,
\end{equation} 
with $I_6=J^{-1}I_5$ and $I_7=-J$. Here, the additional invariant $I_6$ is incorporated due to its physical significance in the representation of ideal dielectric elastomers.
Furthermore, for polyconvex models, by considering the additional invariant $I_7$, the invariant $J$ is included both with a positive and a negative sign. This is motivated by the following: invariants such as ${I_1}= \|\vect{F}\|^2$ are nonlinear functions of the arguments of the polyconvexity condition, cf.~\cref{eq:pc}. Thus, in order to preserve their polyconvexity, the internal energy has to be convex and \emph{non-decreasing} in these invariants, cf.~\cite{klein2022a,klein2023a}. In contrast to that, $J$ is the only invariant directly included in the polyconvexity condition, thus, the potential must not necessarily be non-decreasing in this invariant. This is pragmatically taken into account by adding $I_7=-J$ in the set of arguments for polyconvex models, which allows to formulate potentials which are convex and non-decreasing in all of their arguments without being too restrictive on $J$.

\subsubsection{Transversely isotropic material behavior}

For anisotropic material behavior, structural tensors are introduced to take the materials symmetry into account \cite{Zheng1993}. In the case of the transversely isotropic symmetry group, i.e., for $\mathcal{G}=\mathcal{D}_{\infty h}$ \cite{Ebbing2010}, a single second order structural tensor is sufficient. In this work, the transversely isotropic structural tensor
\begin{equation}\label{eq:struct_ti}
\structTI    =\vect{N}\otimes\vect{N}
\end{equation} 
is applied, where the unitary vector $\vect{N}\in\bbR^3$, $\|\vect{N}\|=1$, denotes the preferred direction of the material. 
With this structural tensor, the polyconvex transversely isotropic invariants
\begin{equation}
\begin{aligned}
    \IoneTI &= \big\|\bF\,\structTI\big\|^2\,,\qquad & \ItwoTI&= \big\|\bH\,\structTI\big\|^{2}\,,\qquad
   \IthreeTI&= \text{tr}\Big(\left(\vect{d}_0\otimes\vect{d}_0\right)\,\structTI\Big)
    \end{aligned}
\end{equation}
can be constructed.
As proposed in \cite{Schroeder2003}, additional polyconvex invariants can be constructed as
\begin{equation}
\begin{aligned}\label{eq:ti_invar_add}
    \IfourTI &=I_1-\IoneTI\,,\qquad & \IfiveTI &= I_2-\ItwoTI\,,\qquad
    \IsixTI &= I_3-\IthreeTI\,.
    \end{aligned}
\end{equation}
Overall, for transverse isotropy, this yields the sets of invariants
\begin{equation}\label{eq:ti_invar}
\vect{\mathcal{I}}^{\text{ti}} = 
\left\{\begin{aligned}
&\big\{\vect{\mathcal{I}}^{\text{iso}},\,\IoneTI,\,\ItwoTI,\,\IthreeTI\big\}&&\in\bbR^{9}\,, &&& (\text{Non-polyconvex models})\\
&\big\{{\vect{\mathcal{I}}}^{\text{iso}},\,\IoneTI,\,\ItwoTI,\,\IthreeTI,\,\IfourTI,\,\IfiveTI,\,\IsixTI\big\}&&\in\bbR^{13}\,, &&& (\text{Polyconvex models})\\
\end{aligned}\right. \,.
\end{equation} 
For the convenience of the reader, an explicit proof of polyconvexity of the additional invariants is given in Appendix \ref{app:pc_add_invs}, see also \cite{Schroeder2003}.
These additional invariants aim to diminish the restriction that polyconvexity poses on the internal energy. In particular, a sufficient condition for polyconvexity is that the internal energy is convex and \emph{non-decreasing} in the invariants. However, as the polyconvex invariants in \cref{eq:ti_invar_add} demonstrate, the non-decreasing condition is not a necessary one. E.g., considering the potential $\widetilde{\mathcal{P}}(\ItwoTI,\,\IfiveTI)$, its partial derivative with respect to invariant $\ItwoTI$ when expressing $\IfiveTI$ as a function of $\ItwoTI$ (see~\cref{eq:ti_invar_add}) yields
\begin{equation}\label{eq:add_inv_example}
    \begin{aligned}
        {\partial_{\ItwoTI} \widetilde{\mathcal{P}}(\ItwoTI,\,\IfiveTI(\ItwoTI))}&={ \partial_{{ \ItwoTI}}\widetilde{\mathcal{P}}(\ItwoTI,\,\IfiveTI)}+{\partial_{\IfiveTI}\widetilde{\mathcal{P}}(\ItwoTI,\,\IfiveTI)}\frac{d\IfiveTI}{d \ItwoTI}={\partial_{\ItwoTI} \widetilde{\mathcal{P}}(\ItwoTI,\,\IfiveTI)}-{\partial_{\IfiveTI} \widetilde{\mathcal{P}}(\ItwoTI,\,\IfiveTI)}\,,
    \end{aligned}
\end{equation}
which demonstrates that including the additional invariant $\IfiveTI$ allows for decreasing potentials in $\ItwoTI$ whenever the condition 
\begin{equation}\label{eq:geq}
    {\partial_{\ItwoTI} \widetilde{\mathcal{P}}(\ItwoTI,\,\IfiveTI)}<{\partial_{\IfiveTI} \widetilde{\mathcal{P}}(\ItwoTI,\,\IfiveTI)}
\end{equation}
holds. In particular, the condition in \cref{eq:geq} can be fulfilled even if the potential is convex and monotonous in all invariants.
Thus, using these additional invariants, at least in parts, allows for decreasing functional relationships of the internal energy in $\{\IoneTI,\,\ItwoTI,\,\IthreeTI\}$, thus providing the constitutive model with more flexibility.
Note that these addtional invariants are not required when non-polyconvex models are considered, since in this case, the internal energy is not subject to convexity or monotonicity restrictions.

\subsection{Invariant-based PANN model}\label{sec:PANN_ces}

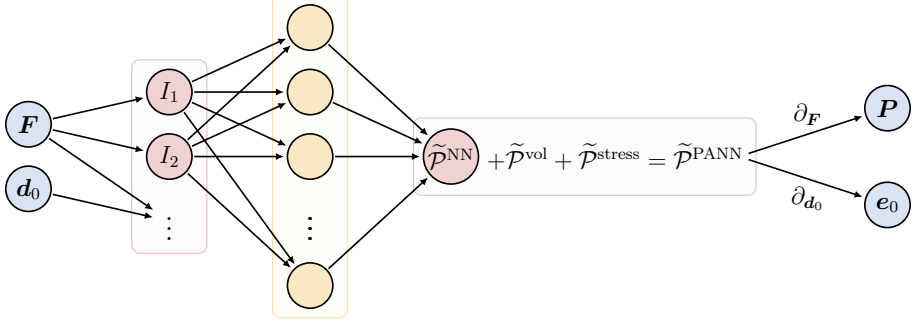
\begin{figure}[t!]
\centering
\resizebox{0.75\textwidth}{!}{
\tikzsetnextfilename{ANN_model}

\begin{tikzpicture}[x=1.6cm,y=1.1cm]
  \large
  \def\NC{6} 
  \def\nstyle{int(\lay<\Nnodlen?(\lay<\NC?min(2,\lay):3):4)} 
  \tikzset{ 
    node 1/.style={node in_out},
    node 2/.style={node inv_pot},
    node 3/.style={node icnn},
  }
  
    \draw[2red!40,fill=2red,fill opacity=0.02,rounded corners=4]
    (1.6,-1.5) --++ (0,3) --++ (0.8,0) --++ (0,-3) -- cycle;
          \draw[2orange!40,fill=2orange,fill opacity=0.02,rounded corners=4]
    (3.1,-2.5) rectangle++ (0.8,5);
              \draw[2red!40,fill=2red,fill opacity=0.02,rounded corners=4]
    (4.6,-0.6) rectangle++ (3.65,1.2);
  
 \node[node 1, outer sep=0.6] (1-1) at (0.5,0.5) {$\vect{F}$};
 \node[node 1, outer sep=0.6] (1-2) at (0.5,-0.5) {$\vect{d}_0$};

 \node[node 2, outer sep=0.6] (3-1) at (2,1) {$I_1$};
 \node[node 2, outer sep=0.6] (3-2) at (2,0) {$I_2$};
 \node[scale=1.2] (3-3) at (2,-1) {$\vdots$};

 \draw[connect arrow] (1-1) -- (3-1);
 \draw[connect arrow] (1-1) -- (3-2);
 \draw[connect arrow] (1-1) -- (3-3);
 \draw[connect arrow] (1-2) -- (3-3);

\def\N{5}
    \foreach \i [evaluate={\y=\N/2-\i+0.5;}] in {1,2,3,5}{ 
     \node[node 3,outer sep=0.6] (4-\i) at (3.5,\y) {};

     \node[scale=1.2] (4-4) at (3.5,-1) {$\vdots$};

    \foreach \j in {1,2}{
    \draw[connect arrow]  (3-\j) -- (4-\i);
    }
    
      }


    \node[node 2, outer sep=0.6] (6-1) at (5,0) {$\widetilde{\mathcal{P}}^{\text{NN}}$};
       \foreach \j in {1,2,3,5}{
    \draw[connect arrow]  (4-\j) -- (6-1);
}

  \node[] (7-1) at (6.75, 0)
  {$+\widetilde{\mathcal{P}}^{\text{vol}}+\widetilde{\mathcal{P}}^{\text{stress}}=\widetilde{\mathcal{P}}^{\text{PANN}}$};

    \node[] (7-2) at (8.07, 0){};

    \node[node 1, outer sep=0.6] (9-1) at (9.65,0.75) {${ \vect{P}}$};
    \draw[connect arrow]  (7-2) -- (9-1);

       \node[node 1, outer sep=0.6] (9-2) at (9.65,-0.75) {${ \vect{e}_0}$};
       
           \draw[connect arrow]  (7-2) -- (9-2) ;
           
          \node[] at (8.8,0.65) {$\partial_{\vect{F}}$};
          \node[] at (8.8,-0.65) {$\partial_{\vect{d}_0}$};
    
  \end{tikzpicture}
} 
\caption{Illustration of the physics-augmented neural network (PANN) constitutive model. Note that the hidden-layer (yellow) of the NN may be multilayered.}   
\label{fig:ANN_model}   
\end{figure}

Formulating electro-elastic constitutive models in terms of invariant-based energy potentials has a sound mechanical motivation. By that, thermodynamic consistency, objectivity and material symmetry conditions are fulfilled by construction, cf.~\cref{sec:const_cond} and \cref{sec:invs}. However, for analytical constitutive models, the last step required to arrive at the final model, i.e., the explicit choice of a model equation, is often more restrictive than it has to be, and mostly heuristically motivated. E.g., the analytical isotropic potential introduced later in \cref{eq:MR_2} is linear in $I_1$ and $I_2$, which is a quite restrictive assumption on the material behavior. 

These limitations associated with analytical constitutive models can be circumvented by by using NNs to represent energy potentials and thus exploiting the extraordinary flexibility that neural networks (NNs) offer \cite{Hornik1991}. Still, the NN is only a part of the overall physics-augmented neural network (PANN) constitutive model given by
\begin{equation}
    \widetilde{\mathcal{P}}^{\text{PANN}}(\boldsymbol{\cI})=\widetilde{\mathcal{P}}^\text{NN}( \boldsymbol{\cI})+\widetilde{\mathcal{P}}^\text{vol}(J)+\widetilde{\mathcal{P}}^{\text{stress}}( \boldsymbol{\cI})\,.
\end{equation}
Here, $\widetilde{\mathcal{P}}^\text{NN}$ denotes the NN part of the internal energy density, which provides the model with flexibility and receives the invariants of the symmetry group under consideration as input. The volumetric growth term $\widetilde{\mathcal{P}}^\text{vol}$ and the stress normalisation term $\widetilde{\mathcal{P}}^{\text{stress}}$ serve to fulfill the volumetric growth condition (see~\cref{eq:growth}) and the normalisation condition (see~\cref{eqn:stress free}), respectively. 
The basic electro-elastic PANN model consisting of $\widetilde{\mathcal{P}}^\text{NN}$ and $\widetilde{\mathcal{P}}^\text{vol}$ was proposed by the authors of the present work in a recent contribution \cite{klein2022b}, while the stress normalisation $\widetilde{\mathcal{P}}^{\text{stress}}$ was originally proposed by \cite{linden2023} for purely mechanical material behavior. 

\subsubsection{NN part of the potential}

Starting with the NN part of the potential $\widetilde{\mathcal{P}}^{\text{NN}}$, for this, feed-forward neural networks (FFNNs) are applied \cite{kollmannsberger2021,aggarwal2018}. In the present work, without loss of generality, FFNNs with a single hidden layer are applied, which have proved to be sufficiently flexible for the representation of energy potentials in many cases \cite{linden2023,klein2022b}. For an introduction to multilayered NN architectures, the reader is referred to \cite[Appendix~A]{linden2023}. Here, the single-layered NN potentials take the form
\begin{equation}\label{eq:e_nn}
    \widetilde{\mathcal{P}}^{\text{NN}}=\vect{W}^2\cdot\SP\!\left(\vect{W}^1\cdot\bcI+\vect{b}\right)\,,
\end{equation}
where $\SP=\log(1+e^x)$ denotes the \emph{Softplus} activation function, which is applied component-wise. The trainable parameters of the NN $\vect{p}=\{\vect{W}^2,\,\vect{W}^1,\,\vect{b}\}$ are given by the weights $\vect{W}^2\in\bbR^m,\,\vect{W}^1\in\bbR^{m\times n}$ and the bias $\vect{b}\in\bbR^m$. Depending on the symmetry group under consideration and whether a polyconvex model is considered or not, the invariants introduced in \cref{sec:invs} are used as inputs $\bcI\in\bbR^n$ for the NN potential. Finally, the hyperparameter $m$ denotes the number of nodes/neurons in the hidden layer. By increasing $m$, the number of trainable parameters increases and the model gains flexibility.

\medskip

Using the \emph{Softplus} activation function has several reasons. First of all, it results in continuous first and second derivatives of the NN potential, which are required for mechanical applications of the potential, cf.~\cref{sec:basics}. More than that, it allows to construct polyconvex NN potentials.
In the case of polyconvex PANN models, polyconvex invariants have to be used and the potential must be constructed as a convex and non-decreasing function, cf.~\cref{sec:invs}. %
By using the convex and non-decreasing \emph{Softplus} activation function and restricting the weights $\vect{W}^1,\,\vect{W}^2$ to be non-negative, the resulting NN potential is a convex and non-decreasing function and can be applied for polyconvex modeling. This special FFNN architecture is referred to as input-convex neural network \cite{Amos2016}.\footnote{Note that, in general, every twice continuously differentiable, convex, and non-decreasing activation function could be applied \cite{klein2022a}.} 
 For explicit convexity proofs of this architecture, the reader is referred to \cite{klein2022a}.
For non-polyconvex PANN models, arbitrary weights can be applied. For polyconvex PANN models with $m$ neurons in the hidden layer, the short notation $\SP^+(m)$ is applied, while for non-polyconvex models, the short notation $\SP(m)$ is applied.

\medskip

Setting aside the nomenclature of machine learning, in \cref{eq:e_nn}, one very important aspect of NNs becomes evident: from a formal point of view, FFNNs are essentially just mathematical functions. In that light, one could consider the NN potential as a classical constitutive model, which uses linear transformations in combination with the \emph{Softplus} function as an ansatz for energy potentials, and which has the weights $\vect{W}^1,\,\vect{W}^2$ and the bias $\vect{b}$ as material parameters. The fundamental difference compared to classical models is that for the NN potential, the flexibility can be immediately increased to basically an arbitrary amount by increasing the number of nodes. At the same time, the powerful optimisation algorithms of the machine learning community still enable a robust and efficient model calibration, i.e., the \emph{training} of the parameters $\vect{p}$ such that the model fits a given dataset, see \cref{sec:num_1}.

In that spirit, it is almost natural to calculate the derivatives of the NN potential in an explicit fashion, as commonly done for analytical models.
For the single-layered NN architecture introduced in \cref{eq:e_nn}, this allows for concise formulations and efficient implementations of the derivatives. The first and second derivative w.r.t.\ the invariants are given by 
\begin{equation}
\begin{aligned}
    \partial_{(\bcI)_i}\widetilde{\mathcal{P}}^{\text{NN}}&=\sum_{a=1}^mW^2_a\frac{e^{h_a}}{1+e^{h_a}}W_{ai}^1\,,\qquad \text{where}\quad
    \vect{h}:=\vect{W}^1\cdot\bcI+\vect{b}\,,
\\
   \partial^2_{(\bcI)_i(\bcI)_j}\widetilde{\mathcal{P}}^{\text{NN}}&= \sum_{a=1}^mW^2_a\frac{e^{h_a}}{\left(1+e^{h_a}\right)^2}W_{ai}^1W_{aj}^1\,.
   \end{aligned}
\end{equation}
The derivatives w.r.t.\ $\vect{F},\,\vect{d}_0$ follow immediately by applying the chain rule.

\subsubsection{Growth term}

The analytical growth term $\widetilde{\mathcal{P}}^\text{vol}$ is chosen as the polyconvex function
\begin{equation}
    \widetilde{\mathcal{P}}^\text{vol}(J):=\left(J+\frac{1}{J}-2\right)^2\,.
\end{equation}

\subsubsection{Normalisation term}

The invariant-based stress normalisation terms proposed by \cite{linden2023} are formulated to comply with all relevant constitutive conditions. In particular, they are polyconvex and fulfill both objectivity and material symmetry conditions.
In the isotropic case, the stress normalization term is given by
\begin{equation}
    \widetilde{\mathcal{P}}^{\text{stress, iso}}
( J):=-\mathfrak{n}J\,,
\end{equation}
where the constant 
\begin{equation}\label{eq:norm_a}
    \mathfrak{n}:=2\left(
    \partial_{I_1}\widetilde{\mathcal{P}}^{\text{NN}}+2\partial_{I_2}\ennISO+\frac{1}{2}\left(\partial_J\widetilde{\mathcal{P}}^{\text{NN}}-\partial_{I_6}\widetilde{\mathcal{P}}^{\text{NN}}\right)
    \right)
    \Bigg\rvert_{\vect{F}=\vect{I},\,\vect{d}_0=\vect{0}}\in\bbR
\end{equation}
is a weighted sum of derivatives of the NN potential w.r.t.\ to the invariants in the reference configuration. For non-polyconvex models, it holds that $\partial_{I_6}\widetilde{\mathcal{P}}^{\text{NN}}=0$.
Apparently, the normalisation term includes only purely mechanical invariants. This is owed to the fact that the contribution of the electro-mechanically coupled invariant $I_5$ to the stress vanishes in the reference configuration, since
\begin{equation}
    \partial_{\vect{F}}I_5\Big\rvert_{\vect{F}=\vect{I},\,\vect{d}_0=\vect{0}} = 2\vect{F}\left(\vect{d}_0\otimes\vect{d}_0\right)\Big\rvert_{\vect{F}=\vect{I},\,\vect{d}_0=\vect{0}}=\vect{0}\,.
\end{equation}
Furthermore, the normalisation condition for the electric field is automatically fulfilled by construction, since
\begin{equation}
    \partial_{\vect{d}_0}I_4\Big\rvert_{\vect{F}=\vect{I},\,\vect{d}_0=\vect{0}}=2\vect{d}_0\Big\rvert_{\vect{F}=\vect{I},\,\vect{d}_0=\vect{0}}=\vect{0}
\end{equation}
and
\begin{equation}
    \partial_{\vect{d}_0}I_5\Big\rvert_{\vect{F}=\vect{I},\,\vect{d}_0=\vect{0}}=2\vect{F}^T\vect{F}\vect{d}_0\Big\rvert_{\vect{F}=\vect{I},\,\vect{d}_0=\vect{0}}=\vect{0}\,.
\end{equation}
Similar reasoning holds for the transversely isotropic terms to follow, where the explicit proofs are left to the interested reader. 

\medskip

Following \cite{linden2023}, the transversely isotropic stress normalization term is given by
\begin{equation}
    \widetilde{\mathcal{P}}^{\text{stress, ti}}(J,\,\IoneTI,\,\ItwoTI):=-\mathfrak{o}J+ \mathfrak{p}\IoneTI+ \mathfrak{q}\ItwoTI\,.
\end{equation}
Here, the constant
\begin{equation}\label{eq:norm_b}
\begin{aligned}
    \mathfrak{o}:=2\Bigg(&
\partial_{I_1}\widetilde{\mathcal{P}}^{\text{NN}}
+\partial_{\IfourTI}\widetilde{\mathcal{P}}^{\text{NN}}+2\left(\partial_{I_2}\widetilde{\mathcal{P}}^{\text{NN}}+\partial_{\IfiveTI}\widetilde{\mathcal{P}}^{\text{NN}}\right)
\\
&+\frac{1}{2}\left(\partial_{J}\widetilde{\mathcal{P}}^{\text{NN}}-\partial_{I_6}\widetilde{\mathcal{P}}^{\text{NN}}\right)+\left(\partial_{\ItwoTI}\widetilde{\mathcal{P}}^{\text{NN}}-\partial_{\IfiveTI}\widetilde{\mathcal{P}}^{\text{NN}}\right)\tr\structTI\Bigg)\Bigg\rvert_{\vect{F}=\vect{I},\,\vect{d}_0=\vect{0}}
\in\bbR
    \end{aligned}
\end{equation}
is a weighted sum of the derivatives of the NN potential w.r.t.\ the invariants in the reference configuration. 
In addition, with the ReLU (rectified linear unit) function denoted as $\RL=\max(0,\,x)$ and the argument
\begin{equation}
    x:=\left(
\partial_{\IoneTI}\widetilde{\mathcal{P}}^{\text{NN}}-\partial_{\IfourTI}\widetilde{\mathcal{P}}^{\text{NN}}-\partial_{\ItwoTI}\widetilde{\mathcal{P}}^{\text{NN}}+\partial_{\IfiveTI}  \widetilde{\mathcal{P}}^{\text{NN}} 
    \right)\Bigg\rvert_{\vect{F}=\vect{I},\,\vect{d}_0=\vect{0}}
    \in\bbR\,,
\end{equation}
the remaining constants are given by 
\begin{equation}\label{eq:norm_pq}
    \mathfrak{p}:=\RL(-x)\in\bbR_{\geq 0},\qquad      \mathfrak{q}:=\RL(x)\in\bbR_{\geq 0}\,.
\end{equation}
For non-polyconvex models, it holds that $\partial_{I_6}\widetilde{\mathcal{P}}^{\text{NN}}=\partial_{\IfourTI}\widetilde{\mathcal{P}}^{\text{NN}}=\partial_{\IfiveTI}\widetilde{\mathcal{P}}^{\text{NN}}=0$.

\section{Material model calibration and evaluation}\label{sec:num_1}

In this section, the PANN models introduced in \cref{sec:PANN} are calibrated and evaluated for different ground truth (GT) models. In \cref{subsec:iso_model}, the PANN model is calibrated to data generated with an analytical isotropic potential, followed by the calibration of the PANN to homogenisation data of a representative volume element (RVE) with a spherical inclusion in \cref{subsec:RVE}. Finally, in \cref{subsec:ROL}, the PANN model is calibrated to a homogenised rank-one laminate. In the last example, extensive investigations on the models prediction of the first and second derivatives of the ground truth electro-elastic potential are carried out.

\subsection{Isotropic phenomenological model}\label{subsec:iso_model}

\subsubsection{Data generation}\label{sec:iso_data}

As a first example, an analytical potential consisting of a purely mechanical Mooney-Rivlin model and a coupled ideal dielectric elastomer is considered as the ground truth model, i.e.,
\begin{equation}\label{eq:MR_2}
\widetilde{\mathcal{P}}^{\text{MR}}(\vect{\mathcal{I}}^{\text{iso}} )  =  \frac{\mu_1}{2}I_1 + \frac{\mu_2}{2}I_2 - \left(\mu_1 + 2\mu_2\right)\log{J} + \frac{\lambda}{2}(J-1)^2 + \frac{I_5}{2\varepsilon J}\,. 
\end{equation}
The material parameters are chosen as
\begin{equation}\label{eq:Iso_Mat_Params}
    \mu_1=\mu_2=0.5,\qquad \lambda=5,\qquad \varepsilon=1 \,.
\end{equation}

For data generation, both the deformation gradient $\vect{F}\in\GL^+(3)$ and the electric displacement field $\vect{d}_0\in\bbR^3$ have to be sampled. 
In the first step, $\vect{F}$ is sampled with the concentric sampling strategy proposed in \cite{kunc2019}, which generates very general yet physically sensible load scenarios, see also \cite[App.~A]{klein2022b} for a brief introduction. Assuming incompressibility, this strategy reduces to the sampling of unit vectors in a 5-dimensional space and combining them with different amplitudes. The former correspond to the deviatoric direction of the deformation, the latter to the deviatoric amplitude. 
Finally, the datasets are generated by assigning each deviatoric direction of $\vect{F}$ a random unit vector for the direction of $\vect{d}_0$, and combining them with all deviatoric amplitudes for $\vect{F}$ as well as different amplitudes for $\vect{d}_0$.
In this example, for the calibration, 10 deviatoric directions are combined with 10 deviatoric amplitudes and 5 amplitudes for $\vect{d}_0$, which overall results in 500 datapoints. For the test dataset, 100 deviatoric directions are combined with 30 deviatoric amplitudes and 10 amplitudes for $\vect{d}_0$, which overall results in 30,000 datapoints.

Overall, in this work, datasets of the form
\begin{equation}\label{eq:dataset}
    \mathcal{D}=\bigg\{\Big(\vect{F}^{1},\,\vect{d}_0^{1};\,\vect{P}^{1},\,\vect{e}_0^{1}\Big),\,\Big(\vect{F}^{2},\,\vect{d}_0^{2};\,\vect{P}^{2},\,\vect{e}_0^{2}\Big)\,,\dotsc\bigg\}
\end{equation}
are considered. 
Since the data is generated synthetic, also values of the energy potential $e$ itself could be included in the dataset. However, previous investigations showed that they would hardly increase the prediction quality of the calibrated models \cite{klein2022a}. Furthermore, values of the potential are usually not available from experimental investigations, thus, an approach which relies on these values would be less general. Note that the scaling strategy described in \cite[App.~A]{klein2022a} is applied to ensure that all electro-mechanical input and output quantities are in the same order of magnitude. 

\subsubsection{Model calibration}\label{sec:iso_calib}

For the model calibration, the NN parameters $\vect{p}$ are optimised to minimise the mean squared error
\begin{equation}\label{MSE}
\begin{aligned}
\mathcal{MSE}(\vect{p})=\frac{1}{\#(\mathcal{D})} \sum_{i} \bigg[
\left\|\vect{P}^{i}-\vect{P}^{\text{PANN}}\left(\vect{F}^{i},\,\vect{d}_0^{i};\,\vect{p}\right)\right\|^2 
+  \left\|\vect{e}_0^{i}-\vect{e}_0^{\text{PANN}}\left(\vect{F}^{i},\,\vect{d}_0^{i};\,\vect{p}\right)\right\|^2 \bigg]\,,
\end{aligned}
\end{equation}
where $\#(\mathcal{D})$ denotes the amount of datapoints in the calibration dataset. This special calibration strategy, where the NN is optimised through its gradients, is referred to as Sobolev training \cite{vlassis2020a,vlassis2022}. In this work, no sample weighting strategy was applied, since it did not lead to improved model qualities.

\medskip

An isotropic polyconvex PANN model is applied, which receives the invariants defined in \cref{eq:invs_iso} as input. The NN architecture consists of one hidden layer with 8 nodes, for which the short notation $\SP^+(8)$ is used. This results in a total of 72 trainable model parameters. The model is implemented and calibrated in TensorFlow 2.6.0 using Python 3.9.13. The optimization is carried out with the ADAM optimiser, where the full batch of training data, TensorFlow's default batch size, 2,000 calibration epochs, and a learning rate of $2\cdot 10^{-3}$ are applied. The model is calibrated three times to account for the influence of randomly initialised parameters and the stochastic nature of the optimisation algorithm.

\subsubsection{Model evaluation}

  \renewcommand{\arraystretch}{1.15}
\begin{table}[t]
 \centering
  \caption{log$_{10}(\mathcal{MSE})$ of the calibrated PANN models for the analytical isotropic potential (ISO), the RVE with spherical inclusion (RVE), and the rank-one laminate (ROL).}
 \begin{tabular}
 {llp{0.6cm}llp{0.6cm}llp{0.03cm}llp{0.03cm}ll}
\toprule
\multicolumn{2}{l}{ISO} && \multicolumn{2}{l}{RVE} && \multicolumn{8}{l}{ROL} \\
  \cmidrule(lr ){1-2}    \cmidrule(lr){4-5}   \cmidrule(lr){7-14} 
\multicolumn{2}{l}{$\SP^+(8)$}& &\multicolumn{2}{l}{$\SP^+(16)$}& &\multicolumn{2}{l}{$\SP^+(8)$} && \multicolumn{2}{l}{$\SP^+(64)$} && \multicolumn{2}{l}{$\SP(8)$} \\
 calib. & test && calib. & test & & calib. & test && calib. & test && calib. & test \\
  \cmidrule(lr){1-2}    \cmidrule(lr){4-5}   \cmidrule(lr){7-8}  \cmidrule(lr){10-11}\cmidrule(lr){13-14}    
 -8.00 & -8.01 && -3.47 &-3.26  && -2.91 & -2.85  && -2.94 & -2.85 & & -4.04 &-3.56  \\
 -7.43 & -7.41 & &-3.40 & -3.30 & & -2.87 &-2.81  && -3.01 & -2.93  && -3.87 &-3.50 \\
  -7.04 & -7.08 &&  -3.29 & -3.32 &  & -2.86 &-2.85 && -2.95 & -2.88 & & -4.32 &-3.79 \\
  \bottomrule
 \end{tabular}
 \label{tab:loss}
\end{table}

In \cref{tab:loss}, the MSEs of the three calibrated models are shown on the left. In all cases, the models show excellent results for both the calibration and the test dataset. Since the function to which the PANN is calibrated is mostly linear in the invariants, cf.~\cref{eq:MR_2}, this excellent result is to be expected, cf.~\cite{linden2023}.
For the finite element analyses conducted in \cref{sec:FEA}, the model with the best test MSE is applied. There, also a more visual evaluation of the model performance is provided.


\subsection{Numerically homogenised RVE with spherical inclusion}\label{subsec:RVE}

\begin{figure}[t]
	\centering
	\begin{tabular}{cc}
		\includegraphics[height=4.45cm]{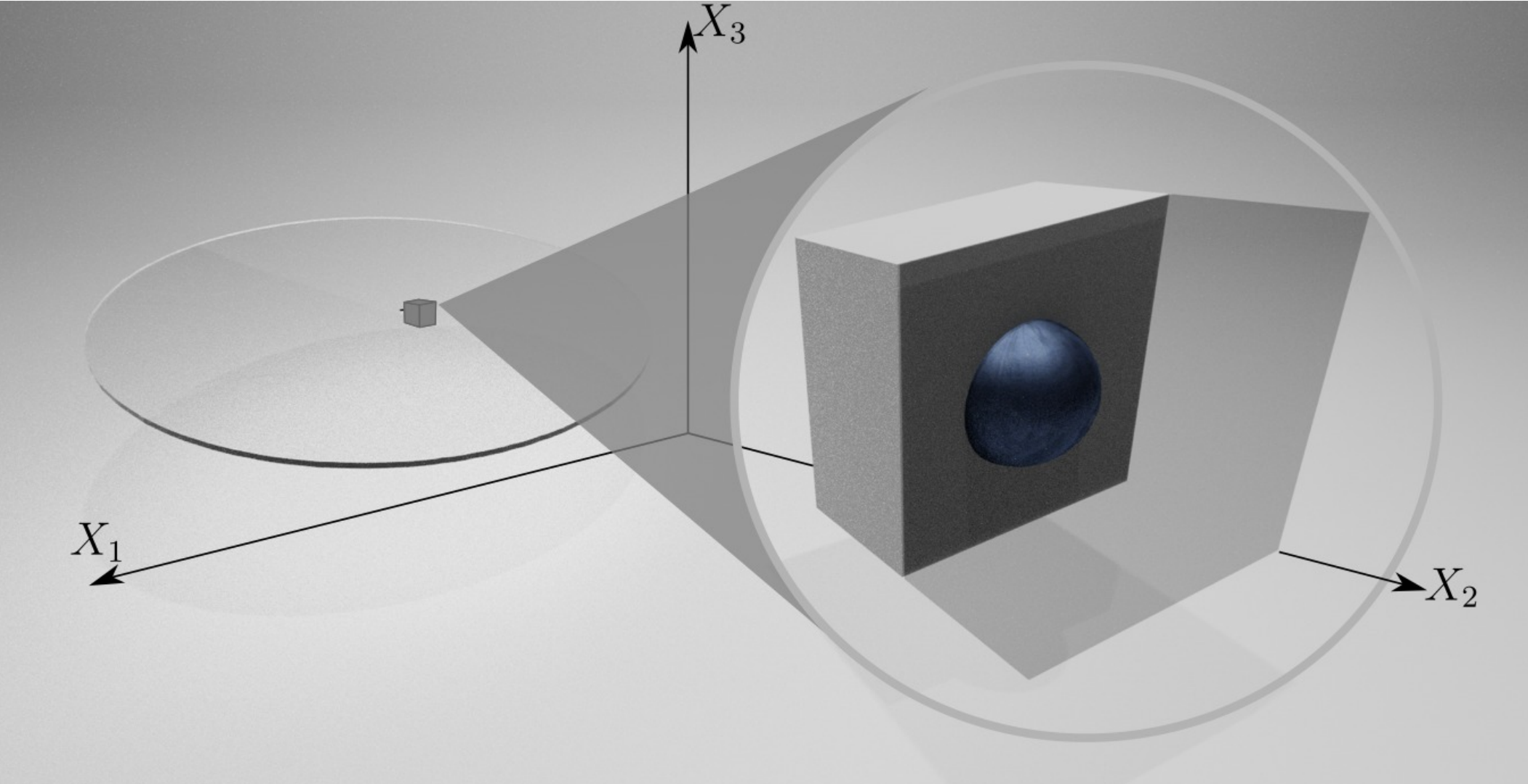} &
		\includegraphics[height=4.45cm]{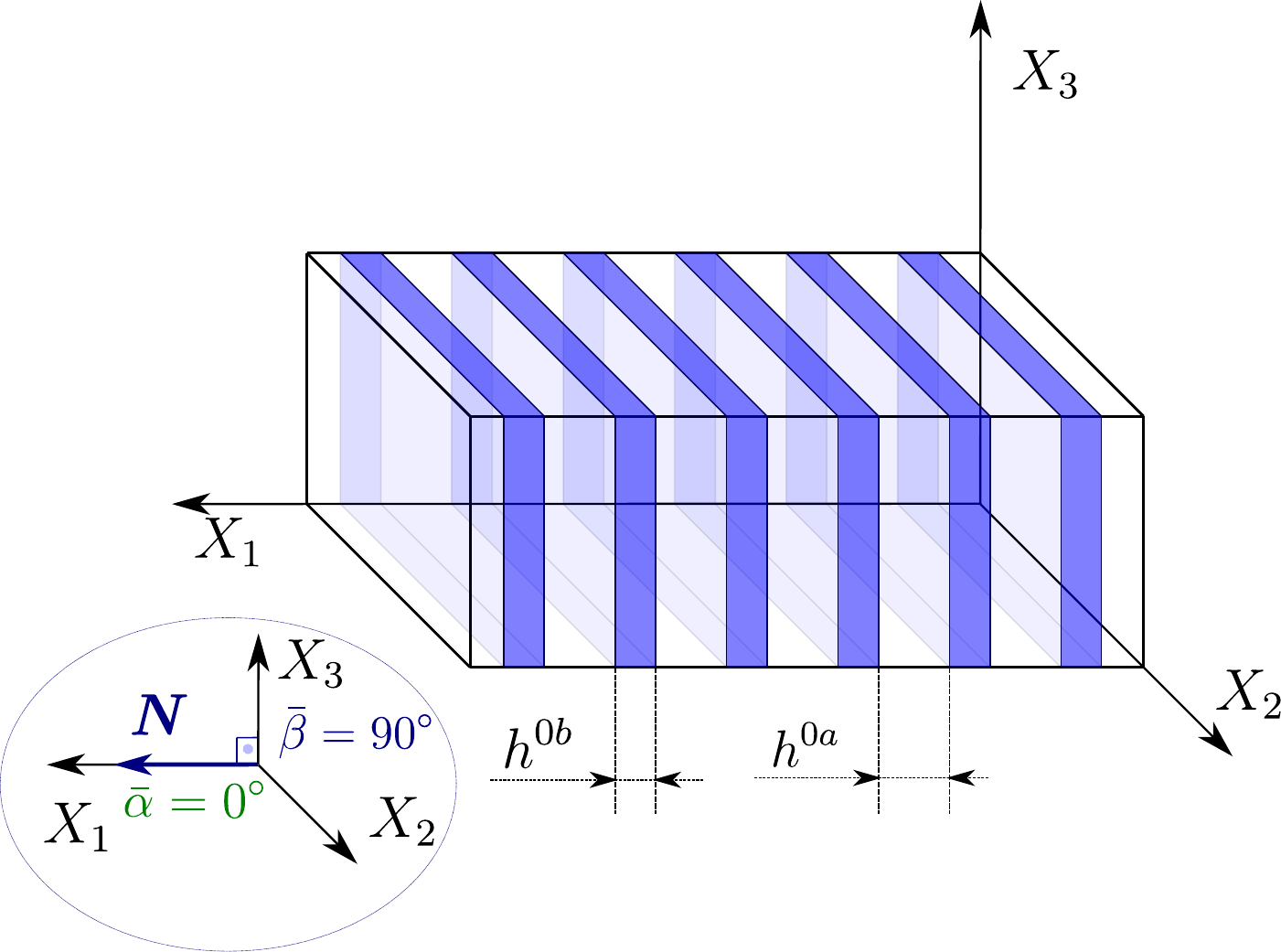}\\
		\small{(a) RVE with spherical inclusion} & \small{(b) Rank-one laminate}
	\end{tabular}
	\caption{Microstructures considered in this work. (a) RVE with spherical inclusion.  (b) Rank-one laminate composite material (particular case with $\vect{N}=[1\,0\,0]^T$).}
	\label{fig:microstructures}
\end{figure}

\subsubsection{Data generation}

Entering the more intricate realm of composite materials, a numerically homogenised RVE with a spherical inclusion is considered, cf.~\cref{fig:microstructures}a.
The material behavior in both phases is characterised by isotropic, Mooney-Rivlin type potentials, i.e.,
\begin{equation}\label{Equation:ROL}
\begin{aligned}	
e^a \left(\vect{F}^a,\vect{d}_0^a \right)=\widetilde{\mathcal{P}}^a(\vect{\mathcal{I}}^a) &= \frac{\mu^a_1}{2}I_1^a + \frac{\mu^a_2}{2}I_2^a- \left(\mu^a_1 + 2\mu^a_2\right)\log{J^a} + \frac{\lambda^a}{2}(J^a-1)^2 + \frac{I_5^a}{2\varepsilon^aJ^a}\,,
\\
e^b \left(\vect{F}^b,\vect{d}_0^b \right)=\widetilde{\mathcal{P}}^b(\vect{\mathcal{I}}^b) &=f_m\left( \frac{\mu^b_1}{2}I_1^b+ \frac{\mu^b_2}{2}I_2^b - \left(\mu^b_1 + 2\mu^b_2\right)\log{J^b} + \frac{\lambda^b}{2}(J^b-1)^2 \right)+ \frac{I_5^b}{2f_e\varepsilon^aJ^b}\,,
\end{aligned}	
\end{equation}
where $a$ refers to the matrix material, $b$ refers to the inclusion, and $f_m,\,f_e$ denotes the mechanical and electrical contrast between the phases. Superscripts on the invariants, energies, and material parameters assign them to one of the phases.
The material parameters are set to
\begin{equation}
    \mu_1^a=\mu_2^a=0.5\,,\qquad \lambda_a=5\,,\qquad \varepsilon^a=1\,,\qquad f_m= f_e=5\,.
\end{equation}
Furthermore, the radius of the inclusion is a quarter of the cell length. The homogenisation of the RVE material is carried out numerically. For details on the homogenisation, the reader is referred to \cite{klein2022b}. The homogenisation provides datasets of the form introduced in \cref{eq:dataset}. 
For data generation, the same strategy as described in \cref{sec:iso_data} is applied, resulting in a calibration dataset consisting of 500 datapoints and a test dataset consisting of 900 datapoints.

\subsubsection{Model calibration}

An isotropic polyconvex PANN model is applied, which receives the invariants defined in \cref{eq:invs_iso} as input. The NN architecture consists of one hidden layer with 16 nodes, for which the short notation $\SP^+(16)$ is used. This results in a total of 144 trainable model parameters. The remaining calibration details are chosen as described in \cref{sec:iso_calib}.

\subsubsection{Model evaluation}\label{sec:RVE_eval}

\begin{figure}[t]
	\begin{center}		
		\begin{tabular}{ c }
			\includegraphics[width=0.99\textwidth]{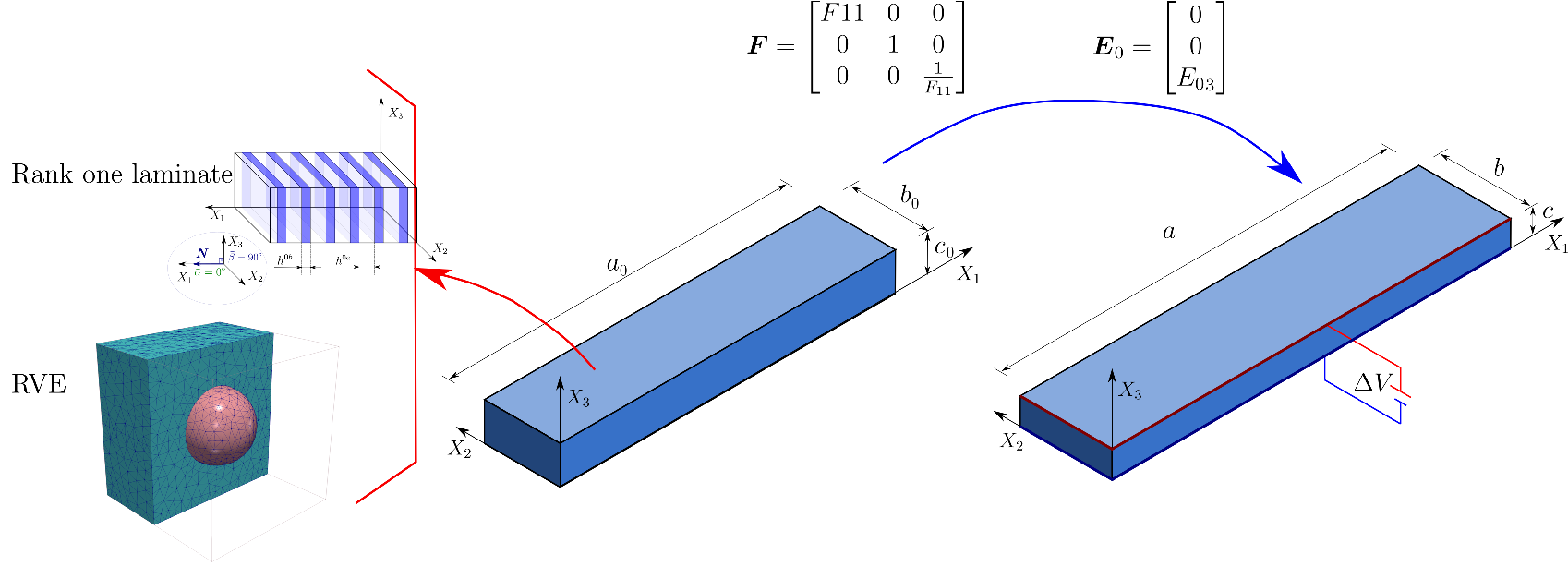}	
		\end{tabular}
	\end{center}
	\vspace{-6mm}
	\caption{Uniform deformation induced in a prism of dimensions $[a_0,b_0,c_0]$ after applying a voltage gradient $\Delta V=-\frac{e_0}{c_0}$, yielding a uniformly deformed prism with dimensions $[a,b,c]$.}
	\label{fig:0D example definition}
\end{figure}

\begin{figure}[t]
    \begin{center}
        \resizebox{0.95\textwidth}{!}{
        \input{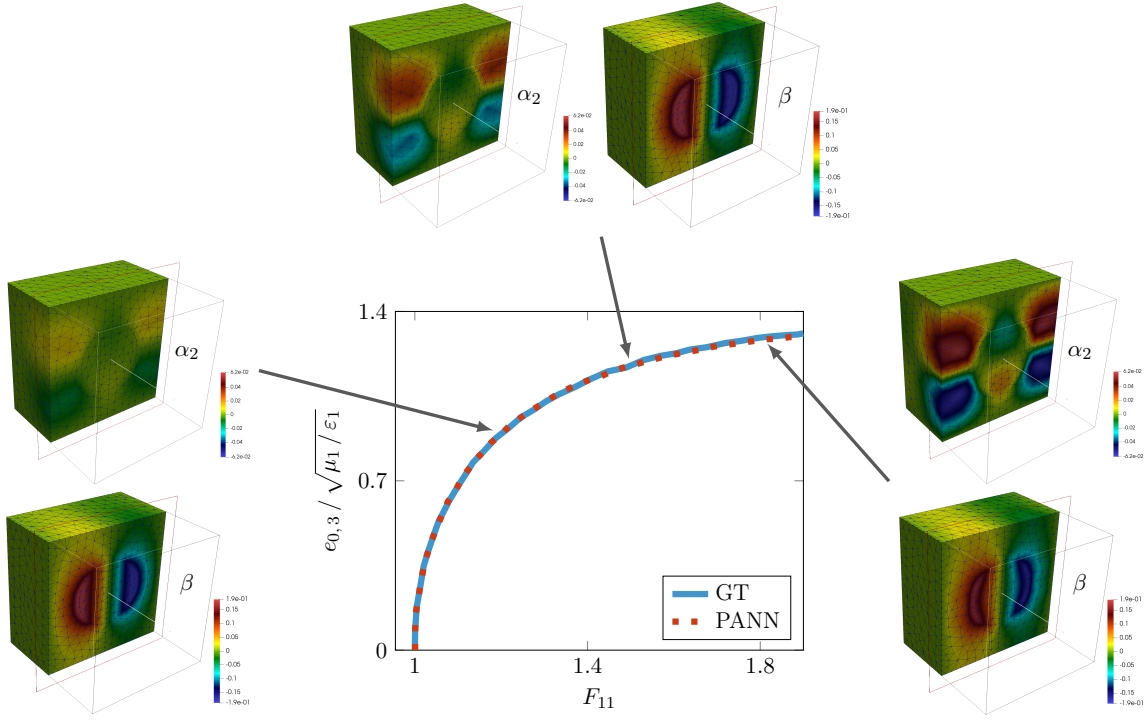}
        }        
    \end{center}
	\caption{Equilibrium path for the PANN model and the RVE with a spherical inclusion. 
	 Contour plot distribution of $\alpha_2$ (mechanical micro-fluctuation) and $\beta$ (micro-fluctuation of the electric potential).
	 The PANN model shows an excellent prediction quality.}
	\label{fig:RVE_results}
\end{figure}

In \cref{tab:loss}, the MSEs of the three calibrated models are shown in the middle. In all cases, the models exhibit excellent results for both calibration and test dataset.
This implies that the RVE has a practically isotropic behavior.
However, for other RVE configurations, e.g., a larger contrast between the phases, it is to be expected that the RVE obtains a more pronounced anisotropy and an isotropic model would not be applicable.
For the following investigation and the FEA conducted in \cref{sec:FEA}, the model with the best test MSE is applied.

While the material behavior of the microstructure in both matrix and inclusion is polyconvex, this does not imply that the effective behavior can be represented by a polyconvex constitutive model, cf.~\cref{subsec:ROL}. 
Still, in this particular example, a polyconvex PANN model is able to represent the homogenised behavior of the RVE, which ensures ellipticity and thus a stable behavior when applying the model in numerical applications. Again, for other RVE configurations such as a larger contrast between the phases, the homogenised behavior of the microstructure would change, and a polyconvex model might not be applicable.

\medskip

To further investigate the performance of the calibrated PANN model, investigations on an equilibrium path are carried out. This can be seen as a study on the Gauss point level for conditions where the deformation is uniformly distributed, cf.~\cref{fig:0D example definition}. 
To conduct this analysis, we introduce the following potential
\begin{equation}\label{eqn:total energy}
{\mathcal{L}}(\vect{F},\,\vect{d}_0,\,p) =   \widetilde{e}(\vect{F},\,\vect{d}_0,\,p) - \vect{e}_0\cdot \vect{d}_0\,,\qquad \widetilde{e}(\vect{F},\,\vect{d}_0,\,p)=e(\vect{F},\,\vect{d}_0) + p\left(J - 1\right)
\end{equation}
Here, $e(\vect{F},\vect{d}_0)$ represents the internal energy density of the electro-active material (whether it is the ground truth or PANN), whereas 
the  second term is responsible for the enforcement of the incompressibility condition  $J=1$, through the introduction of a Lagrange multiplier $p$. 
To become more precise, the load scenario
\begin{equation}\label{eqn:parametrisation}
\vect{F}=\begin{bmatrix}
F_{11}  &  0  &  0\\
0  &  1  &  0\\
0  &  0  &  F_{33}\\
\end{bmatrix}\,,\qquad 
\vect{e}_0=\begin{bmatrix}
0\\
0\\ 
e_0
\end{bmatrix}
\end{equation}
is applied. While the value of $e_0$ is prescribed, the remaining variables $\{F_{11},\,F_{33},\,p,\,\vect{d}_0\}$ are obtained from the stationary conditions of $\mathcal{L}$ in \cref{eqn:total energy}, i.e. 
\begin{equation}
\begin{aligned}
D\mathcal{L}[\delta F_{11}]& = \partial_{\vect{F}}e:D\vect{F}[\delta F_{11}]=0\,,&\qquad
D\mathcal{L}[\delta F_{33}]& = \partial_{\vect{F}}e:D\vect{F}[\delta F_{33}]=0\,,\\ 
D\mathcal{L}[\delta p]&=(J-1)\delta p=0\,,&\qquad
D\mathcal{L}[\delta \vect{d}_0]& = \left(\partial_{\vect{d}_0}e - \vect{e}_0\right)\cdot\delta\vect{d}_0=0\,.\\ 
\end{aligned}
\end{equation}
This results in the nonlinear system of equations
\begin{equation}
\begin{aligned}	
{\mathcal{R}}_{F_{11}}&=[\partial_{\vect{F}}e]_{11}=0\,,& \qquad
{\mathcal{R}}_{F_{33}}&=[\partial_{\vect{F}}e]_{33}=0\,, \\
{\mathcal{R}}_{p}&=J-1=0\,, &\qquad 
\mathcal{R}_{\vect{d}_0}&=\partial_{\vect{d}_0}e - \vect{e}_0=\vect{0}\,,
\end{aligned}
\end{equation}
which is solved numerically.

Figure \ref{fig:RVE_results} illustrates the equilibrium path  for both the ground truth, numerically homogenised RVE composite and its PANN counterpart, where the PANN model shows an excellent prediction quality. Furthermore, for three different snapshots within the equilibrium path, \cref{fig:RVE_results} shows the contour plot of the micro-fluctuations for both mechanics and electric potential obtained by the ground truth model.

\subsection{Analytically homogenised rank-one laminate}\label{subsec:ROL}

\subsubsection{Data generation}\label{sec:ROL_data}

As the last and most challenging example, a rank-one laminate material is considered, cf.~\cref{fig:microstructures}b.
This microstructure consists of two phases, which are repeatedly stacked perpendicular to the preferred direction $\vect{N}$. The material behavior in both phases is again characterised by Mooney-Rivlin type isotropic potentials, i.e.,
\begin{equation}\label{Equation:ROL}
\begin{aligned}	
e^a \left(\vect{F}^a,\vect{d}_0^a \right) =\widetilde{\mathcal{P}}^a(\vect{\mathcal{I}}^a)&= \frac{\mu^a_1}{2}I_1^a + \frac{\mu^a_2}{2}I_2^a- \left(\mu^a_1 + 2\mu^a_2\right)\log{J^a} + \frac{\lambda^a}{2}(J^a-1)^2 + \frac{I_5^a}{2\varepsilon^aJ^a}\,,
\\
e^b \left(\vect{F}^b,\vect{d}_0^b \right)=\widetilde{\mathcal{P}}^b(\vect{\mathcal{I}}^b) &=f_m\left( \frac{\mu^b_1}{2}I_1^b+ \frac{\mu^b_2}{2}I_2^b - \left(\mu^b_1 + 2\mu^b_2\right)\log{J^b} + \frac{\lambda^b}{2}(J^b-1)^2 \right)+ \frac{I_5^b}{2f_e\varepsilon^aJ^b}\,.
\end{aligned}	
\end{equation}
where $a,\,b$ refers to the two different phases and $f_m,\,f_e$ denotes the mechanical and electrical contrast between the phases. Superscripts on the invariants, energies, and material parameters assign them to one of the phases. 
The material parameters are chosen as
\begin{equation}
    \mu_1^a=\mu_2^a=0.5\,,\qquad \lambda^a=5\,,\qquad  \varepsilon^a=1\,,\qquad f_m=f_e=2\,.
\end{equation}
Furthermore, the relative volume of phase $a$ within the composite is chosen as 0.6, and the preferred direction is set to $\vect{N}=0.5\,\big[1,\,1,\,\sqrt{2}\big]$. The homogenisation of the rank-one laminate material is carried out analytically, see \cite{klein2022b} for further details. 

The homogenisation provides datasets of the form introduced in \cref{eq:dataset}.
For data generation, the same strategy as described in \cref{sec:iso_data} is applied, resulting in a calibration dataset consisting of 500 datapoints and a test dataset consisting of 30,000 datapoints. For each datapoint, the ellipticity is numerically checked by examining the positive semi-definiteness of the acoustic tensor, cf.~\cref{eq:acoustic_tensor}. This is possible since the homogenisation not only provides values of the first gradients of the potential, i.e., $\vect{P}$ and $\vect{e}_0$, but also for the second gradients required for calculation of the acoustic tensor. For this, all possible directions $\vect{V}$ of the acoustic tensor are parametrized spherically according to
\begin{equation}\label{eqn:spherical parametrisation}
\vect{V}=\begin{bmatrix}
\cos\theta\sin\psi,&  \sin\theta\sin\psi,&  \cos\psi\end{bmatrix}\,,\qquad \theta\in[0,2\pi]\,,\quad \psi\in[0,\pi]\,.
\end{equation}
In the calibration dataset, a loss of ellipticity is observed for 0.6\% of the datapoints, while for the test dataset, a loss of ellipticity is oberved for 0.16\% of the datapoints.
The loss of ellipticity occurs for datapoints with large mechanical and electrical loads. This suggests that the considered dataset is at the verge of loss of ellipticity, and increased loads would lead to a pronounced non-elliptic behavior of the rank-one laminate.
By not considering the non-elliptic datapoints in the following investigations, the rank-one laminate is only investigated within its elliptic regime.

\subsubsection{Model calibration}\label{sec:ROL_calib}

\begin{figure}[t!]
    \centering
        \begin{subfigure}[b]{0.45\textwidth}   
        \centering 
        \resizebox{!}{0.65\textwidth}{
        \tikzsetnextfilename{ROL_stress_correspondence_pc_31}

\begin{tikzpicture}

\begin{axis}[enlargelimits=false, axis on top, axis equal image, width=7cm,xtick={-2,-1,0,1},ytick={-2,-1,0,1},xlabel={GT},ylabel={PANN}]
\addplot graphics [xmin=-2.15,xmax=1.95,ymin=-2.15,ymax=1.95] {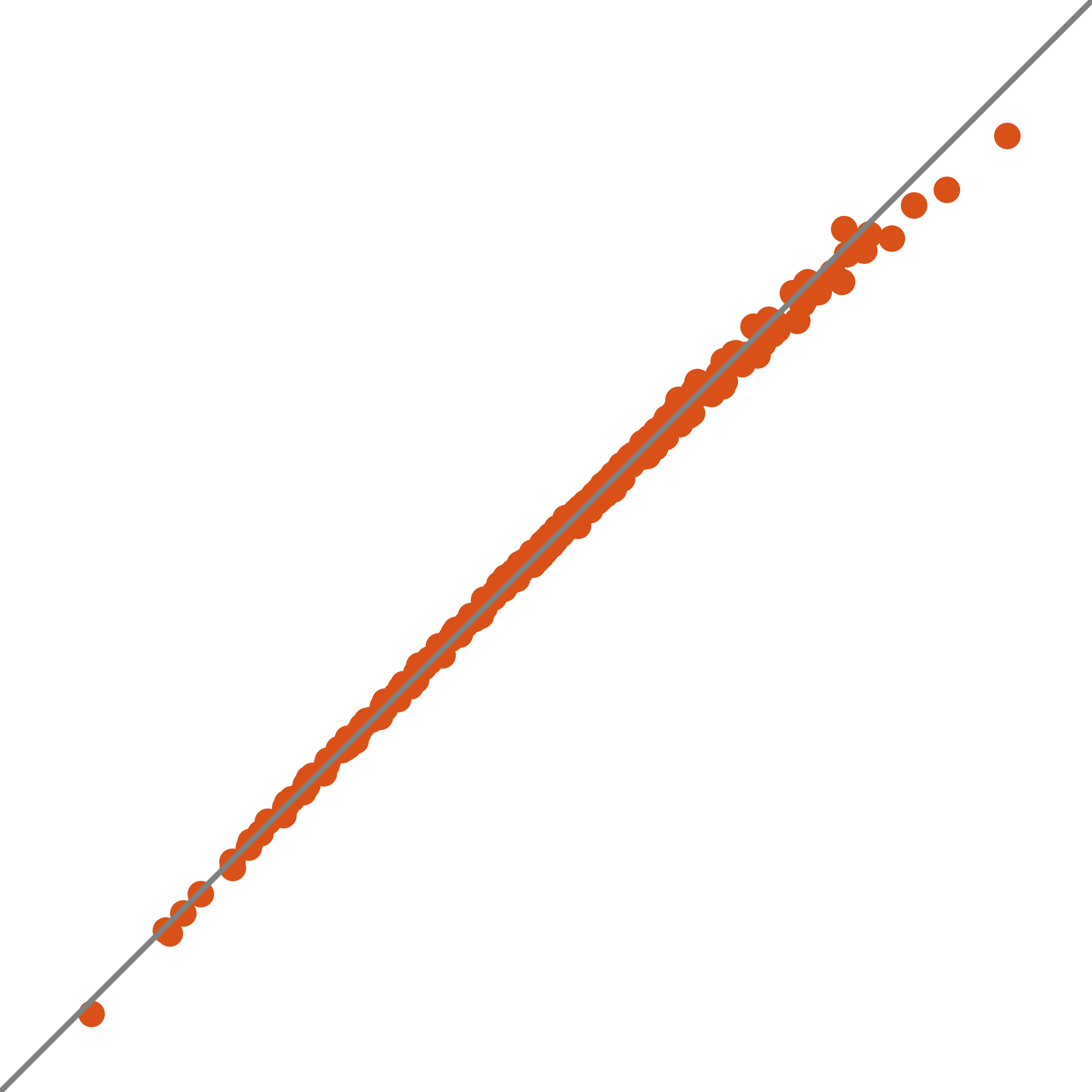};
\end{axis}

\end{tikzpicture}
        }    
        \caption{$\SP^+(64)$}
    \end{subfigure}
         \begin{subfigure}[b]{0.45\textwidth}   
        \centering 
        \resizebox{!}{0.65\textwidth}{
        \tikzsetnextfilename{ROL_stress_correspondence_non_pc_31}

\begin{tikzpicture}

\begin{axis}[enlargelimits=false, axis on top, axis equal image, width=7cm,xtick={-2,-1,0,1},ytick={-2,-1,0,1},xlabel={GT},ylabel={PANN}]
\addplot graphics [xmin=-2.15,xmax=1.95,ymin=-2.15,ymax=1.95] {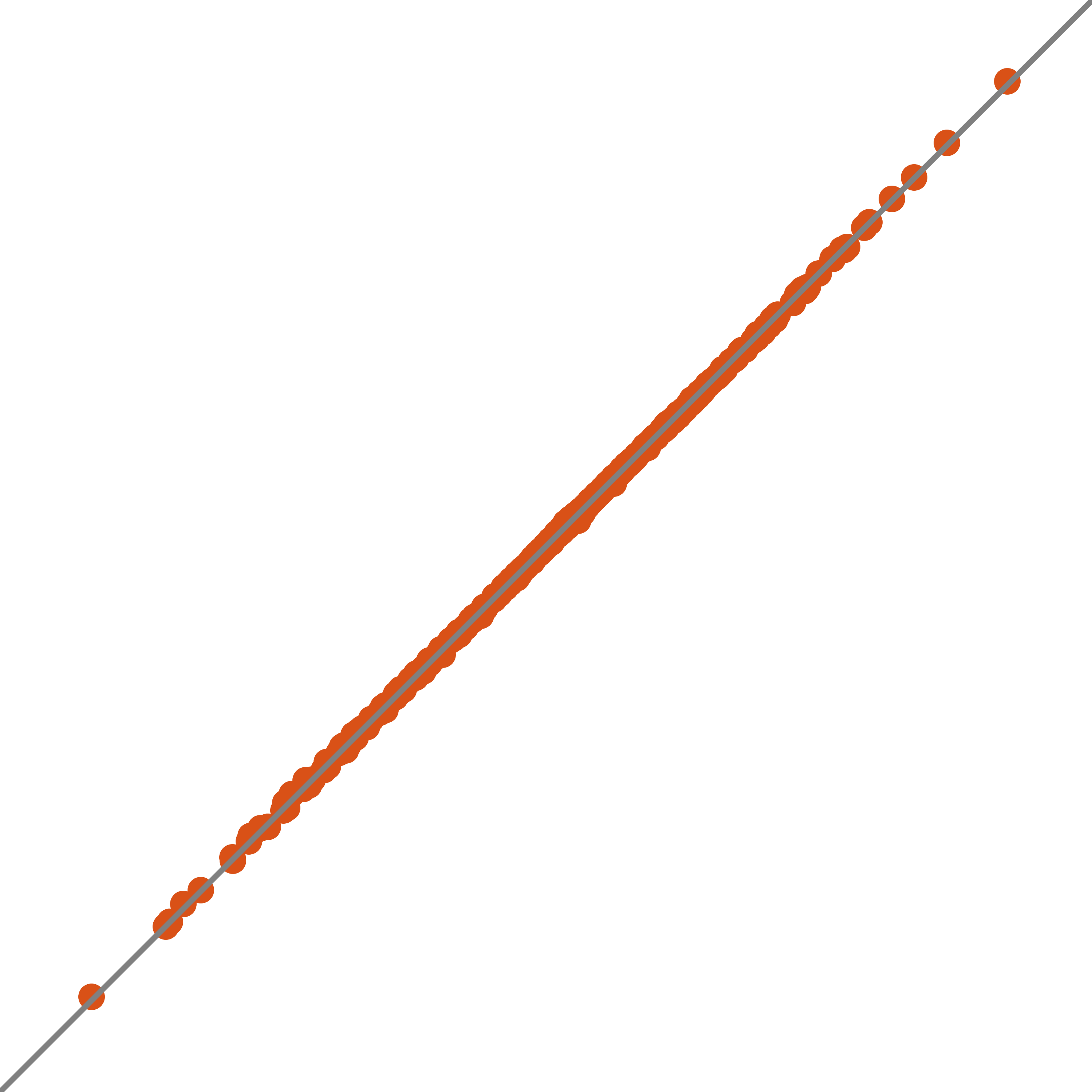};
\end{axis}

\end{tikzpicture}
        }    
        \caption{$\SP(8)$}
    \end{subfigure}
	\caption{Stress correspondence plots for the (a) polyconvex $\SP^+(64)$ and the (b) non-polyconvex $\SP(8)$ PANN models  calibrated to rank-one laminate data, evaluated for the $P_{31}$ stress of the test dataset. The polyconvex model shows some visible deviations from the ground truth model, while the non-polyconvex model shows an excellent performance.}\label{fig:ROL_corr} 
\end{figure}

Three distinct transversely isotropic PANN models are applied, which receive the invariants defined in \cref{eq:ti_invar} as input and apply the preferred direction introduced in \cref{sec:ROL_data}. Two polyconvex models are calibrated, where one has 8 nodes and the other has 64 nodes in the hidden layer. Furthermore, a non-polyconvex model with a single hidden layer and 8 nodes is applied. The models denoted as $\{\SP^+(8),\,\SP^+(64),\,\SP(8)\}$ have $\{120,\,960,\,88\}$ trainable parameters. The calibrations are carried out with the hyperparameters introduced in \cref{sec:iso_calib}, with the only difference that 5,000 calibration epochs are used.

\subsubsection{Model evaluation}\label{sec:ROL_eval}

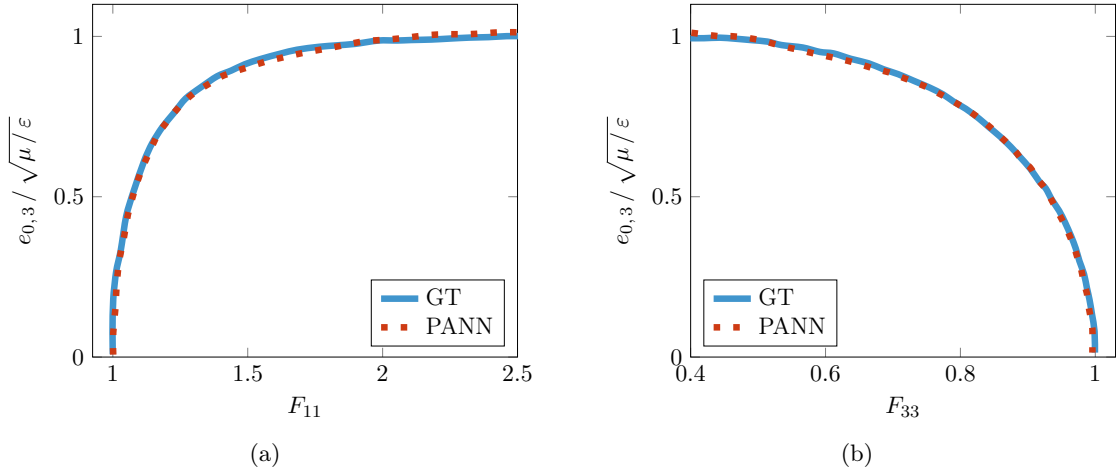
\begin{figure}[t!]
    \centering
    \begin{subfigure}[b]{0.45\textwidth}
        \centering
        \resizebox{!}{0.75\textwidth}{
        \tikzsetnextfilename{ZeroDROL_equilibrium_F11}

\begin{tikzpicture}
\begin{axis}[
        xlabel=$F_{11}$,
        ylabel=$e_{0,\,3}\,/\,\sqrt{\mu\,/\,\varepsilon}$,
        xmin=0.925, xmax=2.5, 
        ymin=0, ymax=1.1, 
        xtick={1,1.5,2,2.5},
        ytick={0,0.5,1},
        legend cell align={left},
        legend pos = south east,
        ]

\addplot[blueR!75, line width=3pt, smooth,mark=none] table[mark=none] {fig/ZeroDROL/F11_GT.txt};
\addplot[redR, line width=3pt, smooth,mark=none,loosely dashed] table[mark=none] {fig/ZeroDROL/F11_PANN.txt};

\addlegendentryexpanded{GT}
\addlegendentryexpanded{PANN}

\end{axis}
\end{tikzpicture}
        }
        \caption{}
    \end{subfigure}
    \hspace{0.15cm}
    \begin{subfigure}[b]{0.45\textwidth}   
        \centering 
        \resizebox{!}{0.75\textwidth}{
        \tikzsetnextfilename{ZeroDROL_equilibrium_F33}

\begin{tikzpicture}
\begin{axis}[
        xlabel=$F_{33}$,
        ylabel=$e_{0,\,3}\,/\,\sqrt{\mu\,/\,\varepsilon}$,
        xmin=0.4, xmax=1.03, 
        ymin=0, ymax=1.1, 
        xtick={0.4,0.6,0.8,1},
        ytick={0,0.5,1},
        legend cell align={left},
        legend pos = south west,
        ]

\addplot[blueR!75, line width=3pt, smooth,mark=none] table[mark=none] {fig/ZeroDROL/F33_GT.txt};
\addplot[redR, line width=3pt, smooth,mark=none,loosely dashed] table[mark=none] {fig/ZeroDROL/F33_PANN.txt};

\addlegendentryexpanded{GT}
\addlegendentryexpanded{PANN}

\end{axis}
\end{tikzpicture}
        }    
        \caption{}
    \end{subfigure}
	\caption{Two $\bF$-components of the equilibrium path for the homogenised rank-one laminate and the non-polyconvex PANN model $\SP(8)$. The PANN mode shows an excellent prediction quality. }\label{fig:equ_ROL}
\end{figure}

\begin{figure}[t!]
    \centering
        \begin{subfigure}[b]{0.45\textwidth}   
        \centering 
        \resizebox{!}{0.63\textwidth}{
        \tikzsetnextfilename{ROL_stress_coeff_pc}

\begin{tikzpicture}

\begin{axis}[enlargelimits=false, axis on top, axis equal image, width=8cm,xtick={1,2,3},ytick={0.375,1.5,2.625},yticklabels={-0.3,-0.15,0},ylabel={$\partial_{\ItwoTI}\,e^{\text{PANN}}-\partial_{\IfiveTI}\,e^{\text{PANN}}$},xlabel={$\ItwoTI$}]
\addplot graphics [xmin=0,xmax=4.5,ymin=0,ymax=3] {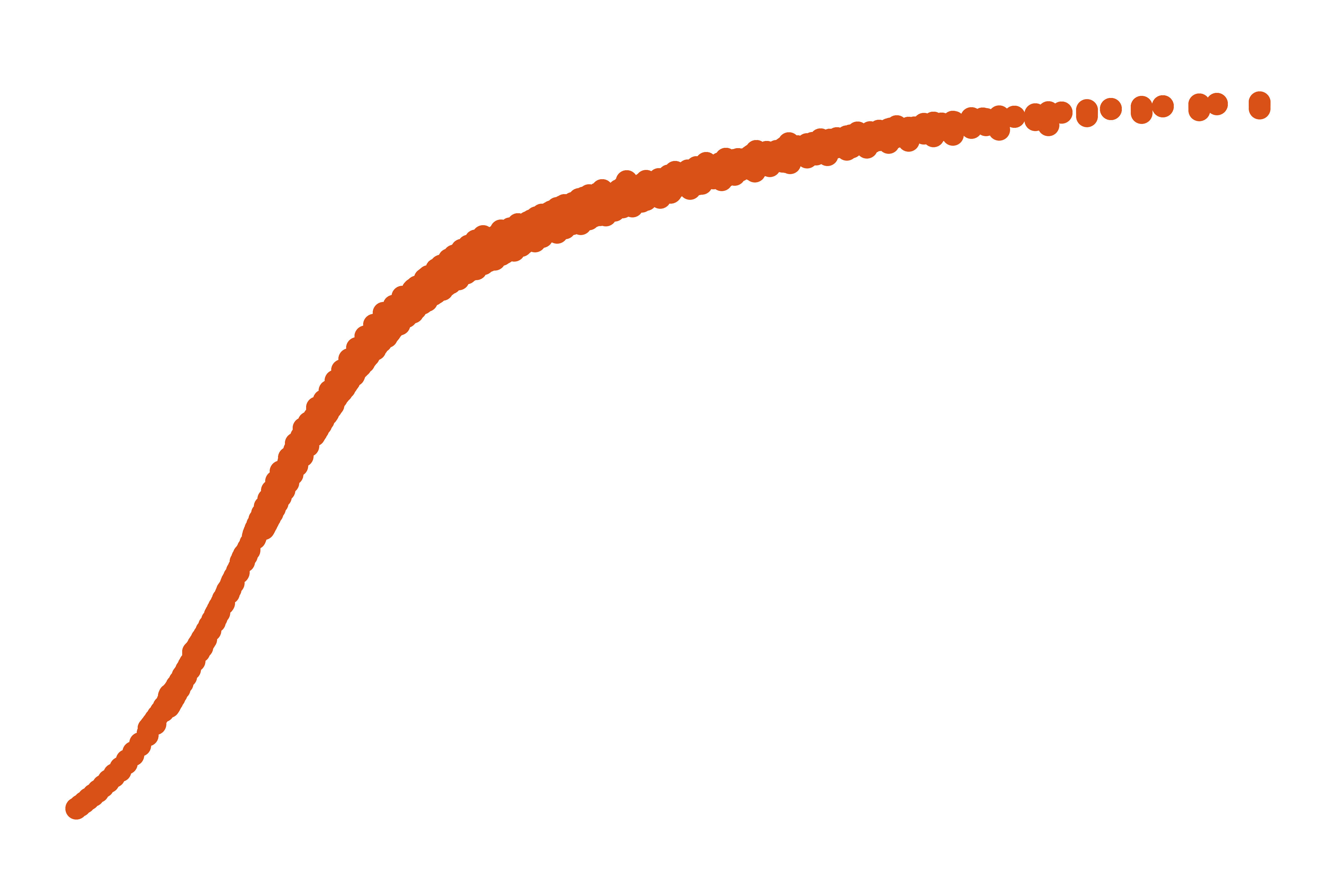};
\end{axis}

\end{tikzpicture}
        }    
        \caption{$\SP^+(64)$}
    \end{subfigure}
    \hspace{0.2cm}
         \begin{subfigure}[b]{0.45\textwidth}   
        \centering 
        \resizebox{!}{0.63\textwidth}{
        \tikzsetnextfilename{ROL_stress_coeff_non_pc}

\begin{tikzpicture}

\begin{axis}[enlargelimits=false, axis on top, axis equal image, width=8cm,xtick={1,2,3},ytick={0.375,1.5,2.625},yticklabels={-1.73,-0.93,-0.12},ylabel={$\partial_{\ItwoTI}\,e^{\text{PANN}}$},xlabel={$\ItwoTI$}]
\addplot graphics [xmin=0,xmax=4.5,ymin=0,ymax=3] {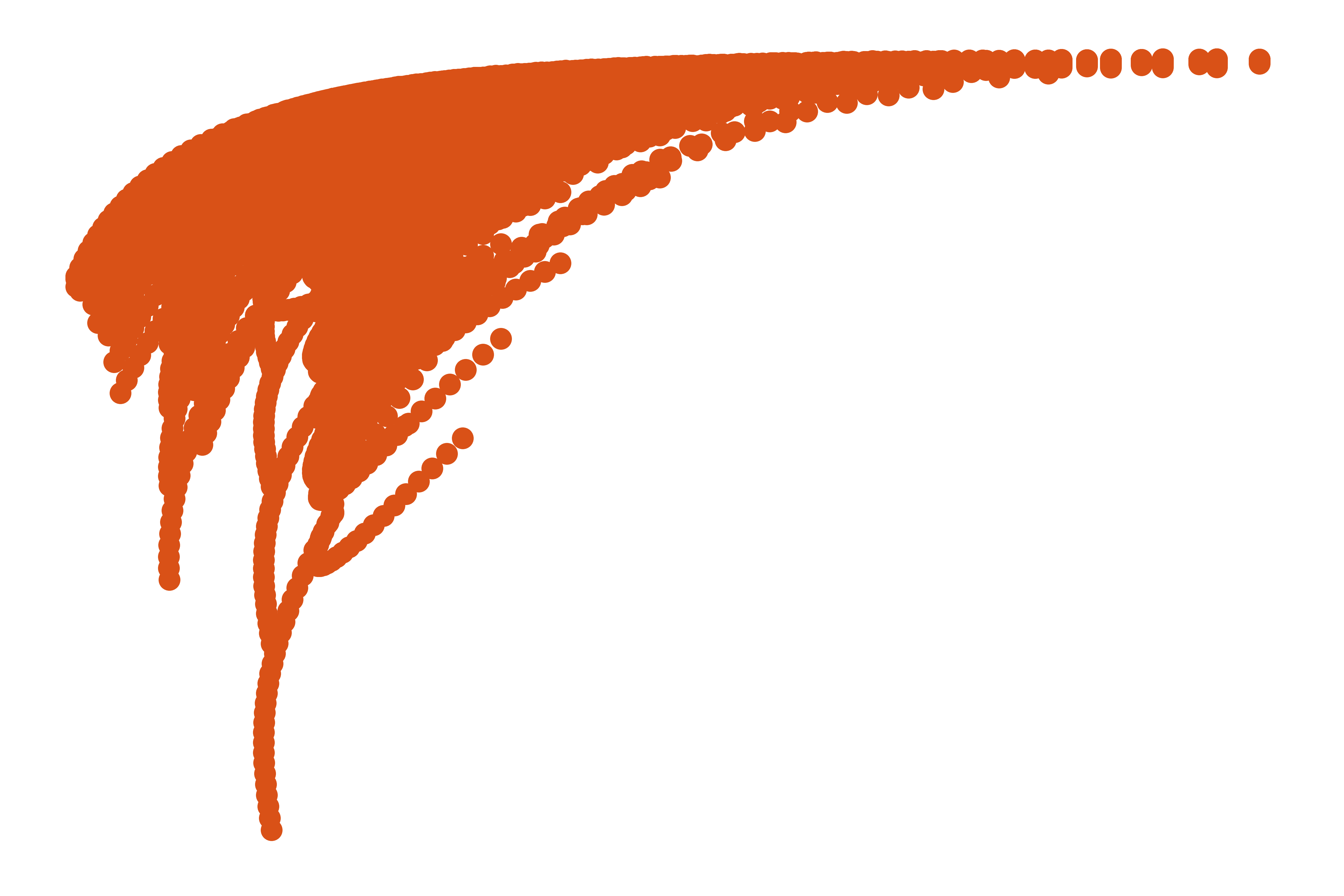};
\end{axis}

\end{tikzpicture}
        }    
        \caption{$\SP(8)$}
    \end{subfigure}
	\caption{Partial derivatives w.r.t.\ the invariants for (a) the polyconvex $\SP^+(64)$ and (b) the non-polyconvex $\SP(8)$ PANN model calibrated to rank-one laminate data, evaluated for the test dataset.
	The non-polyconvex potential takes pronounced negative values for the derivative w.r.t.\ $\ItwoTI$, which the polyconvex model can only partly represent, cf.~\cref{eq:add_inv_example}.
	}\label{fig:ROL_coeff}
\end{figure}

In \cref{tab:loss}, the MSEs of all calibrated models are shown on the right. The polyconvex models show decent predictions for both the calibration and test dataset, where the larger network architecture leads to only very slightly better results. Using a non-polyconvex model results in a remarkably better MSE for both the calibration and test dataset. The best test MSE for the non-polyconvex models of $10^{-3.79}$ is almost an order of magnitude smaller than the best test MSE for the polyconvex models of $10^{-2.93}$. For the following investigations, the polyconvex model with the best test MSE and the non-polyconvex model with the best test MSE are applied.

A more visual investigation of the model performance is provided in \cref{fig:ROL_corr}, where correspondence plots for the $P_{31}$ stress of both the polyconvex and the non-polyconvex model are provided. Again, the polyconvex model shows decent results, but also has some visible deviations from the ground truth model. On the other side, the non-polyconvex model shows an excellent performance. 
Furthermore, the equilibrium path investigation introduced in \cref{sec:RVE_eval} is repeated for the non-polyconvex PANN model $\SP(8)$ and its ground truth counterpart. As visualized in \cref{fig:equ_ROL}, the non-polyconvex model again shows an excellent prediction quality. 
Finally, to anticipate the results of the finite element analysis (FEA) conducted in \cref{sec:FEA}, there it is demonstrated that the prediction quality of the polyconvex PANN model is actually not sufficient for accurate simulation of even moderately complex simulations, cf.~\cref{fig:bending ROL}.
Thus, while on first sight, the prediction quality of the polyconvex PANN model seems decent, it is apparently not accurate enough to be applied in complex FEA.

\medskip

There are good reasons to trace the moderate performance of the polyconvex PANN model not to a badly executed calibration or an insufficient width or depth of the NN, but rather to its inherent mathematical properties. That is, the restrictions that polyconvexity imposes on the model. First of all, the excellent performance of the non-polyconvex model demonstrates that the transversely isotropic invariants introduced in \cref{eq:ti_invar} are sufficient to characterise the rank-one laminate material. However, the polyconvex model does not allow for potentials with arbitrary functional relationships in these invariants, but poses monotonicity and convexity constraints, cf.~\cref{sec:invs}. Thus, if the ground truth behavior includes non-monotonicity or non-convexity, a polyconvex model is not able to represent this. In fact, the non-polyconvex PANN model $\SP(8)$  takes pronounced decreasing functional relationships in several invariants, cf.~\cref{fig:ROL_coeff}, which apparently are required for an accurate representation of the rank-one laminate data. The polyconvex PANN model $\SP^+(64)$  can only partly represent these decreasing functional relationships by stabilising them with other parts of the model. For instance, the decreasing functional relationship in $\ItwoTI$ is stabilised by an increasing functional relationship of the potential in $I_2$, cf.~\cref{eq:ti_invar_add}.

Furthermore, both the small and the large polyconvex PANN model show more or less the same prediction quality, although the larger architecture has eight times as many trainable parameters than the smaller one. This again hints that the loss of flexibility is not a question of too few model parameters, but rather of the loss of flexibility due to the restrictions of polyconvexity.

\medskip

\begin{figure}[t]
	\begin{center}		
		\begin{tabular}{ c l c l}
			\includegraphics[trim={0 6cm 20cm 0},clip,width=0.375\textwidth]{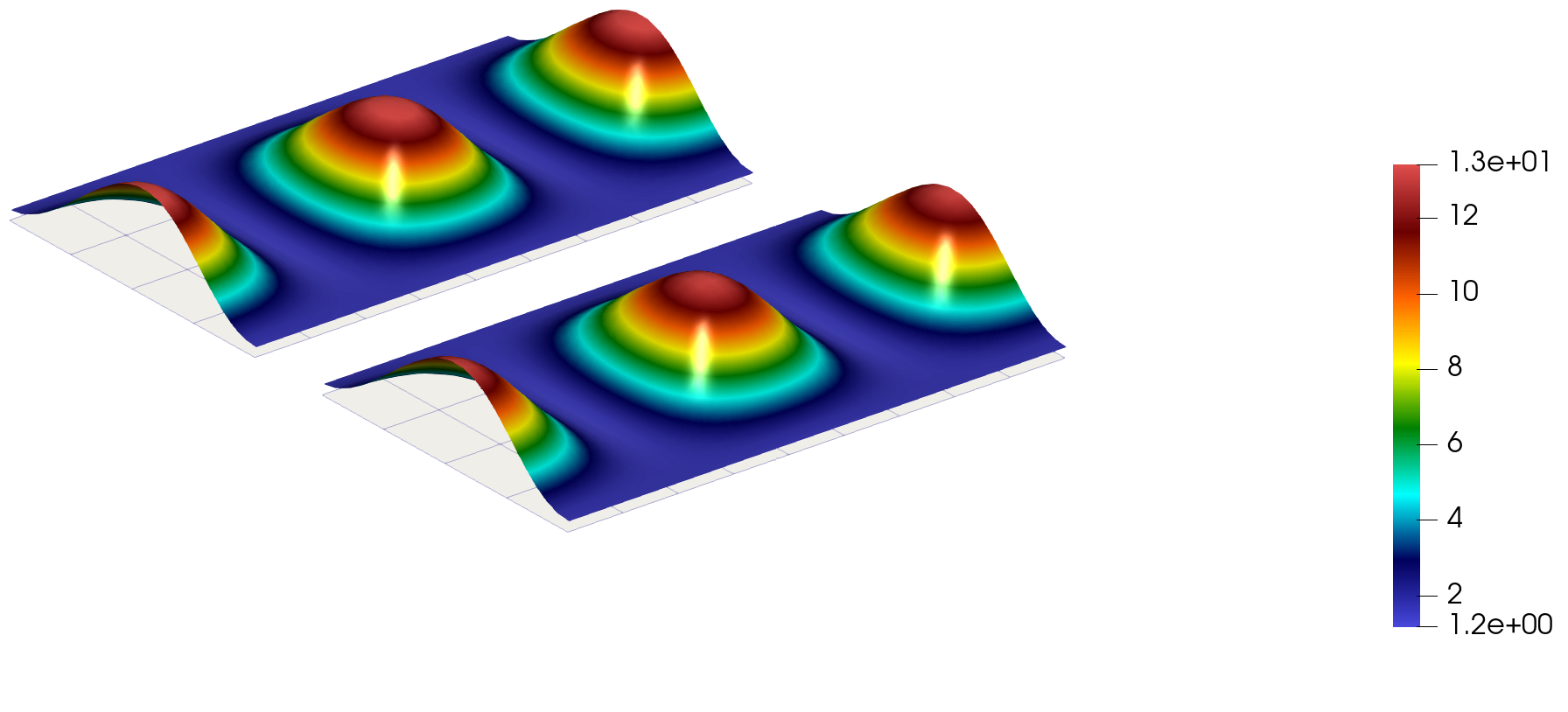} & 
			\includegraphics[trim={56cm 3cm 0 0},clip,width=0.07\textwidth]{fig/ZeroDROL/MinorsROLF11_1_5} & 
			\includegraphics[trim={0 6cm 20cm 0},clip,width=0.375\textwidth]{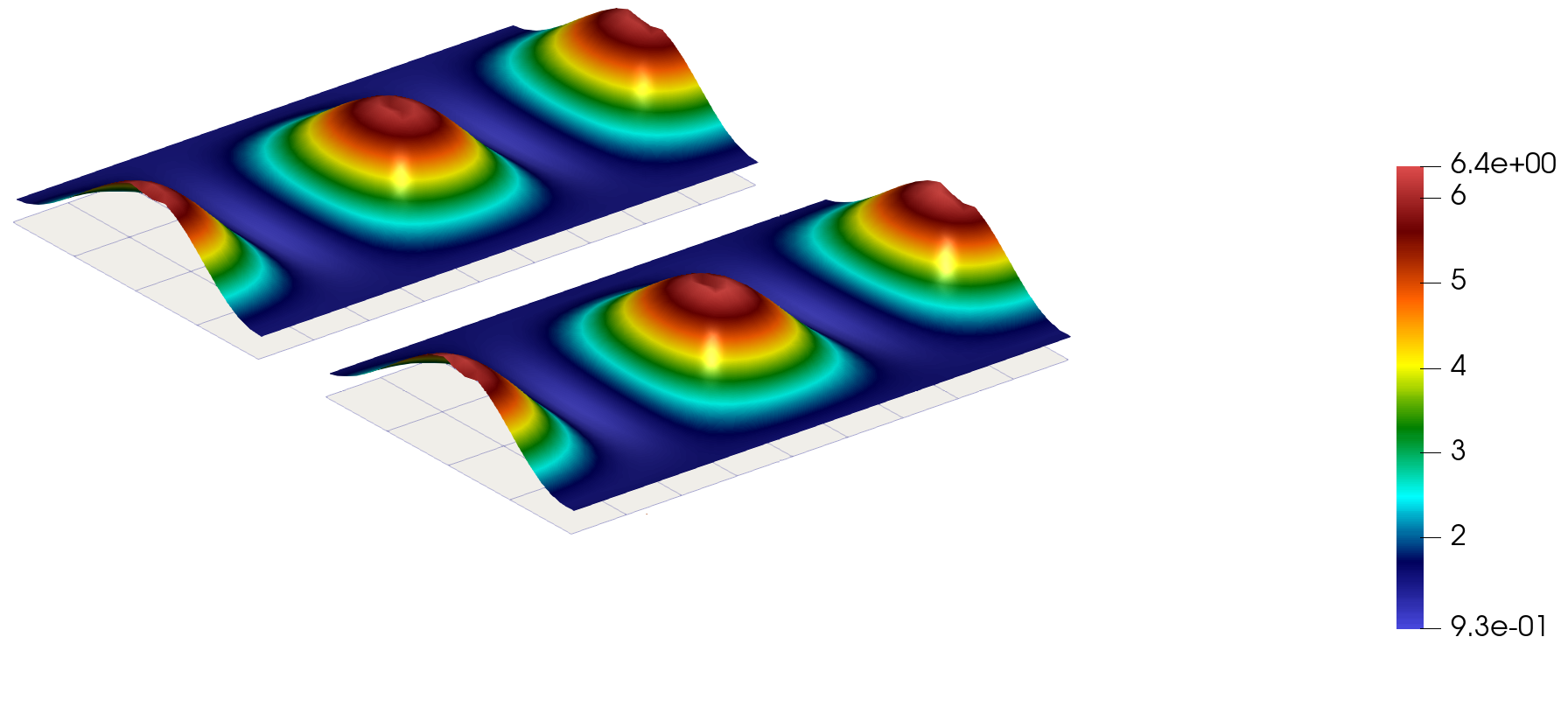}
			&
			\includegraphics[trim={56cm 3cm 0 0},clip,width=0.07\textwidth]{fig/ZeroDROL/MinorsROLF11_2}
			\\
			\small{(a) $F_{11}=1.5$} & & \small{(b) $F_{11}=2$} &
		\end{tabular}
	\end{center}
\vspace{-6mm}
	\caption{Spherical parametrisation of the least of the minors of the acoustic tensor for the homogenised rank-one laminate and the non-polyconvex PANN model $\SP(8)$  at: (a) $F_{11}=1.5$ and (b) $F_{11}=2$. In each subfigure, the left top plot represents the results of the ground truth model, while the right bottom plot represents the results of the PANN model.
}
	\label{fig:minors ROL}
\end{figure}

One of the major drawbacks of non-polyconvex PANN models is that they are not elliptic by construction. This necessitates more investigations on the non-polyconvex PANN models ellipticity as well as its second derivatives, which are closely related to each other, cf.~\cref{eqn:rank one conexity}.

At first, the ellipticity of the calibrated PANN model at two points along the equilibrium path introduced in \cref{sec:RVE_eval} is examined. For this, the least of the minors of the acoustic tensor is examined. Positive semi-definiteness of the acoustic tensors, i.e., non-negativity of the least of the minors, is equivalent to the ellipticity of the internal energy density, cf.~\cref{eq:acoustic_tensor}. For this, all possible directions $\vect{V}$ of the acoustic tensor are spherically parametrized as introduced in \cref{eqn:spherical parametrisation}.
In \cref{fig:minors ROL}, the results for both the PANN model and its ground truth counterpart are visualized. All minors of the models are non-negative, meaning the models are elliptic at the examined points. Furthermore, the PANN model predicts the same values of the least of the minors as its ground truth counterpart. Finally, the ellipticity of the PANN model is examined for all datapoints of the test dataset. Again, the PANN model shows the same behavior in terms of ellipticity as its ground truth counterpart, and is elliptic for all datapoints. Apparently, in the calibration process, the PANN model has adopted the ellipticity of the dataset. 

\medskip

\begin{figure}[t]
	\begin{center}		
		\begin{tabular}{ c l c l c l}
			\includegraphics[trim={0 0 18cm 0},clip,width=0.21\textwidth]{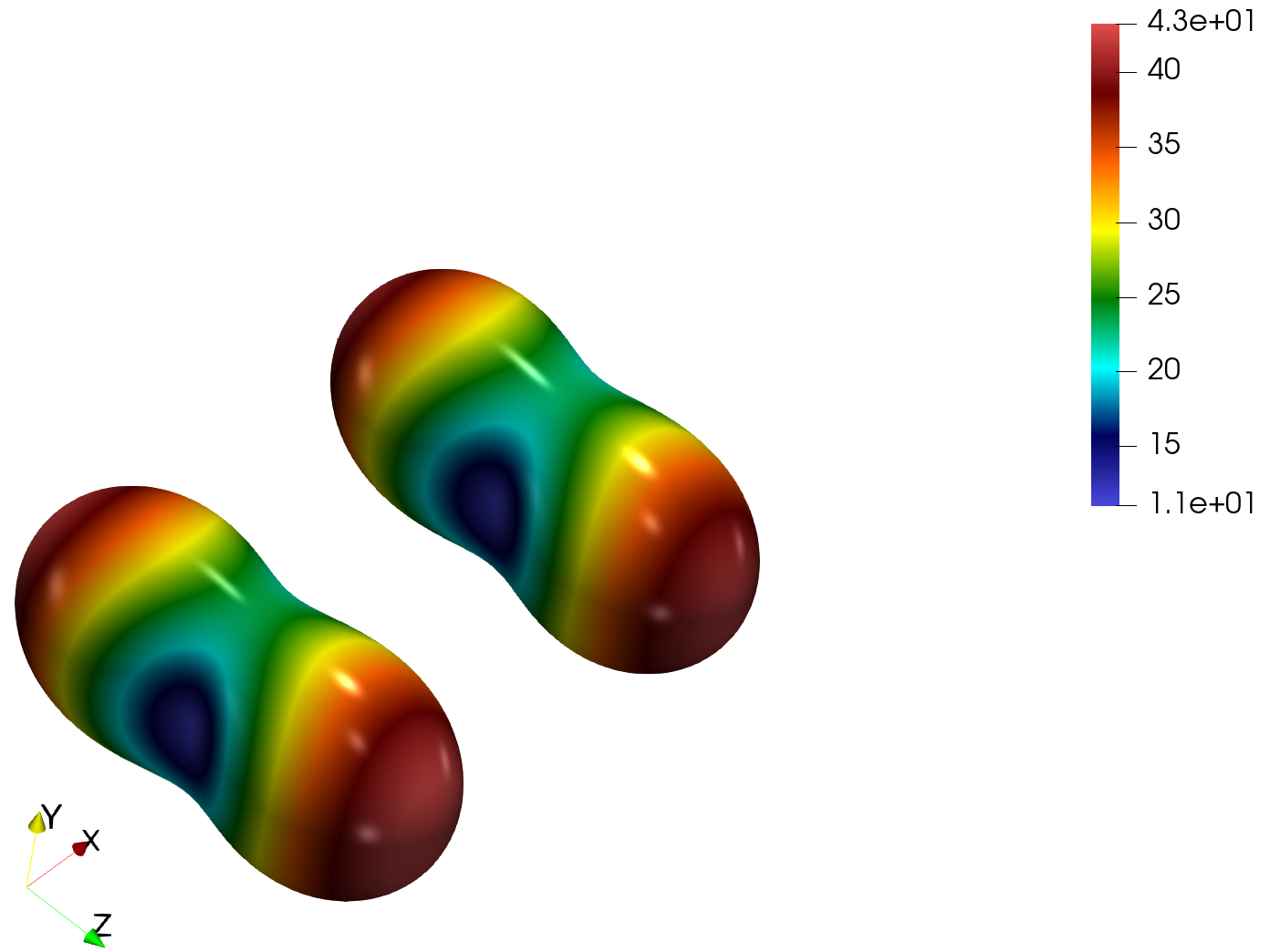}& 
			\includegraphics[width=0.07\textwidth]{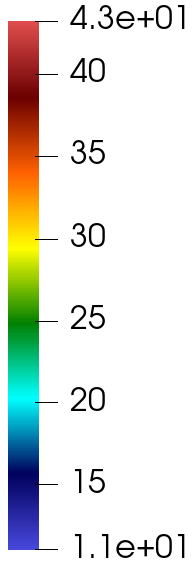}& 		
			\includegraphics[trim={0 0 25cm 0},clip,width=0.21\textwidth]{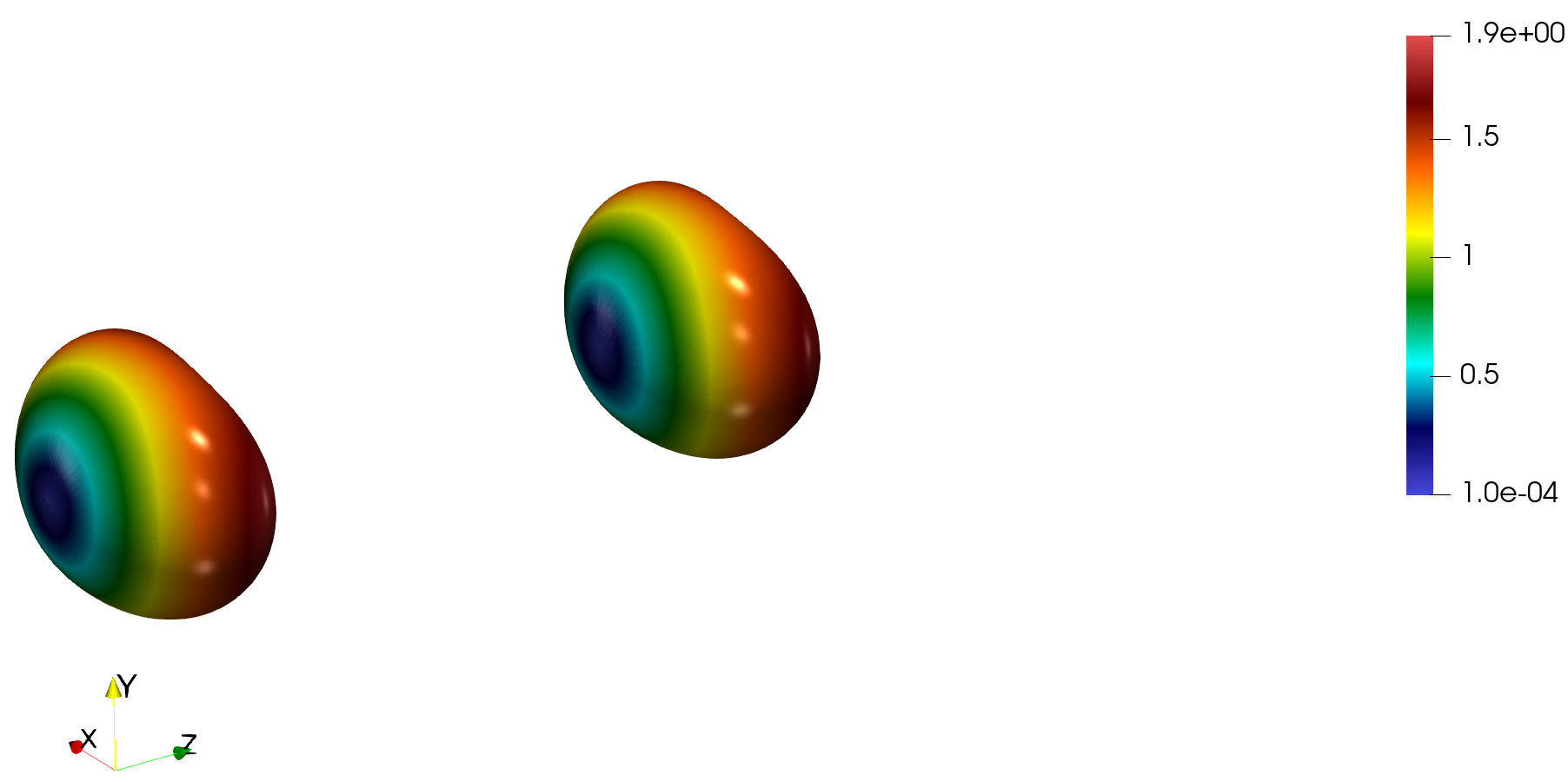}& 
			\includegraphics[width=0.07\textwidth]{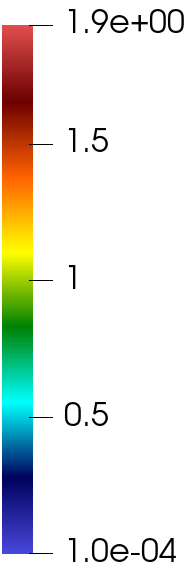} &		
	        \includegraphics[trim={0 0 20cm 0},clip,width=0.21\textwidth]{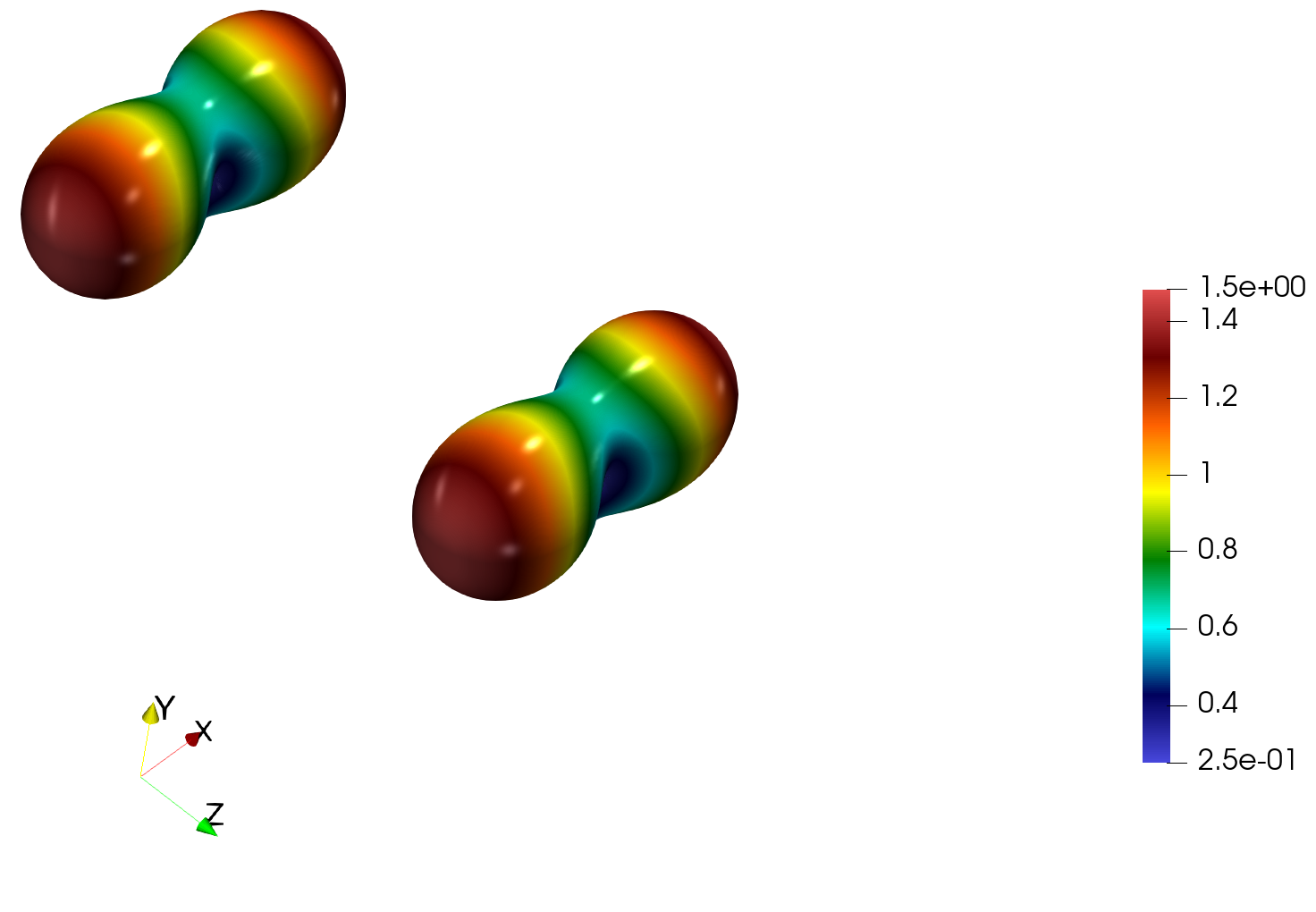}&
            \includegraphics[width=0.07\textwidth]{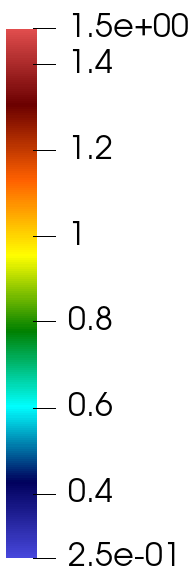}\\
\small{(a) $\widetilde{\mu}$ at $F_{11}=1.5$} &&
\small{(b) $\widetilde{q}$ at $F_{11}=1.5$} &&
\small{(c) $\widetilde{\theta}$ at $F_{11}=1.5$} & \\
\\
			\includegraphics[trim={0 0 16cm 0},clip,width=0.21\textwidth]{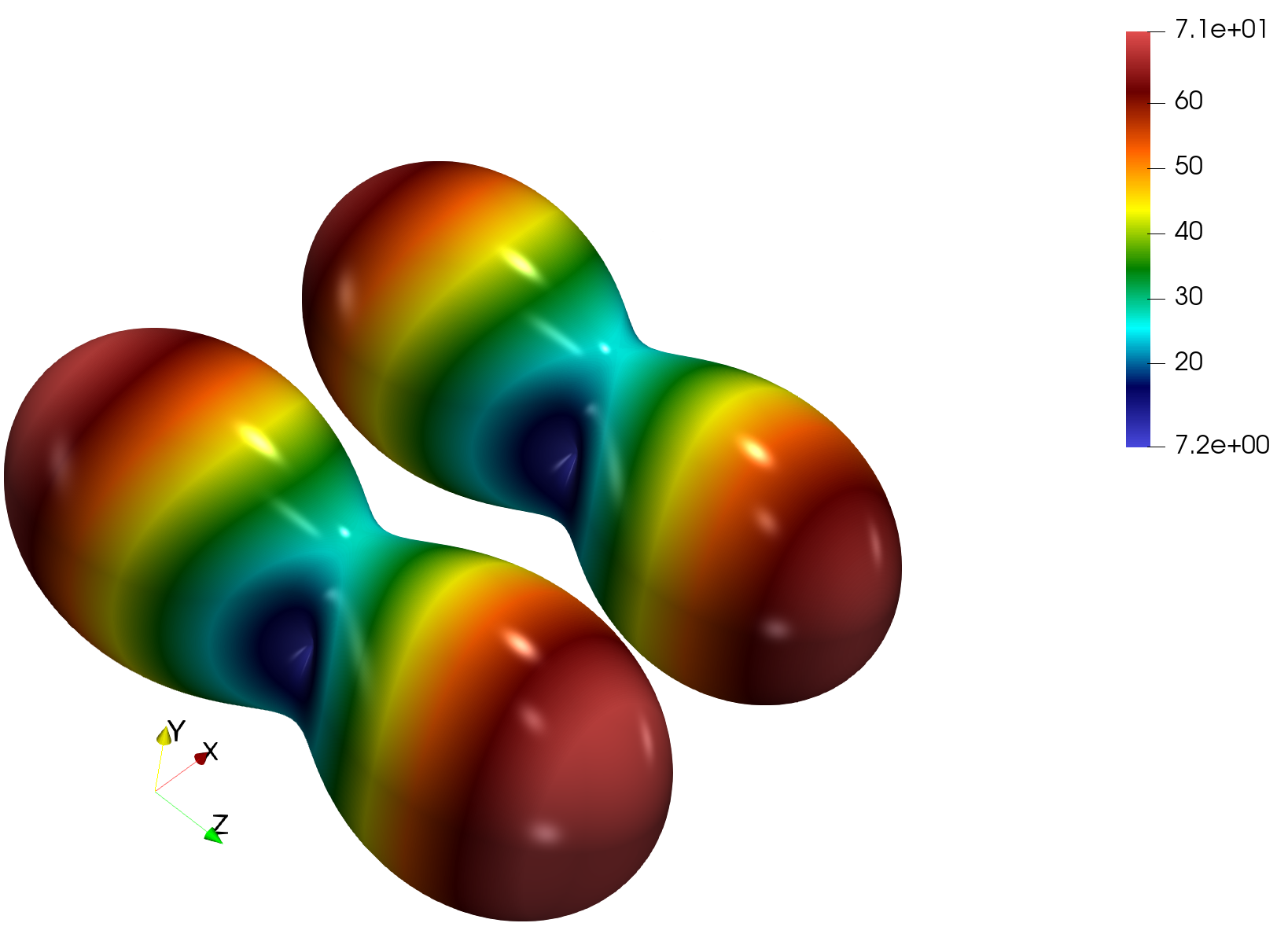} & 
			\includegraphics[width=0.07\textwidth]{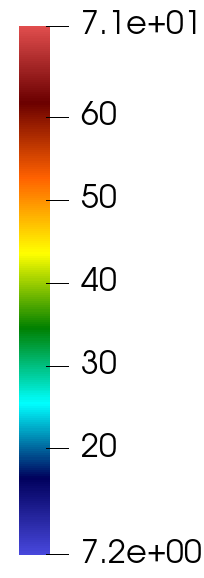}&
            \includegraphics[trim={0 0 25cm 0},clip,width=0.21\textwidth]{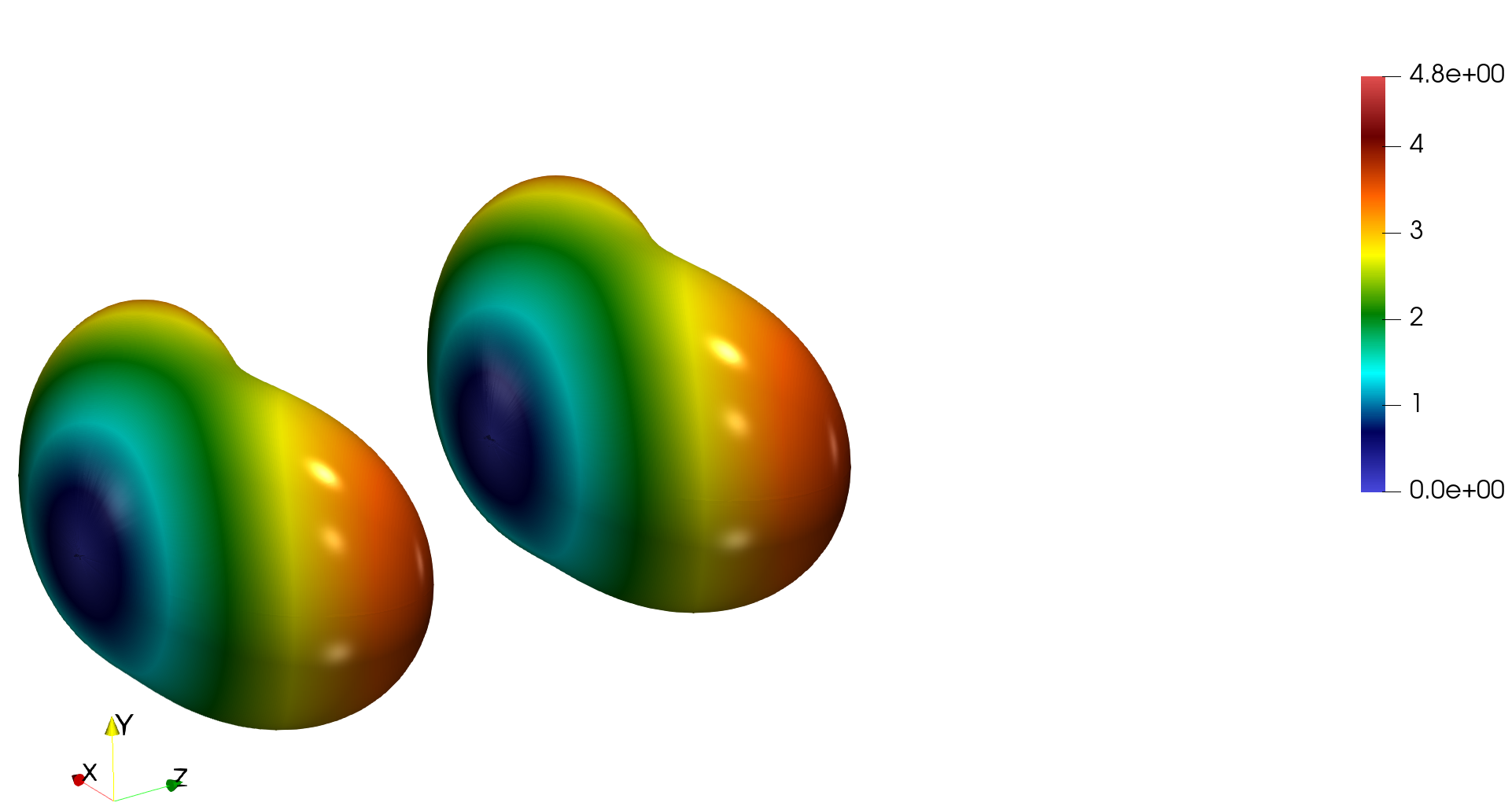}&
            \includegraphics[width=0.07\textwidth]{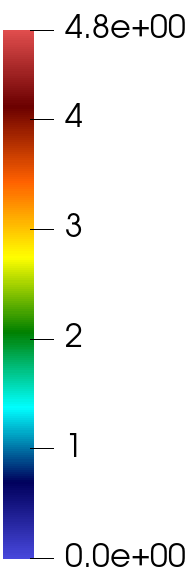}&
            \includegraphics[trim={0 0 20cm 0},clip,width=0.21\textwidth]{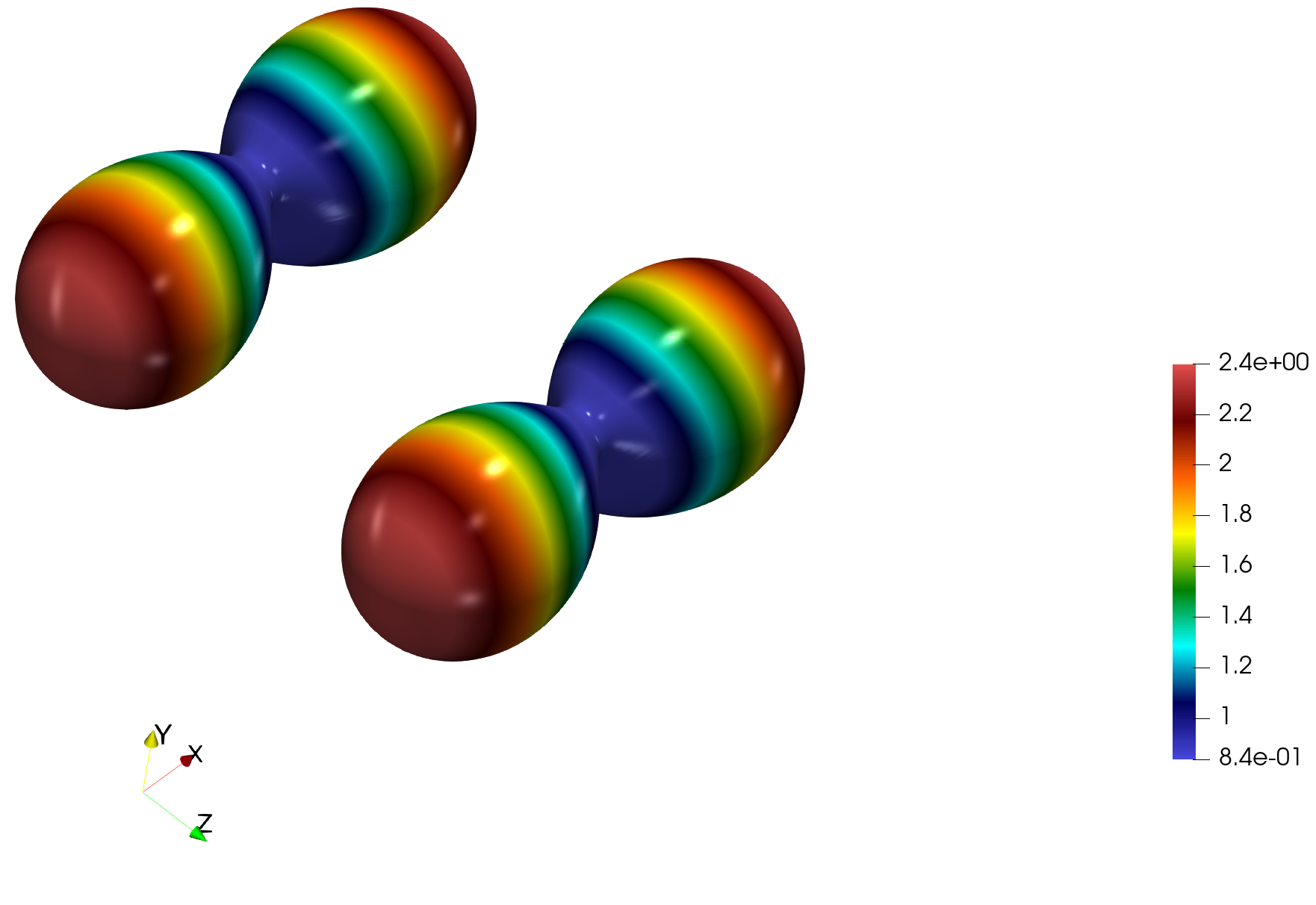} &
            \includegraphics[width=0.07\textwidth]{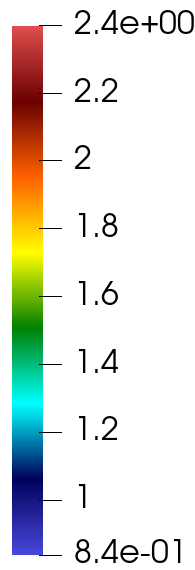}
			\\
\small{(d) $\widetilde{\mu}$ at $F_{11}=2$} &&
\small{(e) $\widetilde{q}$ at $F_{11}=2$} &&
\small{(f) $\widetilde{\theta}$ at $F_{11}=2$} &
		\end{tabular}
	\end{center}
\vspace{-6mm}
	\caption{Spherical parametrisation of the shear modulus $\widetilde{\mu}$, the piezoelectric modulus $\widetilde{q}$, and the dielectric modulus $\widetilde{\theta}$ at  $F_{11}=1.5$ and  $F_{11}=2$. 
	In each subfigure, the left figure represents the results of the ground truth model, while the right figure represents the results of the PANN model.}
	\label{fig:tangent operators ROL}
\end{figure}

Lastly, the shear modulus $\widetilde{\mu}$, the piezoelectric modulus $\widetilde{q}$ and the inverse of the dielectric modulus $\widetilde{\theta}$ are examined, which are closely related to the second gradients of the internal energy density and are defined as 
\begin{equation}\label{eqn:shear modulus}
\begin{aligned}	
\widetilde{\mu}&=\partial_{\vect{FF}}^2e:\vect{e}(\theta,\psi)\otimes\vect{e}(\theta,\psi)\otimes\vect{e}(\theta,\psi)\otimes\vect{e}(\theta,\psi)\,,\\
\widetilde{q}&=\partial_{\vect{d}_0\vect{F}}^2e:\vect{e}(\theta,\psi)\otimes\vect{e}(\theta,\psi)\otimes\vect{e}(\theta,\psi)\,,\\
\widetilde{\theta}&=\partial_{\vect{d}_0\vect{d}_0}^2e:\vect{e}(\theta,\psi)\otimes\vect{e}(\theta,\psi)\,,
\end{aligned}
\end{equation}
where $\vect{e}(\theta,\psi)$ is spherically parameterized in terms of $\theta$ and $\psi$ according to \cref{eqn:spherical parametrisation}.
In \cref{eqn:shear modulus}, the operator ''$:$'' indicates contraction of the corresponding tensors along all axes.
Again, the moduli are evaluated for two points along the equilibrium path introduced in \cref{sec:RVE_eval} for both the non-polyconvex PANN model and its ground truth counterpart. The results in \cref{fig:tangent operators ROL} show an excellent performance of the PANN model in the prediction of the material moduli. This demonstrates that the PANN model is able to learn the second derivatives of the ground truth potential, although only being calibrated on the first gradients.

\medskip

To conclude, the non-polyconvex PANN model is able to excellently predict the first gradients of the ground truth energy potential, i.e., $\vect{P}$ and $\vect{e}_0$. But more than that, it is also able to learn the second derivatives of the ground truth model, although only being calibrated on the first gradients. This is essential when applying the model in numerical applications such as the finite element method, where both first and second gradients of the energy potential are required, cf.~\cref{sec:fe}.
In particular, the second derivatives are closely related to ellipticity and thus the stability of numerical simulations, cf.~\cref{eqn:rank one conexity}. Given a sufficiently large elliptic dataset, the PANN model can learn to be elliptic, and polyconvexity is not necessarily required.
Still, this does not ensure that the model is elliptic for all relevant load scenarios. Thus, ``only if it is inevitable, only when more flexibility is required, should the structure of the model be weakened'' \cite{klein2022b} and a non-polyconvex model be applied. After all, polyconvexity is the only practically applicable way of ensuring ellipticity of the model by construction and to be absolutely sure about its fulfillment.

\section{Nonlinear finite element analysis}\label{sec:FEA}

In this section, finite element analysis is conducted with the PANN models calibrated for isotropic material behavior in \cref{sec:iso_calib} and for transversely isotropic material behavior in \cref{sec:ROL_calib}, as well as with the corresponding ground truth models. In \cref{sec:act}, large electrically induced complex deformations are simulated, followed by the simulation of electrically induced wrinkles in \ref{sec:wrinkles}.


\begin{figure}[t!]
	\begin{center}
\includegraphics[width=0.6\textwidth]{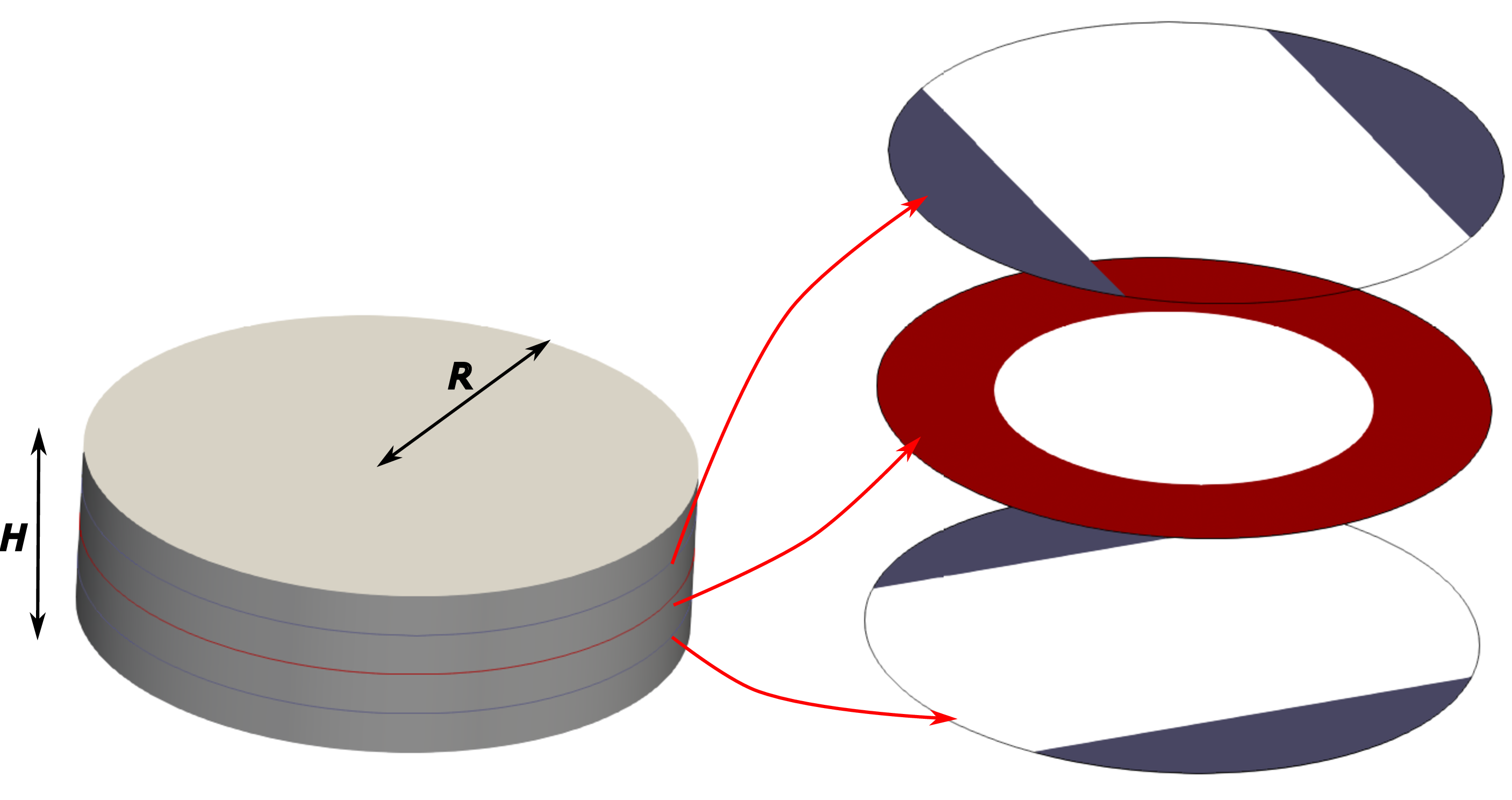}
		\end{center}
\vspace{-4mm}
\caption{Electrically induced actuation with an isotropic material. Device with $R=35$ mm, $H=0.83$ mm and four layers of equal thickness with three electrodes of alternating polarity. 
With the coordinate system $(X,\,Y)$ lying in the radial plane of the device, the top bottom electrodes are located at $|Y|>0.6R$ and $|X|>0.6R$, respectively. The electrode in the middle comprises the region $\sqrt{X^2+Y^2}>0.6R$. In the middle plane of the device, the region $\sqrt{X^2+Y^2}<0.01R$ is completely fixed.}	
\label{fig:the square}
\end{figure}

\subsection{Electrically induced actuation}\label{sec:act}

\subsubsection{Isotropic material}\label{sec:act_iso}

As a first example, a complex electrically induced actuation is simulated for isotropic material behavior. Geometry and boundary conditions of the device are visualized and detailed in \cref{fig:the square}. The distribution of the electrodes, i.e., the positions where Dirichlet boundary conditions are prescribed for the electric potential, is a simplified version of the design proposed in \cite{Ortigosa_Electrodes}. Therein, making use of topology optimisation, this device was designed to attain a saddle-like electrically induced deformation. 
For the finite element discretisation, $Q2$ (tri-quadratic) hexahedral elements are applied, with  a mesh comprising 220,887 degrees of freedom for the displacement field and 73,629 for the electric potential. As constitutive models, the analytical Mooney-Rivlin type isotropic potential and the corresponding polyconvex PANN model ($\SP^+(8)$) introduced in \cref{subsec:iso_model} are applied. The maximum total voltage difference applied between adjacent electrodes (which corresponds to the load factor $\lambda=1$) is set to $\Delta V\sqrt{\varepsilon/\mu_1}=2.95\times 10^{-4}$, with $\varepsilon,\,\mu_1$ as defined in \cref{sec:iso_data}.

\medskip

\begin{figure}[t]
\begin{center}
\begin{tabular}{cc@{\hskip 0.04\textwidth}cc}
\includegraphics[width=0.234\textwidth]{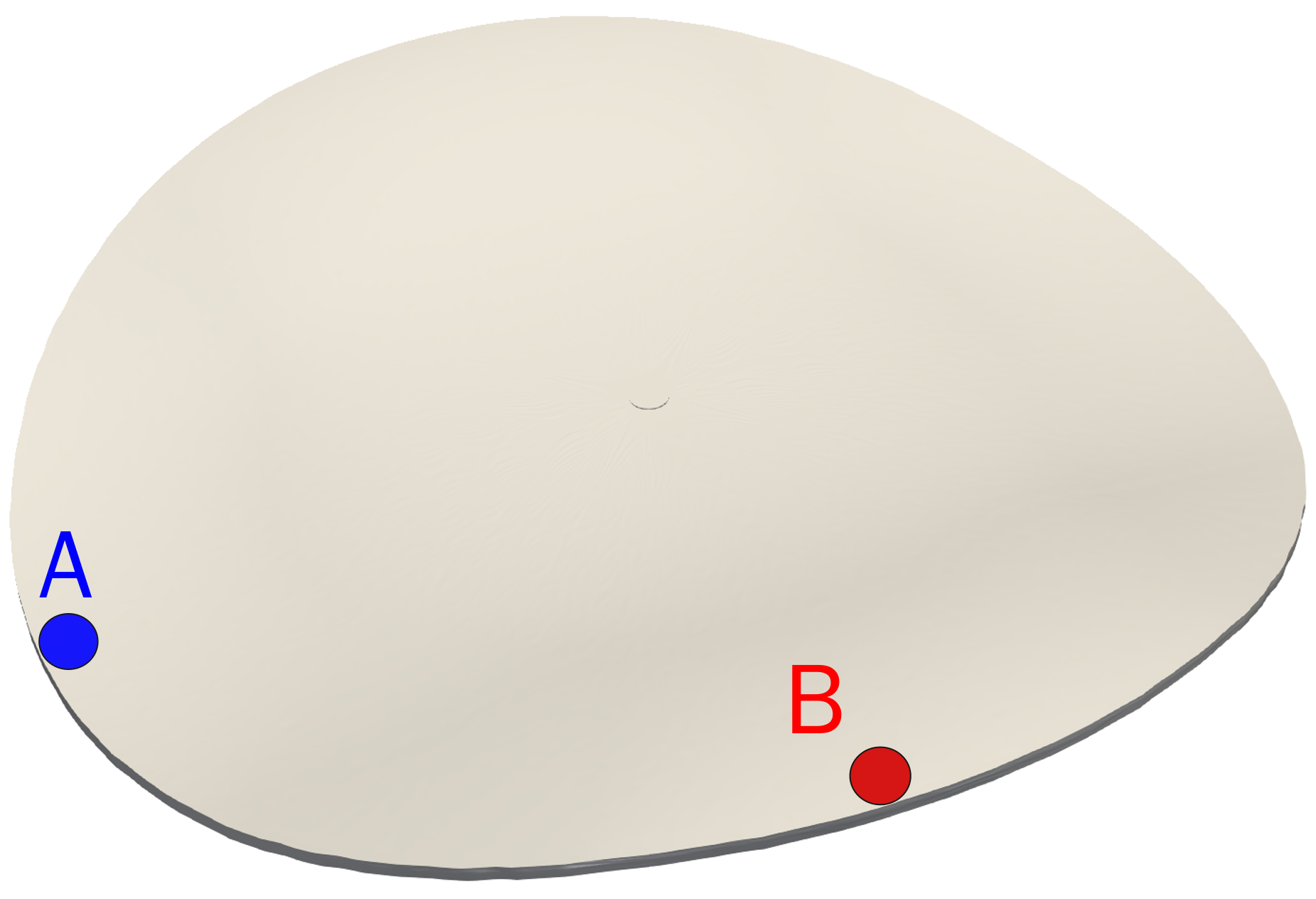}&
\includegraphics[width=0.198\textwidth]{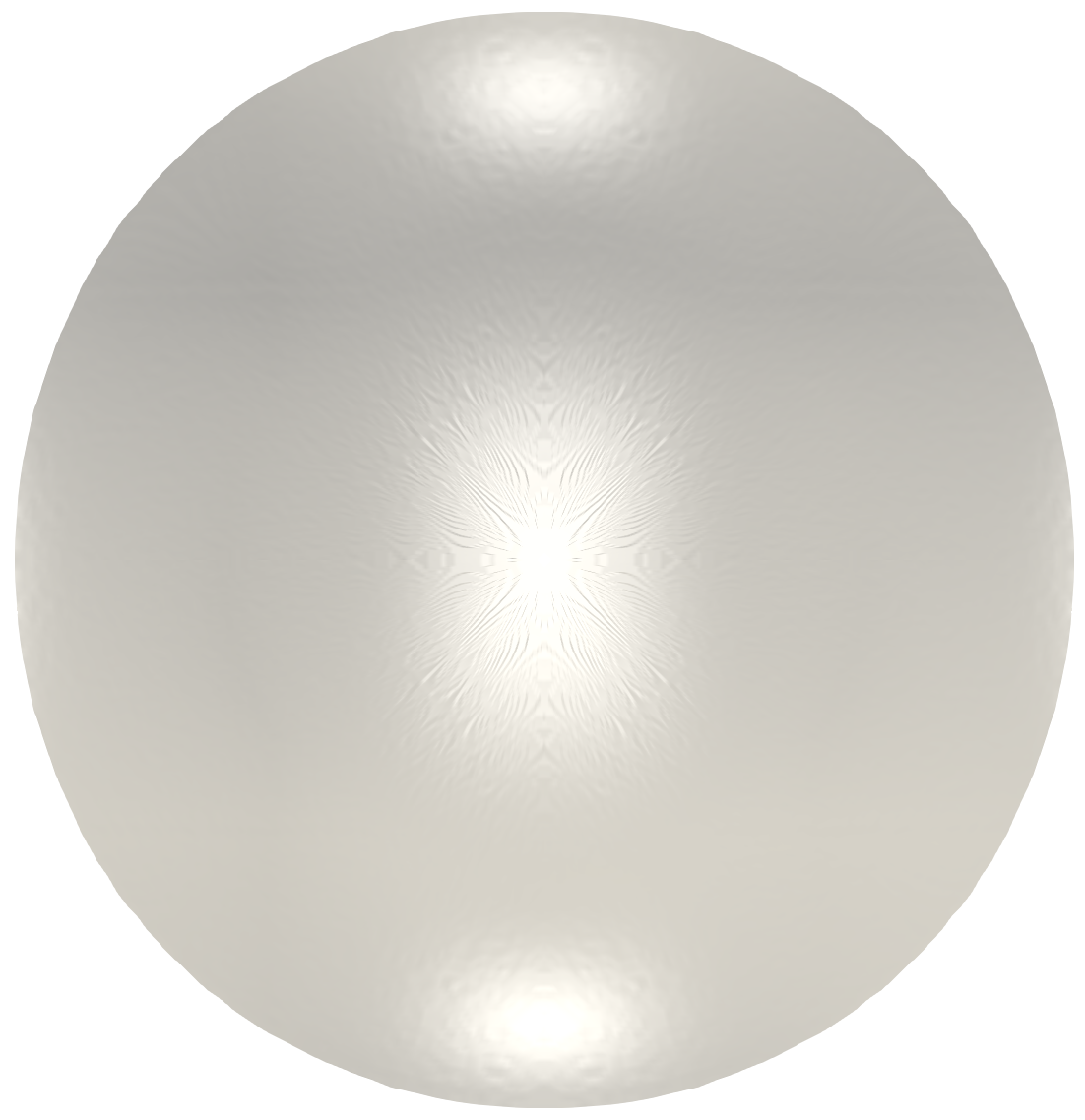}&
\includegraphics[width=0.234\textwidth]{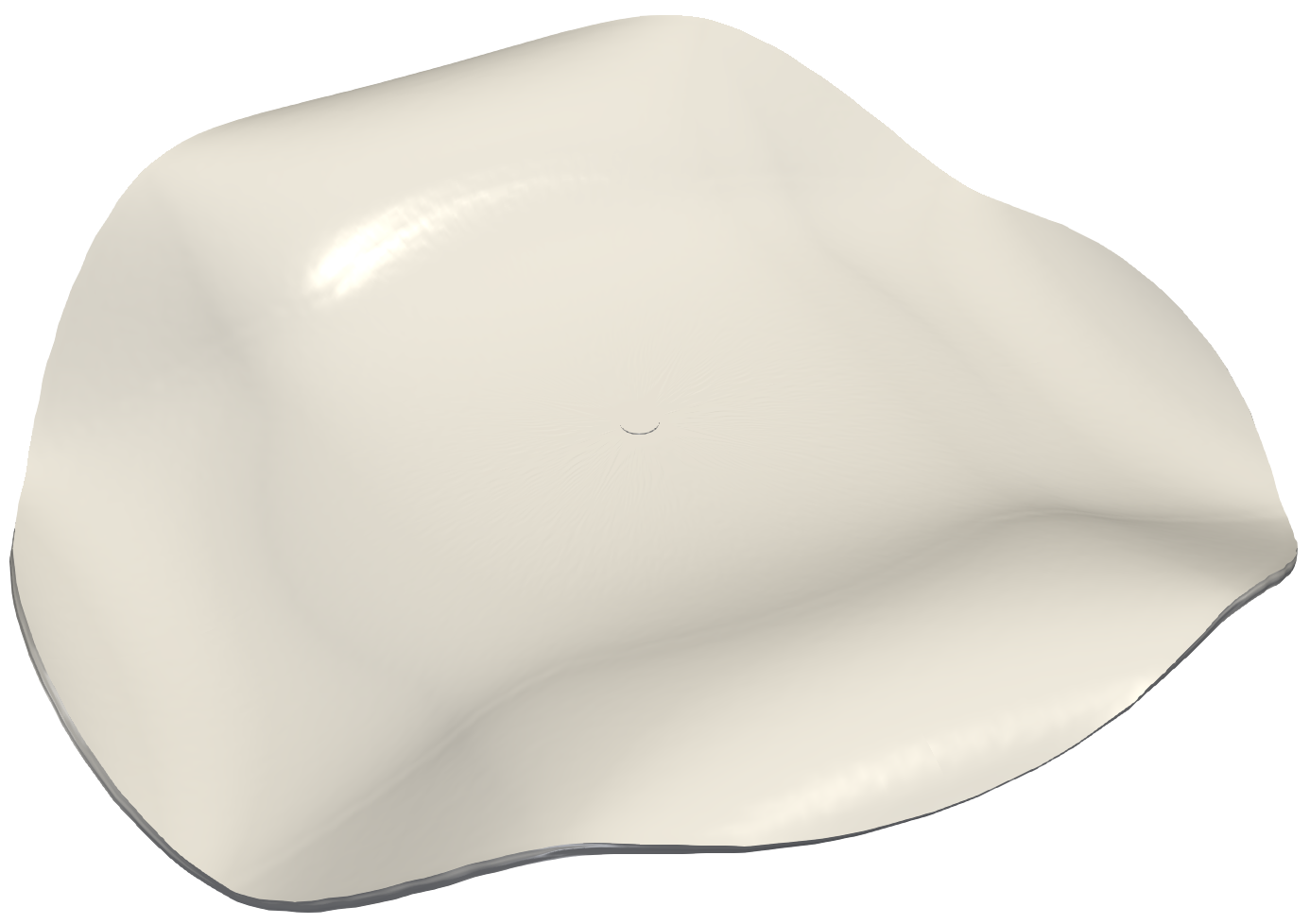}&
\includegraphics[width=0.198\textwidth]{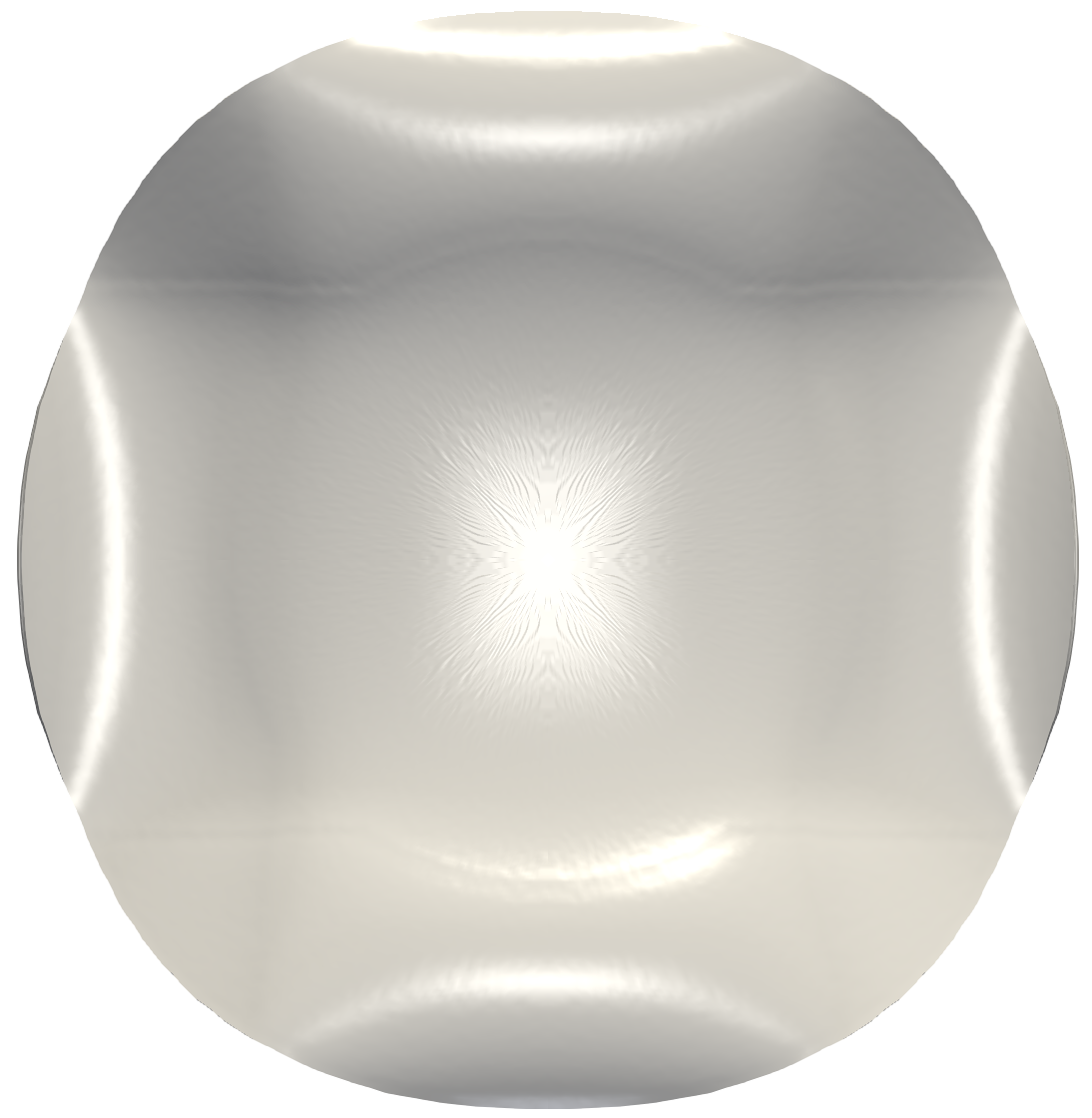}\\
\multicolumn{2}{c}{\small (a) $\lambda=0.22$} & \multicolumn{2}{c}{\small (b) $\lambda=0.44$}\\[1ex]
\includegraphics[width=0.234\textwidth]{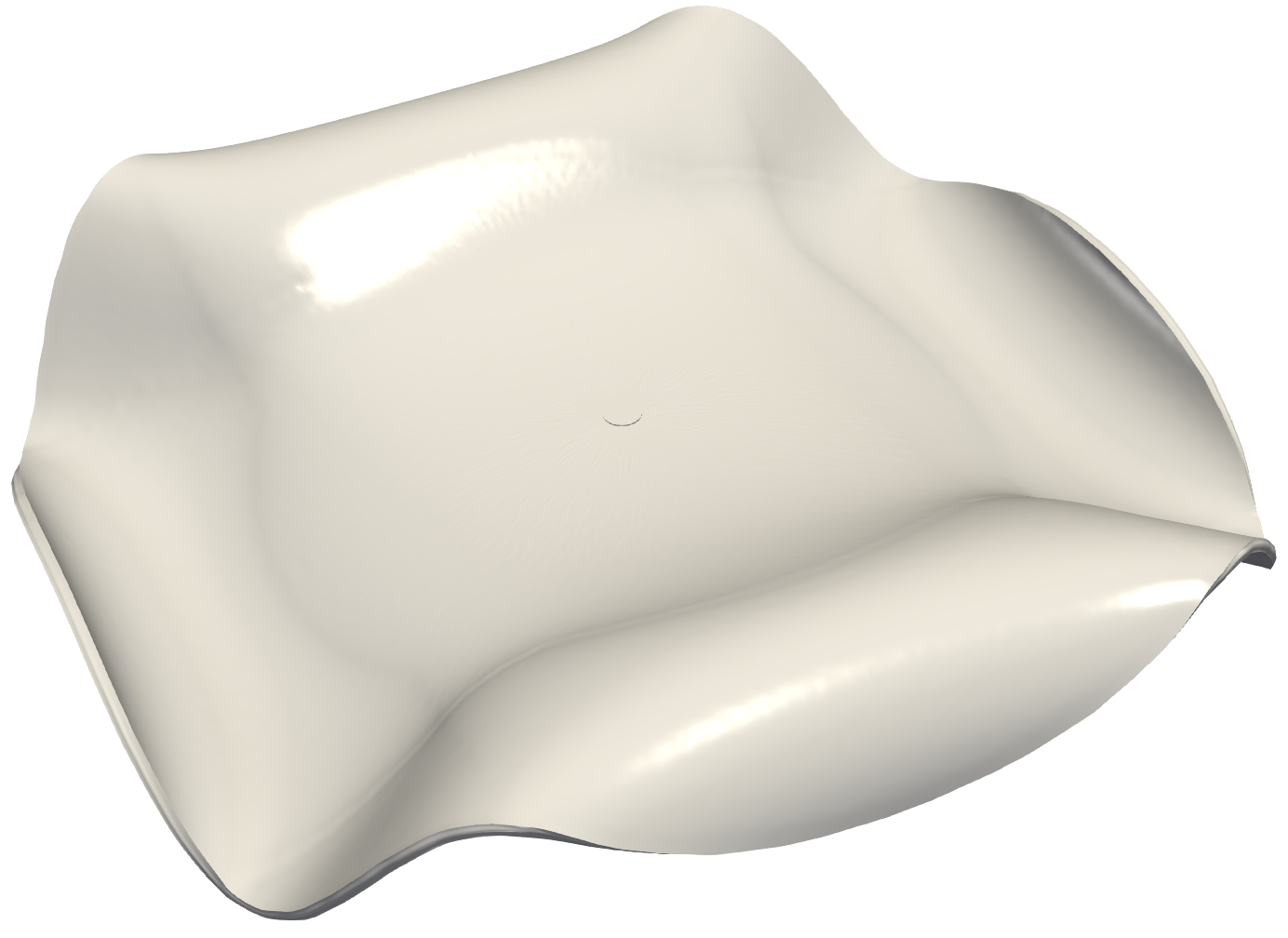}&
\includegraphics[width=0.198\textwidth]{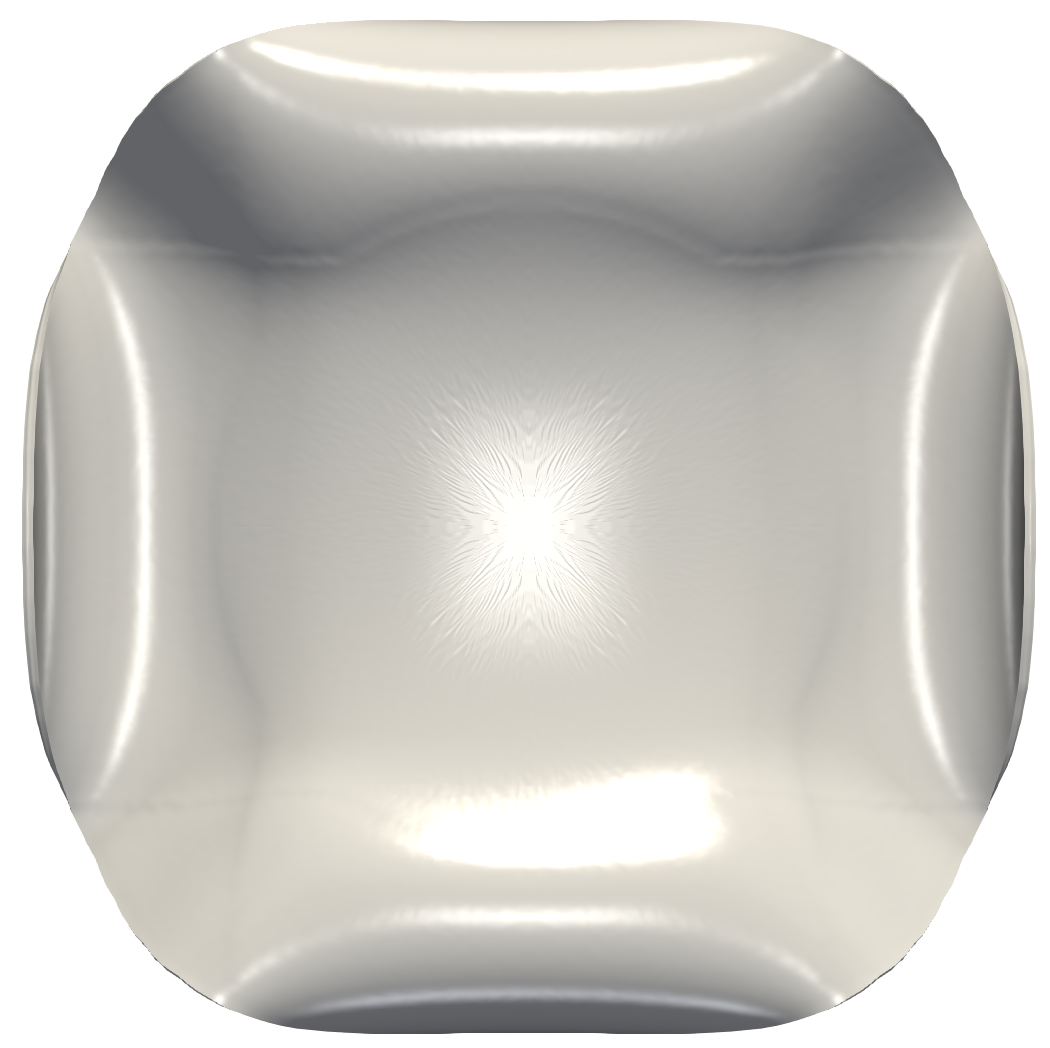}&
\includegraphics[width=0.266\textwidth]{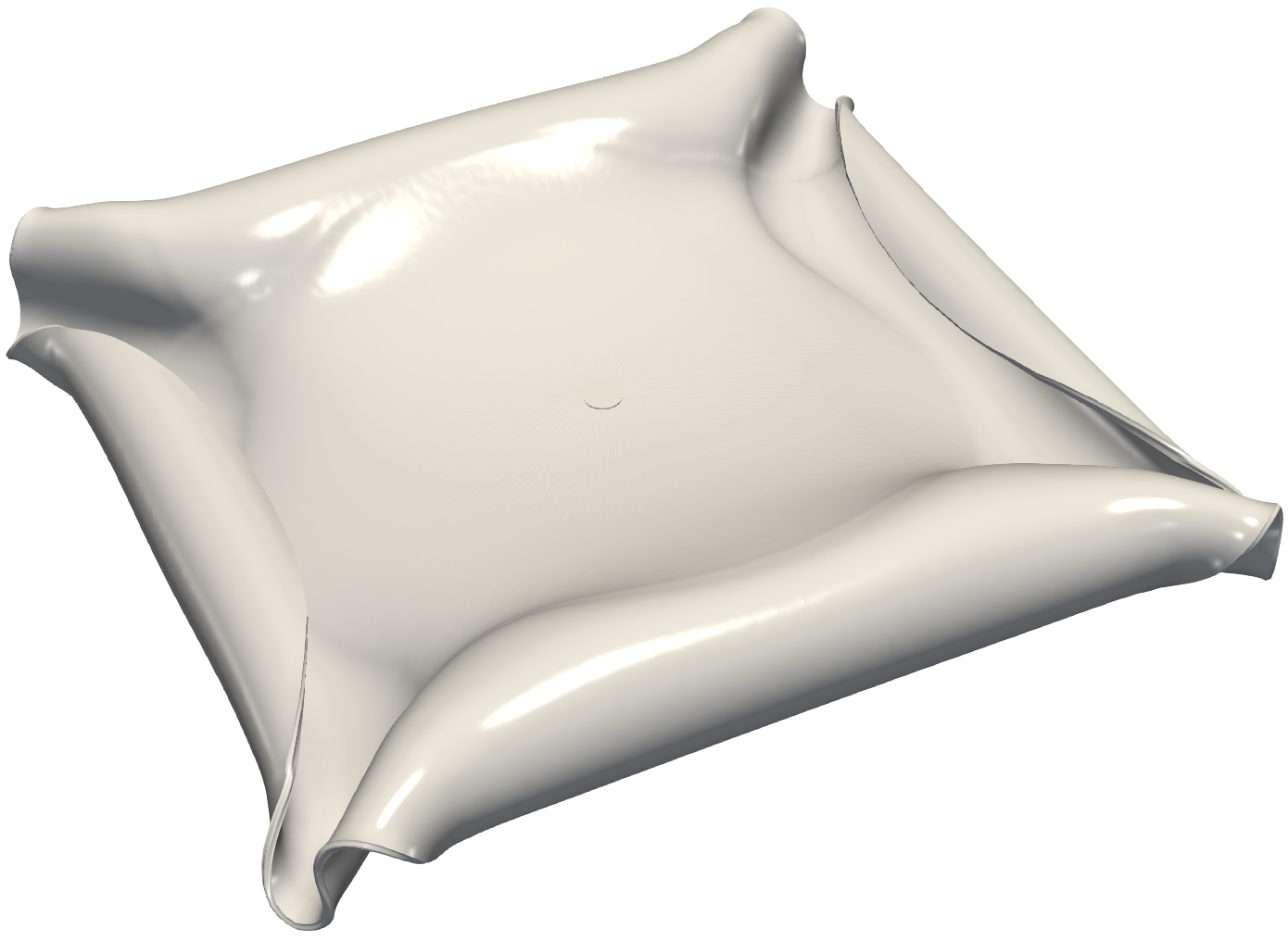}&
\includegraphics[width=0.20\textwidth]{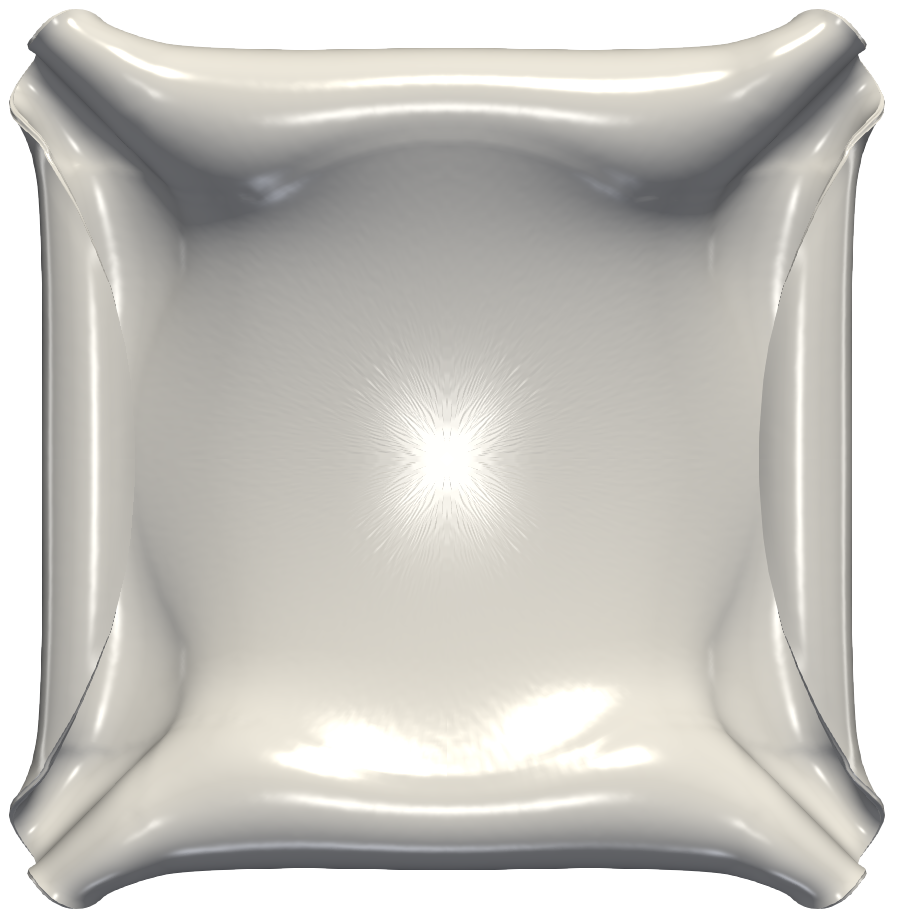}\\
\multicolumn{2}{c}{\small (c) $\lambda=0.56$} & \multicolumn{2}{c}{\small (d) $\lambda=1.00$}\\
\end{tabular}

\end{center}
\vspace{-6mm}
\caption{Electrically induced actuation with an isotropic material. Deformation of the device for various values of the load factor $\lambda$, obtained by FEA using the polyconvex PANN model. The device experiences a considerable change of shape, ranging from saddle-like deformations to a square-like shape.}\label{fig:isotropic wrinkles 1}
\end{figure}

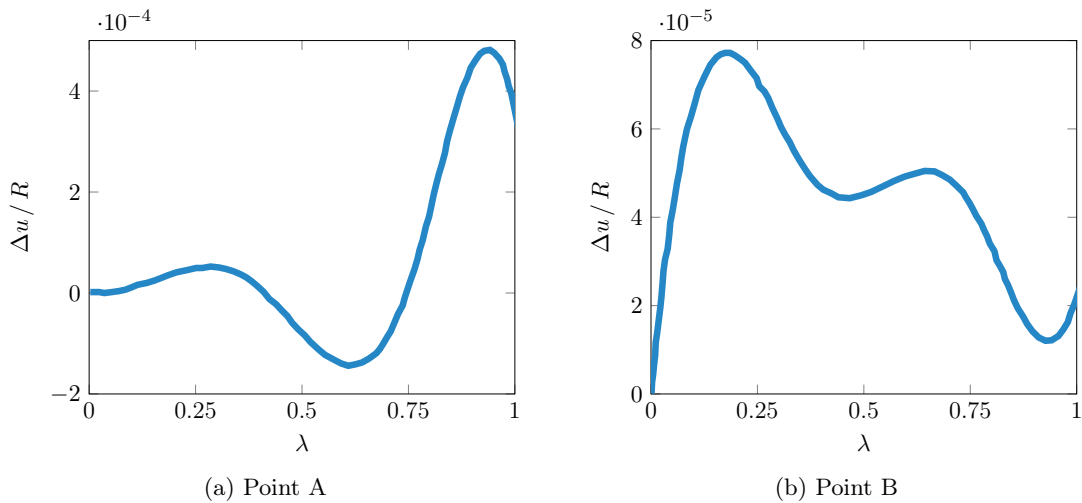
\begin{figure}[t!]
    \centering
    \begin{subfigure}[b]{0.45\textwidth}
        \centering
        \resizebox{!}{0.8\textwidth}{
        \tikzsetnextfilename{iso_wrinkles_pointA}

\begin{tikzpicture}
\begin{axis}[
        xlabel=$\lambda$,
        ylabel=$\Delta u \,/\,R$,
        xmin=0, xmax=0.59, 
        ymin=-2e-4, ymax=5e-4, 
        xtick={0,0.1475,0.295,0.4425,0.59},
        xticklabels={0,0.25,0.5,0.75,1},
        legend cell align={left},
    ]

\addplot[blueR!85, line width=3pt,mark=none] table[mark=none] {fig/IsotropicWrinkles/iso_actuation_point_A.txt};

\end{axis}
\end{tikzpicture}
        }
        \caption{Point A}
    \end{subfigure}
    \begin{subfigure}[b]{0.45\textwidth}   
        \centering 
        \resizebox{!}{0.8\textwidth}{
        \tikzsetnextfilename{iso_wrinkles_pointB}

\begin{tikzpicture}
\begin{axis}[
        xlabel=$\lambda$,
        ylabel=$\Delta u \,/\,R$,
        xmin=0, xmax=0.59, 
        ymin=0, ymax=8e-5, 
        xtick={0,0.1475,0.295,0.4425,0.59},
        xticklabels={0,0.25,0.5,0.75,1},
        legend cell align={left},
    ]

\addplot[blueR!85, line width=3pt, mark=none] table[mark=none] {fig/IsotropicWrinkles/iso_actuation_point_B.txt};

\end{axis}
\end{tikzpicture}
        }    
        \caption{Point B}
    \end{subfigure}
\caption{Electrically induced actuation with an isotropic material. Difference in displacement $\Delta u=u^{\text{GT}}-u^{\text{PANN}}$ between the isotropic ground truth (GT) material model and the PANN model at points A and  B.}\label{fig:isotropic wrinkles 2}
\end{figure}

In \Cref{fig:isotropic wrinkles 1}, the electrically induced deformation is visualized for various values of the load factor $\lambda$. Clearly, for the value of voltage difference for which this device was designed ($\lambda=0.22$), the device shows a saddle-like deformation. Upon elevating this value, the device exhibits a distinctive folding behavior in proximity to regions where the top and bottom electrodes are concentrated. The consequence of this electromechanical deformation is readily observed when inspecting the device from a top-down perspective. Remarkably, the initial circular geometry undergoes a transformation, ultimately manifesting in a square shape. In \cref{fig:isotropic wrinkles 2}, the norm of the difference between the displacement obtained by the ground truth model and its PANN counterpart at the two points $A$ and $B$ are visualized (cf.~\cref{fig:isotropic wrinkles 1}), scaled with the radius of the device. The difference between both models is extremely small, which indicates the extraordinary accuracy of the isotropic PANN model even for this complex electrically induced deformation pattern.
Furthermore, this case shows the remarkable stability of the PANN model when applied in FEA, which is required to simulate the highly challenging electrically induced changes of shape considered here.

\subsubsection{Rank-one laminate}\label{sec:act_ROL}

\begin{figure}[t]
	\begin{center}
		\begin{tabular}{c@{\hskip 0.04\textwidth}c@{\hskip 0.04\textwidth}c}
			\includegraphics[width=0.33\textwidth]{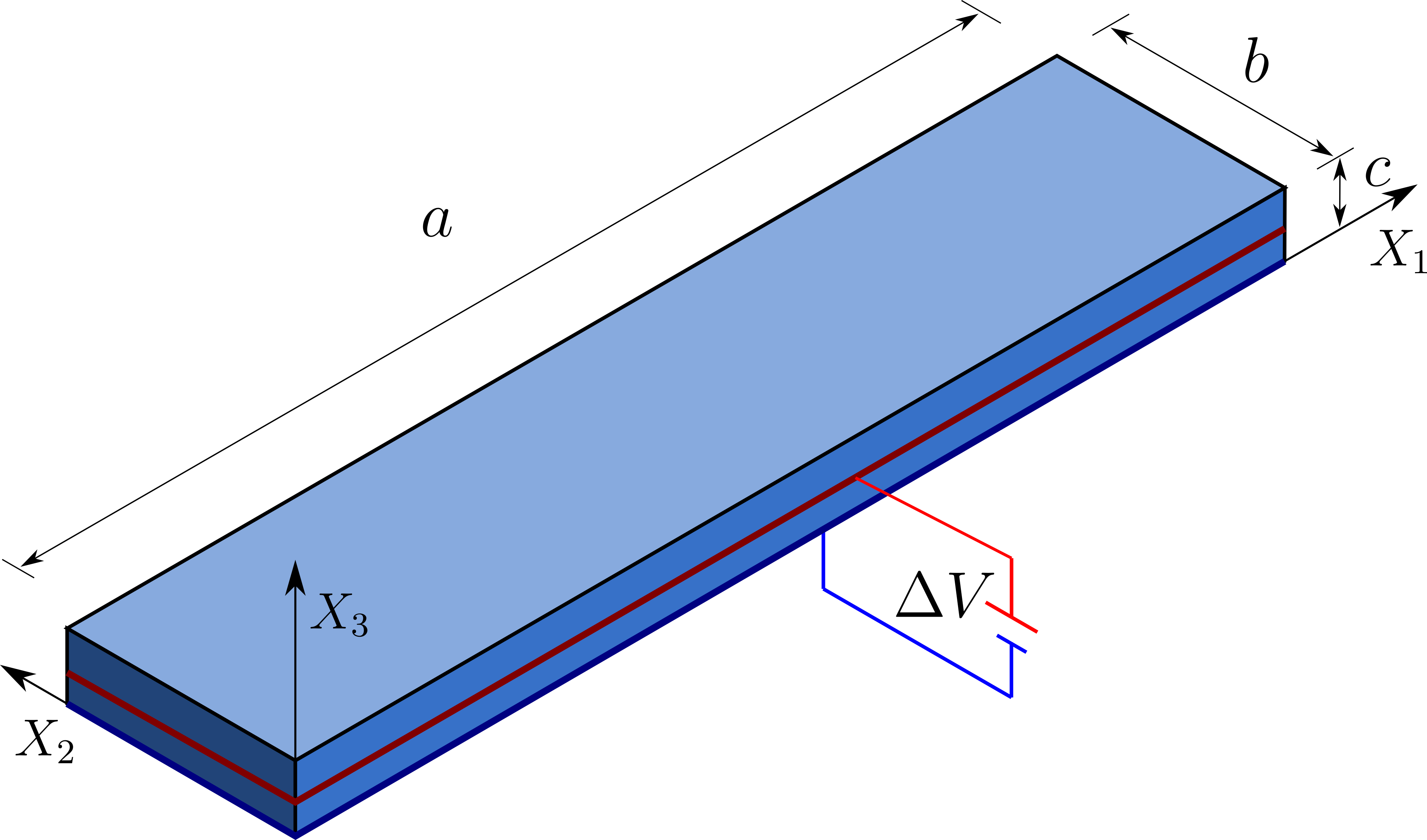}&
			\includegraphics[width=0.27\textwidth]{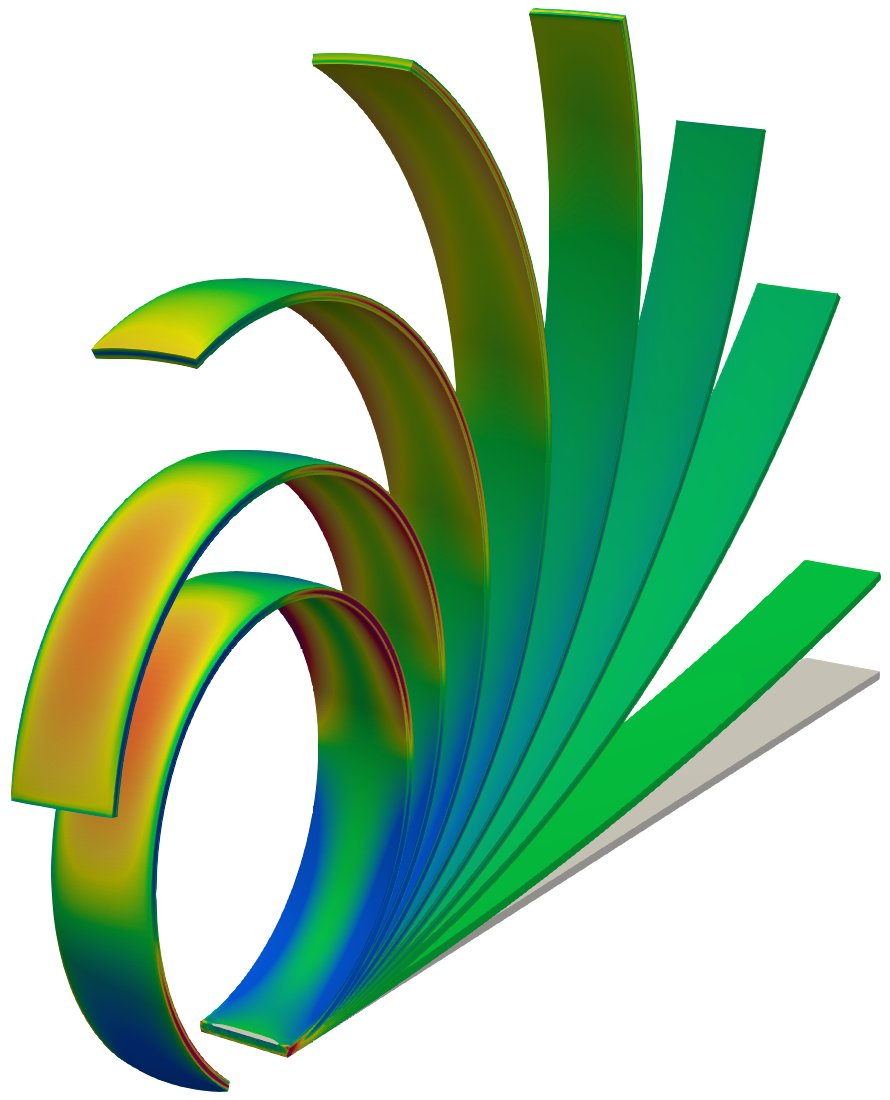}&
			\includegraphics[width=0.27\textwidth]{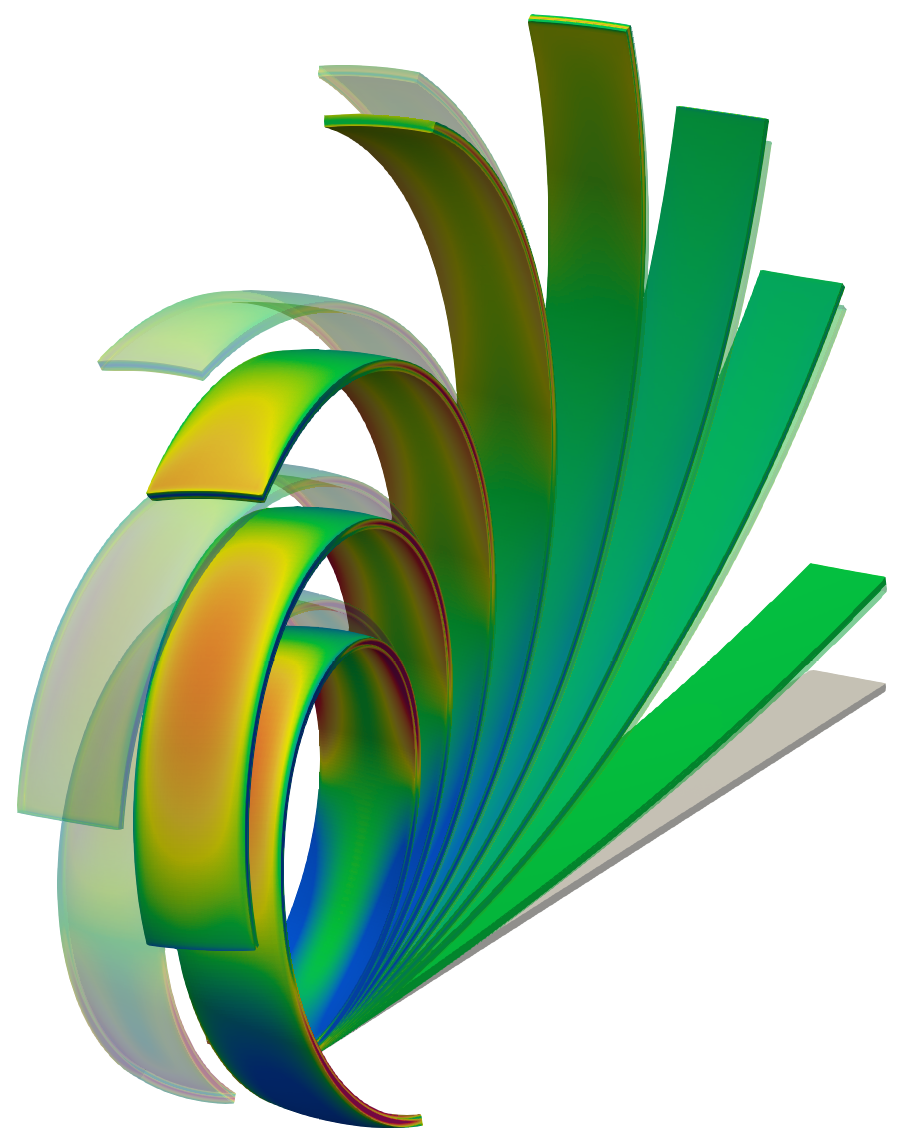}			\\
			&\multicolumn{2}{c}{\includegraphics[width=0.35\textwidth]{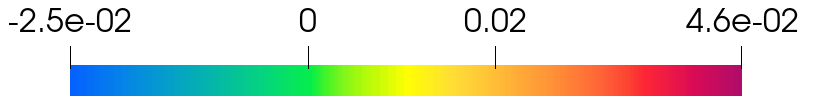}} \\
			\small{(a) Problem setup} &
			\small{(b) Non-polyconvex PANN} &
			\small{(c) Polyconvex PANN} 
		\end{tabular}
	\end{center}
\vspace{-6mm}
	\caption{Electrically induced actuation with a rank-one laminate material. (a) Geometry and electrical boundary conditions. Displacements fixed at $X_1=0$. $\{a,b,c\}=\{120,10,1\}$ (mm). (b-c) Contour plot distribution for hydrostatic pressure (N$\cdot \text{mm}^{-2}$) for various load factors. Shaded plot show the results for the ground truth model while unshaded plots show (b) the non-polyconvex PANN model and  (c) the polyconvex PANN model. While the simulation with the non-polyconvex PANN model perfectly coincides with that of the ground truth model, the polyconvex PANN model shows large deviations.}\label{fig:bending ROL} 
\end{figure}

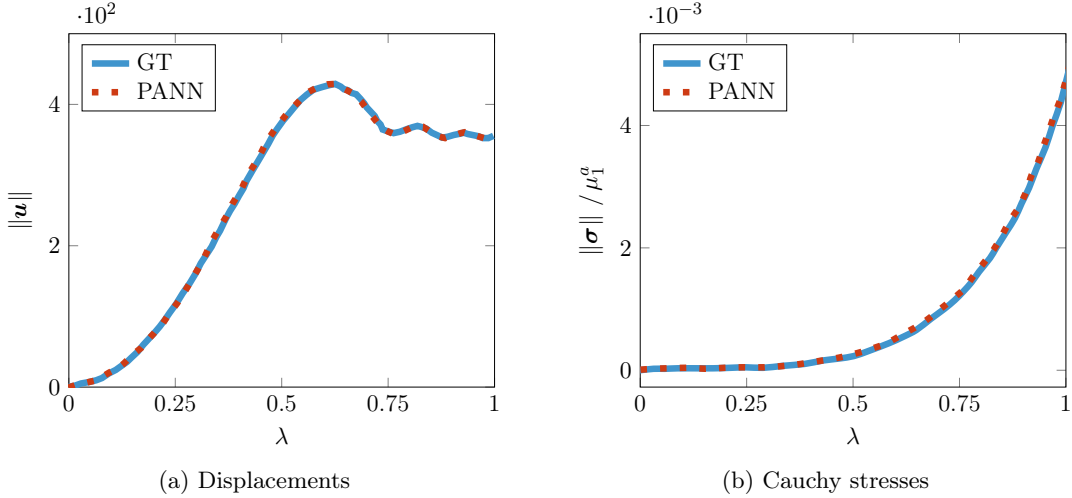
\begin{figure}[t!]
    \centering
    \begin{subfigure}[b]{0.45\textwidth}
        \centering
        \resizebox{!}{0.8\textwidth}{
        \tikzsetnextfilename{BendingTorsionROL_norm_disp}

\begin{tikzpicture}
\begin{axis}[
        xlabel=$\lambda$,
        ylabel=$\norm{\vect{u}}$,
        xmin=0, xmax=0.65, 
        ymin=0, ymax=500, 
        xtick={0,0.1625,0.325,0.4875,0.65},
        xticklabels={0,0.25,0.5,0.75,1},
        ytick={0,200,400},
        legend cell align={left},
        legend pos = north west,
        scaled y ticks=base 10:-2
        ]

\addplot[blueR!75, line width=3pt,mark=none] table[mark=none] {fig/BendingTorsionROL/disp_GT.txt};
\addplot[redR, line width=3pt,mark=none,loosely dashed] table[mark=none] {fig/BendingTorsionROL/disp_PANN.txt};

\addlegendentryexpanded{GT}
\addlegendentryexpanded{PANN}

\end{axis}
\end{tikzpicture}
        }
        \caption{Displacements}
    \end{subfigure}
    \begin{subfigure}[b]{0.45\textwidth}   
        \centering 
        \resizebox{!}{0.8\textwidth}{
        \tikzsetnextfilename{BendingTorsionROL_norm_stress}

\begin{tikzpicture}
\begin{axis}[
        xlabel=$\lambda$,
        ylabel=$\norm{\vect{\sigma}}\,/\,\mu_1^a$,
        xmin=0, xmax=0.65, 
        ymin=-2.75e-4, ymax=5.5e-3, 
        ytick={0,2e-3,4e-3},
        xtick={0,0.1625,0.325,0.4875,0.65},
        xticklabels={0,0.25,0.5,0.75,1},
        legend cell align={left},
        legend pos = north west
    ]

\addplot[blueR!75, line width=3pt, smooth,mark=none] table[mark=none] {fig/BendingTorsionROL/stress_GT.txt};
\addplot[redR, line width=3pt, smooth,mark=none,loosely dashed] table[mark=none] {fig/BendingTorsionROL/stress_PANN.txt};

\addlegendentryexpanded{GT}
\addlegendentryexpanded{PANN}

\end{axis}
\end{tikzpicture}
        }    
        \caption{Cauchy stresses}
    \end{subfigure}
\caption{Electrically induced actuation with a rank-one laminate material. Evolution of the $L^2$-norm of the (a) displacements and (b) Cauchy stresses for the ground truth model and the non-polyconvex PANN model.}\label{fig:bending error}
\end{figure}

In this example, the large electrically induced deformation of a beam is simulated for a rank-one laminate material. Geometry and boundary conditions are visualized in \cref{fig:bending ROL}a. For the finite element discretisation, $Q2$ (tri-quadratic) hexahedral elements are applied, with  a mesh comprising $60\times 5\times 2$ elements in $X_1$-, $X_2$- and $X_3$-directions, respectively. As constitutive models, the analytically homogenised rank-one laminate as well as a polyconvex ($\SP^+(64)$) and a non-polyconvex PANN ($\SP(8)$) model as introduced in \cref{subsec:ROL} are applied. The maximum total voltage difference (which corresponds to the load factor $\lambda=1$) is set to $\Delta V\sqrt{\varepsilon/\mu_1}=1.3$, with $\varepsilon,\,\mu_1$ as defined in \cref{sec:ROL_data}. 

\medskip

In \Cref{fig:bending ROL}, the results of the simulations with the non-polyconvex and the polyconvex PANN models are compared to the results obtained with the ground truth model for various load increments. While the simulation with the non-polyconvex PANN model perfectly coincides with that of the ground truth model, the polyconvex PANN model show deviations for larger load factors. This coincides with the observations made in \cref{sec:ROL_eval}, where the polyconvex PANN model also showed a moderate performance at large strains and stresses when being calibrated to rank-one laminate data.
In \cref{fig:bending error}, a more quantitative analysis of the performance of the non-polyconvex PANN model compared to the ground truth model is conducted, where the $L^2$-norms of the displacements and Cauchy stress tensors are compared. Both results are in agreement with the results in \cref{fig:bending ROL}, and again, the non-polyconvex PANN model shows excellent performance in predicting the ground truth model.
Importantly, both PANN models allow for stable numerical simulation of this large electrically induced deformation. This is particularly remarkable for the non-polyconvex PANN model, which does not fulfill the ellipticity condition by construction, and thus, does not necessarily lead to stable numerical simulations. This again demonstrates that even a non-polyconvex PANN model can ``learn'' a stable numerical behavior when being calibrated to suitable and sufficient data.

\subsection{Electrically induced wrinkles}\label{sec:wrinkles}

\begin{figure}[t]
	\begin{center}
		\includegraphics[width=0.35\textwidth]{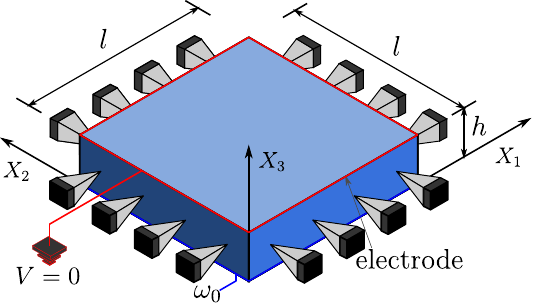}
	\end{center}
\vspace{-6mm}
	\caption{Electrically induced wrinkles. Geometry and boundary conditions for a square plate with side lengths of $60$ mm and thickness of $1$ mm.  A force per unit undeformed volume of value $9.8\times 10^{-2}$ is acting along the $X_3$-axis in the positive direction.}\label{fig:electrical wrinkles config}
\end{figure}

In the last example, electro-mechanically induced instabilities are examined, which result in the development of massive wrinkling. For this, a square plate completely fixed at its boundaries is considered, with the voltage grounded at the upper surface of the plate and a surface charge of $\omega_0$ applied at the lower surface of the plate, cf.~\cref{fig:electrical wrinkles config}. This example was previously also analysed in other works by the authors \cite{Ortigosa_Electrodes}. 
For the finite element discretisation, $Q2$ (tri-quadratic) hexahedral elements are applied with $80\times 80\times 2$ elements in $X_1$-, $X_2$- and $X_3$-directions. This simulation is carried out with both the Mooney-Rivlin type analytical isotropic model and its polyconvex PANN counterpart, cf.~\cref{subsec:iso_model}, as well as with the rank-one laminate material and its non-polyconvex PANN counterpart, cf.~\cref{subsec:ROL}. In the isotropic case, a surface charge of $\omega_0=7/\sqrt{\mu_1\varepsilon}$ is applied, while for the rank-one laminate material, a surface charge of $\omega_0=3.2/\sqrt{\mu_1^a\varepsilon^a}$ is applied.

\medskip

\begin{figure}[t!]
	\begin{center}
		\begin{tabular}{c c}
			\includegraphics[width=0.47\textwidth]{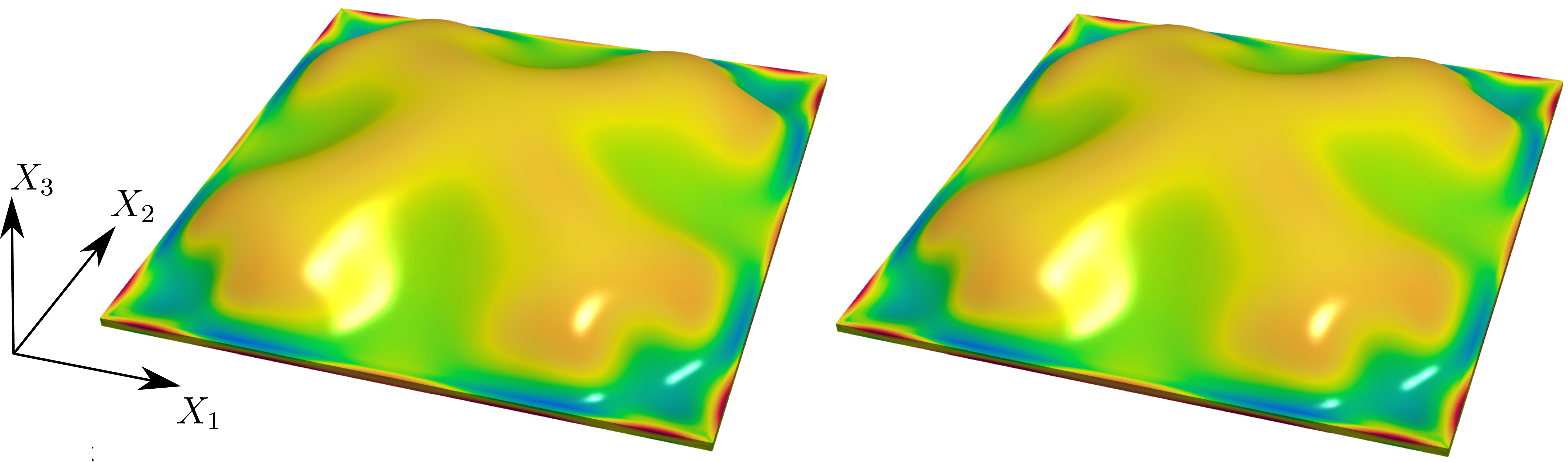}&
			\includegraphics[width=0.47\textwidth]{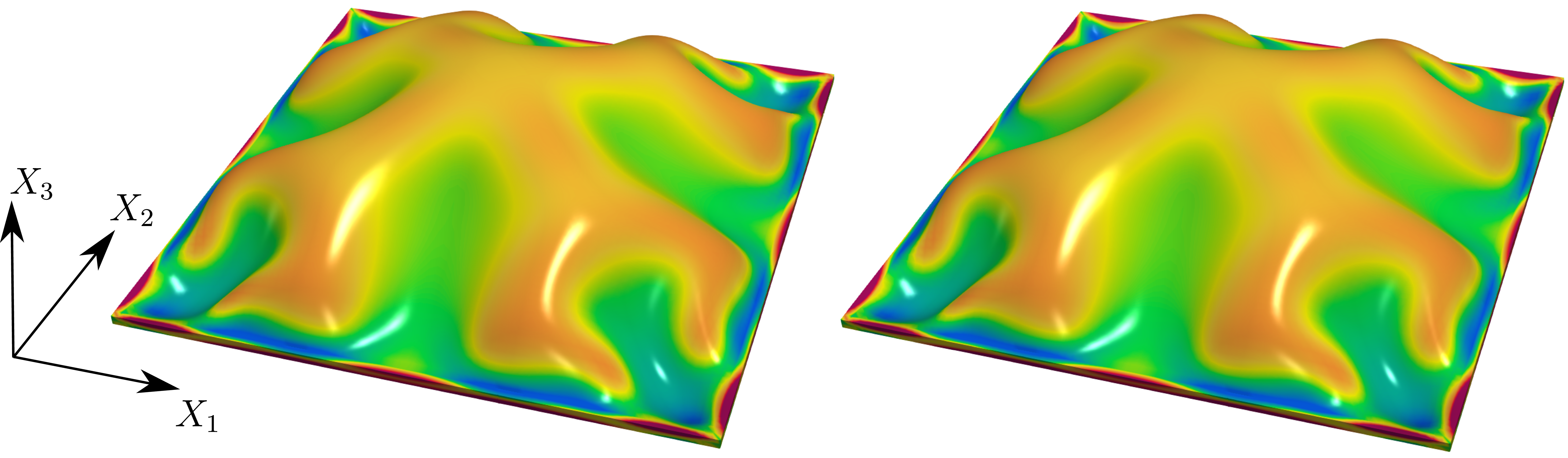}\\
			\small{(a) $\lambda=0.34$}  &  \small{(a) $\lambda=0.51$}\\
			\multicolumn{2}{c}{\includegraphics[width=0.35\textwidth]{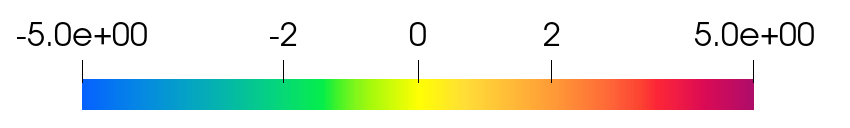}}\\
			\includegraphics[width=0.47\textwidth]{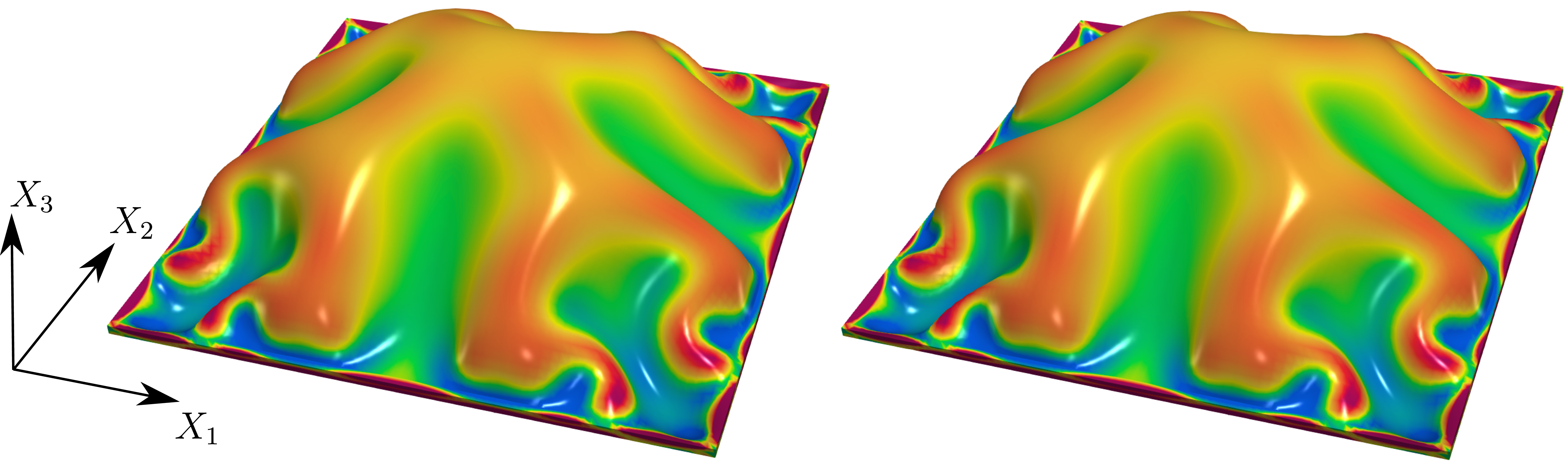}&						
			\includegraphics[width=0.47\textwidth]{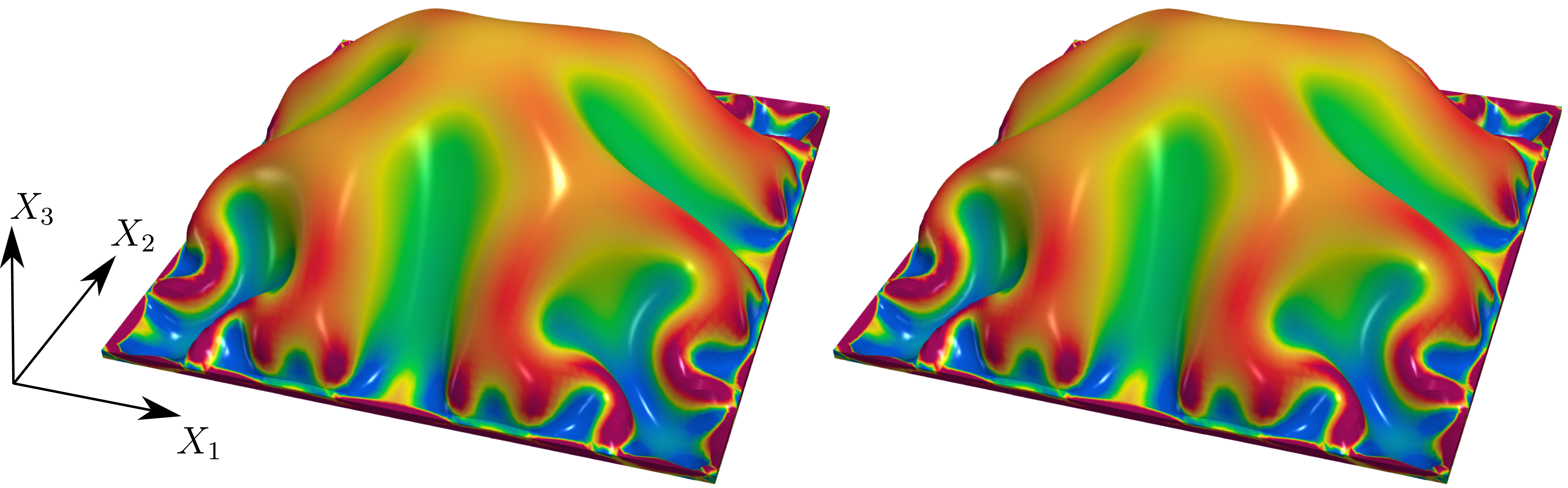}\\
\small{(c) $\lambda=0.71$} &\small{(d) $\lambda=1.00$}		
\end{tabular}

	\end{center}
	\vspace{-6mm}
	\caption{Electrically induced wrinkles with an isotropic material. Contour plot distribution of $\sigma_{13}$ for different load factors $\lambda$. In each subfigure, the left plot represents the ground truth model, while the right plot represents the results obtained by the isotropic PANN model.}\label{fig:wrinkles Isotropic}
\end{figure}

\begin{figure}[t!]
	\begin{center}
		\begin{tabular}{c c}
			\includegraphics[width=0.47\textwidth]{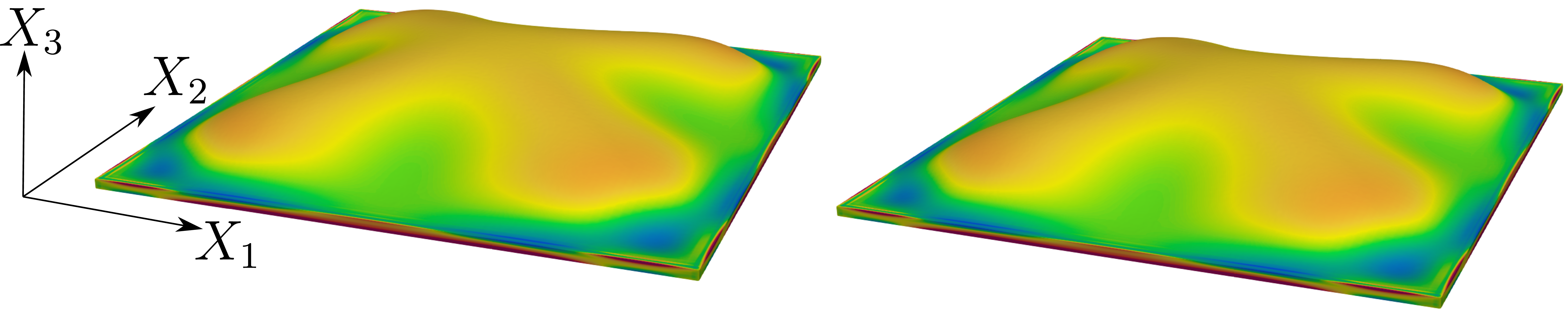}&
			
\includegraphics[width=0.47\textwidth]{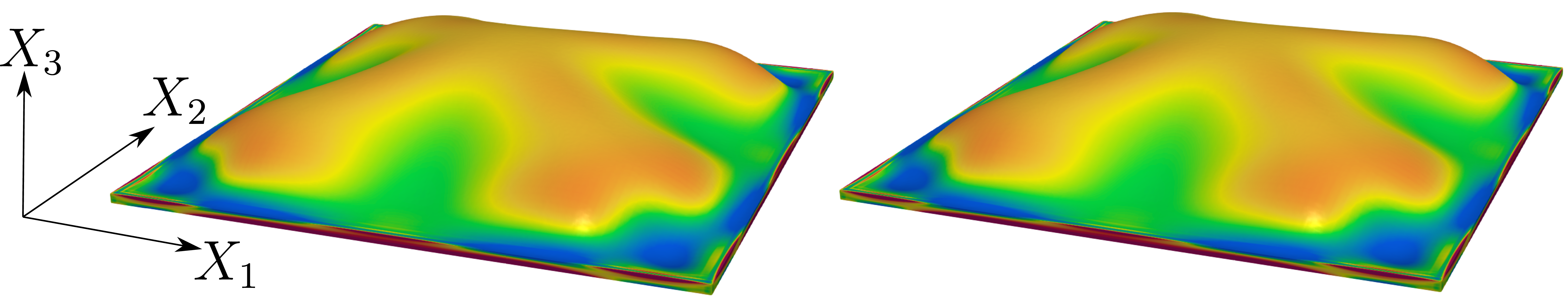}\\
			\small{(a) $\lambda=0.38$}  &  \small{(a) $\lambda=0.50$}\\
			\multicolumn{2}{c}{\includegraphics[width=0.35\textwidth]{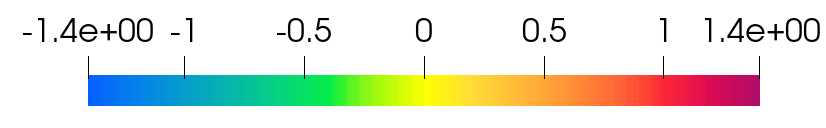}}\\
			\includegraphics[width=0.47\textwidth]{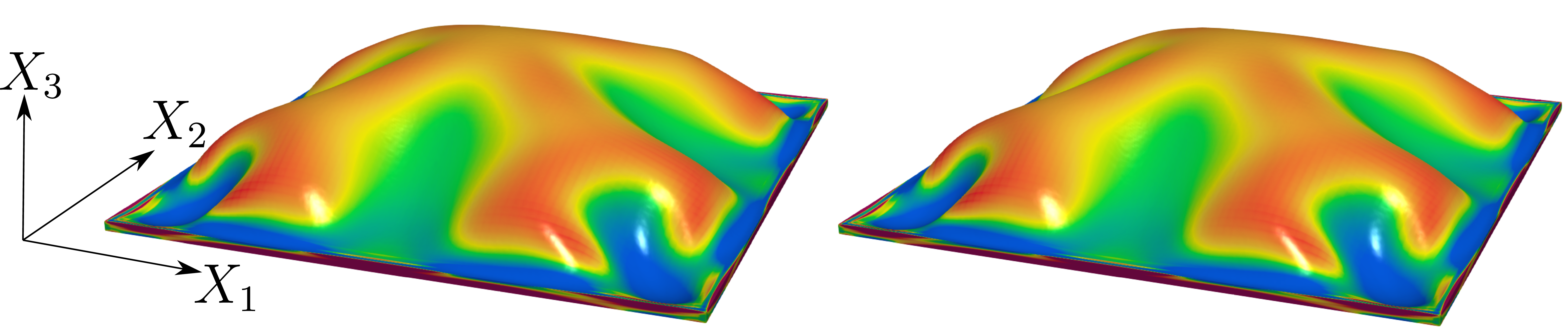}&			
			\includegraphics[width=0.47\textwidth]{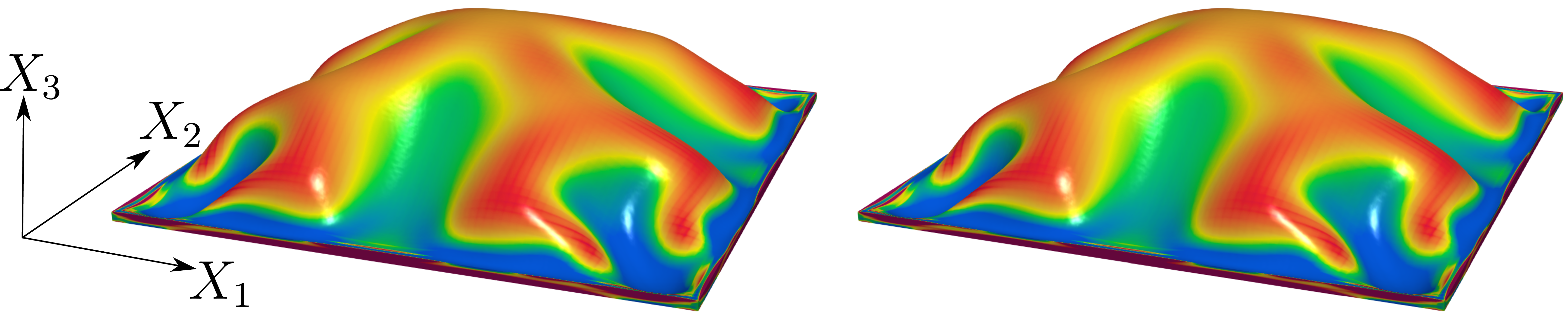}\\
			\small{(c) $\lambda=0.75$} &\small{(d) $\lambda=1.00$}
		\end{tabular}
	
	\end{center}
	\vspace{-6mm}
	\caption{Electrically induced wrinkles with a rank-one laminate material. Contour plot distribution of $\sigma_{13}/\mu_1^a$ for different load factors $\lambda$. In each subfigure, the left plot represents the ground truth model, while the right plot represents the results obtained by the transversely isotropic PANN model.}\label{fig:wrinkles ROL}
\end{figure}

In \Cref{fig:wrinkles Isotropic,fig:wrinkles ROL}, the deformed plates are visualized for different load factors for the isotropic and the rank-one laminate case, respectively. In both cases, the PANN substitute model shows an excellent performance and the simulation results received with the PANN and the ground truth models practically coincide. In particular, both the PANNs and the ground truth models yield the same wrinkling patterns.
Again, simulating such instability scenarios requires an extraordinary numerical stability. In that regard, both the isotropic polyconvex PANN as well as the transversely isotropic, non-polyconvex PANN provide extraordinary accuracy, robustness, and reliability for the highly nonlinear simulations.

\section{Conclusion}\label{sec:conc}

In the present work, the applicability of physics-augmented neural network (PANN) constitutive models for complex electro-elastic finite element analysis (FEA) is demonstrated.
For this, boundary value problems inspired by engineering applications of composite electro-elastic materials are considered. Including large electrically induced deformation and electrically induced instabilities, these are very challenging scenarios. The PANN model shows excellent performance in both prediction quality and numerical stability in all cases, rendering it applicable for the simulation of complex engineering scenarios. 

\medskip

First of all, in \cref{sec:num_1}, polyconvex isotropic PANN models are calibrated to both data generated with an analytical isotropic potential and a RVE with a spherical inclusion. The PANN model shows excellent performance in representing these materials. In \cref{sec:FEA}, FEA is carried out with the polyconvex isotropic PANN model calibrated to the analytical potential. The PANN is able to excellently predict the ground truth material behavior for a complex electrically induced deformation in \cref{sec:act_iso}, as well as electrically induced wrinkling in \cref{sec:wrinkles}. Furthermore, in all cases, the FEA carried out with the PANN shows an excellent numerical stability, which is required to simulate aforementioned scenarios.

\medskip

Furthermore, in \cref{sec:num_1}, both a polyconvex and a non-polyconvex transversely isotropic PANN model are calibrated to rank-one laminate data. The polyconvex model is not able to represent the rank-one laminate material behavior, which can be traced back to the restrictions that polyconvexity imposes on the model. Thus, a non-polyconvex PANN model is applied. This model, in turn, has an excellent prediction quality. 
However, this model does not fulfill the ellipticity condition by construction, which is important for numerical stability when applying the model in FEA. Thus, extensive investigations on the second gradients of the PANN potential are carried out, which are closely related to its ellipticity. There it becomes evident that, although the model is only calibrated on the first gradients of the electro-elastic potential, it also learns to excellently predict the second gradients of the ground truth potential. Thus, given that the ground truth model is elliptic, a non-polyconvex model can learn to be elliptic in the calibration process, thus leading to stable numerical simulations. 
The non-polyconvex model is applied in FEA including a large electrically induced actuation in \cref{sec:act_ROL} and electrically induced wrinkling in \cref{sec:wrinkles}. In both cases, the model shows an excellent prediction quality, as well as an extraordinary numerical stability.

\medskip

With an efficient and robust simulation tool for microstructured electro-elastic materials at hand, future work will be concerned with further analysis on multiscale simulation as well as topology optimisation \cite{Ortigosa_Electrodes}.


\vspace*{3ex}

\noindent
\textbf{Conflict of interest.} The authors declare that they have no conflict of interest.
\vspace*{1ex}

\noindent
\textbf{Acknowledgment.} D.K.~Klein and O.~Weeger acknowledge the financial support provided by the Deutsche Forschungsgemeinschaft (DFG, German Research Foundation, project number 492770117), the Graduate School of Computational Engineering at TU Darmstadt, and Hessian.AI. R. Ortigosa and J. Mart\'inez-Frutos acknowledge the support of  grant PID2022-141957OA-C22 funded by MCIN/AEI/10.13039/501100011033 and by ``RDF A way of making Europe". The authors also acknowledge the support provided by the Autonomous Community of the Region of Murcia, Spain through the programme for the development of scientific and technical research by competitive groups (21996/PI/22), included in the Regional Program for the Promotion of Scientific and Technical Research of Fundacion Seneca - Agencia de Ciencia y Tecnologia de la Region de Murcia. 

\vspace*{1ex}

\noindent
\textbf{Data availability.}
The authors provide access to the complete simulation data required to reproduce the results through the public GitHub repository \url{https://github.com/CPShub/sim-data}.

\appendix
\numberwithin{equation}{section} 
\section{Polyconvexity of the additional transversely isotropic invariants}\label{app:pc_add_invs}

\begin{theorem}\label{theorem:ti}
With $\structTI$ as defined in \cref{eq:struct_ti}, the twice continuously differentiable function
\begin{equation}
    f:\bbR^{3\times 3}\rightarrow \bbR,\quad \vect{A}\mapsto f(\vect{A})=\norm{\vect{A}}^2-\norm{\vect{A}\vect{G}^{\text{ti}}}^2
\end{equation}
is convex in $\vect{A}$.
\end{theorem}
\begin{proof}
To show convexity of $f$ in $\vect{A}$, we have to show the positive semi-definiteness of the Hessian of $f$, which is equivalent to the non-negativity of the following expression:
\begin{equation}\label{eq:conv_proof_1}
    \delta\vect{A}:\frac{\partial^2 f(\vect{A})}{\partial\vect{A}\partial\vect{A}}:\delta\vect{A}=\norm{\delta\vect{A}}^2-\norm{\delta\vect{A}\vect{G}^{\text{ti}}}^2\,.
\end{equation}
With $\norm{\vect{G}^{\text{ti}}}=1$, the non-negativity of \cref{eq:conv_proof_1} follows directly from the Cauchy-Schwarz inequality
\begin{equation}
    \norm{\delta\vect{A}\vect{G}^{\text{ti}}}^2\leq\norm{\delta\vect{A}}^2\norm{\vect{G}^{\text{ti}}}^2=\norm{\delta\vect{A}}^2\,.
\end{equation}
\end{proof}
\begin{corollary}
The twice continuously differentiable invariants $\IfourTI,\,\IfiveTI$ as defined in \cref{eq:ti_invar_add} are polyconvex, which immediately follows by setting $\vect{A}=\{\vect{F},\,\vect{H}\}$ in \cref{theorem:ti}. 
\end{corollary}
\begin{theorem}
With $\structTI$ as defined in \cref{eq:struct_ti}, the invariant
\begin{equation}
    \IfiveTI:\bbR^{3}\rightarrow \bbR,\quad \vect{d}_0\mapsto \IfiveTI(\vect{d}_0)=\tr\left(\vect{d}_0\otimes\vect{d}_0\right)-\tr\left(\left(\vect{d}_0\otimes\vect{d}_0\right)\structTI\right)
\end{equation}
is convex in the components of $\vect{d}_0$.
\end{theorem}
\begin{proof}
The Hessian 
\begin{equation}
\begin{aligned}
        \frac{\IfiveTI(\vect{d}_0)}{\partial \vect{d}_0 \partial \vect{d}_0}=2(\bI- \structTI)\,.
        \end{aligned}
\end{equation}
is positive semi-definite \cite[Lemma~A.12]{Schroeder2003}.
\end{proof}

\renewcommand*{\bibfont}{\footnotesize}
\printbibliography

\end{document}